\documentclass{statsoc}

\usepackage[a4paper]{geometry}
\usepackage{graphicx}
\usepackage[textwidth=8em,textsize=small]{todonotes}
\usepackage{amsmath}
\usepackage{natbib}
\usepackage{amssymb}
\usepackage{url}
\usepackage{mathtools}
\usepackage[linesnumbered,ruled,vlined]{algorithm2e}
\usepackage{multirow}

\usepackage{fancyheadings}

\newcommand{\nats}{\mathbb{N}}
\newcommand{\vecJ}{\operatorname{vec}}

\title[The economic response to COVID-19 using GNARX Models]{Quantifying the economic response to COVID-19 mitigations and death rates via forecasting Purchasing Managers' Indices using Generalised Network Autoregressive models with exogenous variables}

\author[G.\ P.\ Nason and J.\ L.\ Wei]{Guy P. Nason and James L. Wei\thanks{{\em Address:} Dept of Maths, 180 Queen's Gate, London, SW7 2AZ, UK {\tt james.wei19@ic.ac.uk}}}
\address{Imperial College, London, UK.}

\begin{document}

\thispagestyle{fancy}
\lhead{{\small \tt [To be read before The Royal Statistical Society at the Society’s 2021 annual conference held in Manchester on Wednesday, September 8th, 2021, the President, Professor Sylvia Richardson, in the Chair]}}

\begin{abstract}
Knowledge of the current state of economies, how they respond to  COVID-19 mitigations and indicators,
and what the future might hold \textcolor{black}{for them} is important.
We use recently-developed generalised network autoregressive (GNAR) models, using trade-determined networks, to model and forecast the
{\color{black} Purchasing Managers' Indices for a number of countries. We use
networks that link countries where the links themselves, or their weights,
are determined by the degree of export trade between the countries.}
We extend \textcolor{black}{these models} to include node-specific time series exogenous variables (GNARX models),
using this to incorporate COVID-19 mitigation stringency indices and COVID-19 death rates into our analysis.
The highly parsimonious GNAR models considerably outperform vector autoregressive models in terms of mean-squared forecasting error and our GNARX models themselves outperform GNAR ones.
Further mixed frequency modelling predicts the extent to which that the UK economy will be affected by harsher, weaker or no interventions.
\end{abstract}

\section{Introduction}
The severe impact of COVID-19 on global economic conditions cannot be understated, with UK real 
gross domestic product (GDP)
contracting by a record 20\% in the second quarter (Q2) of 2020 alone.
Timely estimates of economic activity through the current turbulent period and forecasts of the impact of government virus mitigation strategies will be crucial for evaluating the costs and benefits of different policy interventions.

This article introduces new methods
to forecast Purchasing Managers' Indices (PMIs) during the COVID-19 pandemic  and
quantifies the potential impact of government virus mitigation strategies.
PMIs are survey-based leading indicators of the direction of economic activity.
We choose to model PMIs as they are used extensively for the nowcasting of slower-moving headline measures of economic activity, such as GDP.
See, for example,~\citet{d2012survey}. 
Our analysis builds on the class of generalised network autoregressive (GNAR) models \citep{knight2016,knight2019generalised},
which model multivariate time series in a highly parsimonious fashion by associating the dependencies between constituent series with an underlying network structure. Hence, GNAR models can characterise a rich diversity of dependence structures, whilst
incurring a lower risk of overfitting when compared to classical vector autoregression (VAR) models, translating
into superior forecasting performance.
Indeed, our work below demonstrates that GNAR models outperform VAR models by 33\% based on mean-squared forecasting error for
forecasting PMIs from historical  PMI data.

A goal here is to model PMIs in terms of historical PMIs,
{\color{black} PMIs from neighbouring countries via our network model and}
 (i) stringency measures of government intervention policies enacted to mitigate
 the spread of COVID-19 and (ii) the monthly number of confirmed COVID-19 deaths per million.
 {\color{black} Incorporating information about past PMIs from neighbouring countries via a network
 structure improves
 forecasting as it is extra relevant information. Our network model uses countries' PMIs on the
 network nodes (vertices)
 and the strength of link between two countries (nodes) increases with increasing
 quantities of exports between the two. Figure~\ref{fig:GNARgraph} (below) shows
 an example of one of the networks we use. In addition,}
 the incorporation of multiple exogenous time-dependent
 node-specific regressors is both new to GNAR modelling and
 practically essential for properly understanding the potential influence of (i) and (ii).
 Our new extension is named GNARX, where X indicates incorporation of  fully time-dependent
 exogenous variables, which is \textcolor{black}{in} line with similar naming conventions, such as VAR to VARX.

We structure our paper as follows.
Section~\ref{sec:math-setup} introduces the new GNARX specification and
characterises it as a specially-constrained linear regression problem. Asymptotic properties of the corresponding estimators follow from this. Section~\ref{sec:experiments} provides details of our multivariate time series and network data and
discusses two experiments: the first compares out-of-sample forecasting performance of 
GNAR models (i.e.\ without using external regressors) with a VAR model, and the second evaluates the GNARX model estimates when the stringency indices and COVID-19 death rates are incorporated.
Section~\ref{sec:scenario}
uses our new GNARX modelling ability to analyse UK economic conditions under three different scenarios pertaining to either tightening, easing or
unchanged intervention policies over the next six months.
Section~\ref{sec:conclusion} concludes with a brief discussion of the
experimental results and avenues for future research.

\section{The GNARX mathematical model and notation}\label{sec:math-setup}

{\em Definition.}
Let our multivariate {\color{black} real-valued} time series {\color{black} $\{ Y_{i, t}\}$} be observations collected on variable $i \in \mathcal{K} = \{ 1, \ldots, N\}$ at
time $t=1, \ldots, T \in \nats$, where $N \in \nats$ is the number of variables.
We sometimes write the multivariate series as the
vector time series {\color{black} $\{ \mathbf{Y}_t\} = (Y_{1, t}, \ldots, Y_{N, t} )^\intercal$}, where $\intercal$ denotes transpose, for
$t=1, \ldots, T$. Now let  $\mathcal{G}$ be a graph consisting of a set of
vertices/nodes, $\mathcal{K}$, and an edge set denoted by $\mathcal{E}$.
Let $i\leftrightsquigarrow j\in\mathcal{E}$ denote the presence of an undirected edge between nodes $i$ and $j$ and $i\rightsquigarrow j\in\mathcal{E}$ denote the existence of a directed edge from $i$ to $j$, for $i, j \in \mathcal{K}$.
For network time series a graph vertex $i\in\mathcal{K}$ corresponds to time series  variable
{\color{black} $\{ Y_{i,t} \}$}. 
Let {\color{black} $\{ X_{h,i,t} \}$} be the $h$-th {\color{black} real-valued stationary}
exogenous node-specific regressor series for node $i$ at time $t$, for $h= 1, \ldots, H \in \nats$,
the number of such regressors.

We introduce the GNARX$(p, \mathbf{s}, \textbf{p}')$ model,
where $( p, \mathbf{s},\textbf{p}' ) \in\mathbb{N}\times\mathbb{N}_0^p\times\mathbb{N}_0^H$, by
\begin{equation}\label{eqn:GNARX}
    Y_{i,t}=\sum_{j=1}^{p}\left(\alpha_{i,j}Y_{i,t-j} + \sum_{r=1}^{s_j}  \beta_{j,r} \sum_{q\in\mathcal{N}^{(r)}(i)}\omega_{i,q}Y_{q,t-j}\right)+
        \sum_{h=1}^H
    \sum_{j'=0}^{p'_h}
    \lambda_{h,j'} X_{h,i,t-j'}+u_{i,t},
\end{equation}
where
\textcolor{black}{the parameters $\alpha_{i,j}, \beta_{j,r}, \lambda_{h,j'}\in \mathbb{R}$ for all $i, j, r, h, j'$,}
$p$ is the autoregressive order of the model and also maximum order of neighbour time lags,
$p'_h$ (the $h$th element of $\mathbf{p}'$) denotes the maximum lag of the $h$th exogenous regressor involved in the model,
and $s_j$ (the $j$th element of $\mathbf{s}$, which we sometimes write as $[s]$) is the maximum stage of neighbour dependence for time lag $j$. For example, $s_1=2$ implies that nodes are dependent on the first lags of its first and second stage neighbours in the  network $\mathcal{G}$.
{\color{black} The set $\mathcal{N}^{(r)}(i)$ is the $r$-th stage neighbourhood set 
of node $i \in \mathcal{G}$ defined by
\begin{equation}
\mathcal{N}^{(r)} (i) = \mathcal{N} \{ \mathcal{N}^{(r-1)} (i) \} / [ \{ \cup_{q=1}^{r-1} \mathcal{N}^{(q)} (i) \}
	\cup \{ i \} ],
\end{equation}
for $r= 2, 3, \ldots$, $\mathcal{N}^{(1)} = \mathcal{N}$, where
$\mathcal{N}(A) = \{ j \in \mathcal{K}/A : i  \rightsquigarrow j ; i \in A\}$}.
The $\omega_{i,q}\in[0,1]$ in~\eqref{eqn:GNARX} are network connection weights, typically set as the inverse of some prior notion of distance between nodes, as in~\cite{knight2019generalised} using
a similar approach to that used in the lifting scheme described by~\citet{jansen2009multiscale}.
{\color{black} Sometimes we consider the \mbox{global-$\alpha$} model where
$\alpha_{i, j} = \alpha_j$, i.e.\ use the same $\{ \alpha_j \}_{j=1}^p$ sequence for each node. 
Consequently,  the \mbox{local-$\alpha$} model is where the $\alpha_{i, j}$ sequence is
different for each node $i$.}

 The GNARX model cannot be captured by the standard GNAR model of~\citet{knight2019generalised} by incorporating the exogenous regressors
 into $\mathbf{Y}_t$, as $p$ and $p'_h$ may differ. The network vector autoregression model proposed by~\citet{zhu2017network} does incorporate dependence on node-specific covariates, but unlike model~\eqref{eqn:GNARX},  they do not vary with time {\color{black} and have limitations on other parameters/quantities
 in the network. Full specifications for both of these models can be found in  the Supplementary
 Material Section~1.}
 
{\color{black} GNARX models share the same advantages with respect to missing data as GNAR models,
especially when compared to vector autoregression-based approaches. This is because, e.g., VAR models
try to estimate parameters that link one specific variable to another. If, for example, one of those variables
disappears for a significant block of time, then it can be difficult to estimate the parameter associated
with any of its pairings. By contrast, in a GNARX model, a variable pair is just a nearest
neighbour and every such pair in the data contributes to the `nearest neighbour' parameter,
one of the $\beta_{j ,r}$. So, even if one node suffers from significant missingness, there will be usually
many other pairs so that the respective parameter is still well-estimated. More details on this, particularly
on how this works for $r$th-stage neighbours, can be found in Section~2.6 of~\citet{knight2019generalised}. }

{\em GNARX Parsimony.}
The GNARX$( p, \mathbf{s}, \textbf{p}')$ model with a single exogenous regressor ($H=1$) contains a total of $M=Np+\sum_{j=1}^ps_j+p'+1$ parameters.
By comparison, the corresponding vector autoregression model with exogenous variables (VARX) containing $p$ lags of the modelled series
{\color{black} $\{ \mathbf{Y}_t \}$}, and $p'$ lags of the exogenous series {\color{black}  $\{ \mathbf{X}_{t} \}= \{(X_{1,t},...,X_{N,t})^\intercal \}$}, can be given in the reduced form by 
\begin{equation}\label{eqn:VAR-form}
    \mathbf{Y}_t=\phi_1\mathbf{Y}_{t-1}+ \cdots +\phi_p\mathbf{Y}_{t-p}+\Lambda_0\mathbf{X}_{t}+\Lambda_1\mathbf{X}_{t-1}+ \cdots +\Lambda_{p'}\mathbf{X}_{t-p'}+\mathbf{u}_t,
\end{equation}
which contains $P = N^2 (p+p' + 1)$  parameters.
The parsimonious nature of GNARX compared to VARX is clear when parameter growth is $\mathcal{O} (N)$ for the former and
$\mathcal{O} (N^2)$ for the latter. Indeed,
it has been suggested by~\citet{bernanke2005measuring} that vector autoregressions are not often used
for macroeconomic datasets that contain more than six to eight time series.
The parsimony of GNARX, combined with the flexibility to model dynamic networks representing multiple covariates, ensures our methodology is readily applicable to many applications involving the forecasting of multivariate time series.

\textcolor{black}{{\em GNARX Stationarity.} Under the assumption of stationarity of $\left\{X_{h,i,t}\right\}$ for all $h, i$ and parameter constraints
\begin{equation}
\sum_{j=1}^p \left(|\alpha_{i,j}|+\sum_{c=1}^C\sum_{r=1}^{s_j}|\beta_{j,r,c}|\right)<1\;\;\; \forall i\in 1, \ldots, N,
\end{equation}
the GNARX process $\left\{\mathbf{Y}_t\right\}$ is stationary. We prove this in Section~2
of the Supplementary Material. 
} For  simplicity, we will assume $H=1$ for the remainder of this section only, replacing $X_{h,i,t}$ with $X_{i,t}$ and $\textbf{p}'$ with $p'$. Extending the following \textcolor{black}{results} to multiple external node-specific regressors is straightforward.

{\em Precise model definition and theoretical statistical properties.}
With the $H=1$ assumption, the GNARX$(p, \mathbf{s}, p')$ model can be expressed as a vector autoregression of the form~\eqref{eqn:VAR-form},
but where coefficient matrices are restricted so that $\phi_k=\text{diag}\left\{\alpha_{i,k}\right\}+\sum_{r=1}^{s_k}\beta_{k,r}W^{(r)}$, $[W^{(r)}]_{l,m}=w_{l,m}\mathbb{I}\left\{m\in\mathcal{N}^{(r)}(l)\right\}$, and $\Lambda_{k}=\lambda_{k}I_N$. In matrix form, the model can be written as
$Y = BZ + U$,
where $Y=[\mathbf{Y}_{p^*+1}, \ldots, \mathbf{Y}_{T}]$, $p^*= \max{(p, p')}$, $B=[\phi_1, \ldots ,\phi_p, \Lambda_0, \Lambda_1, \ldots ,\Lambda_{p'}]$,  $Z=[\mathbf{Z}_{p^*}, \ldots,\mathbf{Z}_{T-1}]$, $\mathbf{Z}_{t}^\intercal =[\mathbf{Y}^\intercal_t, \ldots, \mathbf{Y}^\intercal_{t-p+1},
 \mathbf{X}^\intercal_{t+1}, \ldots, \mathbf{X}^\intercal_{t-p'+1}]$, and $U=[\mathbf{u}_{p^*+1},...,\mathbf{u}_T]$. Additional regressors can be incorporated by concatenating the corresponding coefficient and data matrices to $B$ and $\mathbf{Z}_t$, respectively.

Following \citet{lutkepohl2005new}, the coefficient restrictions on $B$ that result from the GNARX assumptions are
$\underset{P \times1}{\vecJ(B)}= R \gamma$,
where 
$R$ is the $P\times M$ model matrix, defined in  the Supplementary Material Section~4,
$\underset{M\times 1}{\gamma} =(\gamma_1^\intercal, \ldots,\gamma_p^\intercal, \lambda_0, \ldots, \lambda_{p'})^\intercal$,
with GNAR parameters that we are interested in given by
$\gamma_k=(\alpha_{1,k}, \ldots,\alpha_{N, k},\beta_{k,1}, \ldots,\beta_{k,s_k})^\intercal$
and all elementary vectors are of length $N$.  With the global alpha assumption,
$\alpha_{i,k}=\alpha_k$ for all $i=1, \ldots, N$, we have $\gamma_k=(\alpha_k,\beta_{k,1}, \ldots, \beta_{k,s_k})^\intercal$
and $A = L$.

Given the VAR matrix representation and linear coefficient restrictions from above,
\citet{lutkepohl2005new} proposes the feasible generalised least squares
{\color{black} (FGLS)} GNAR parameter estimator to be
\begin{equation}\label{eqn:FGLS}
    {\hat\gamma}^{FGLS}=\left\{ R^\intercal \left(Z Z^\intercal \otimes{\hat\Sigma}_u^{-1}\right)R\right\}^{-1}R^\intercal \left(Z\otimes{\hat\Sigma}_u^{-1}\right) \vecJ(Y),
\end{equation}
where the maximum likelihood estimator of $\Sigma_u$ is 
\begin{equation}\label{eqn:Sigma-hat}
    {\hat\Sigma}_u=T^{-1}\left(Y-\hat{B}Z\right)\left(Y-\hat{B}Z\right)^\intercal,
\end{equation}
and where $\hat{B}$ satisfies $\vecJ (\hat B)=R\hat\gamma$ and $\hat\gamma$ is the ordinary least squares
estimator of $\gamma$. In the case of missing observations, the corresponding residuals are set to zero. \textcolor{black}{Under the conditions of stationarity of $\left\{\textbf{Y}_t\right\}_{t=1, \ldots,T}$ and $\left\{\textbf{X}_t\right\}_{t=1, \ldots, T}$, consistency} and asymptotic
normality are proved in \textcolor{black}{Section}~2 of the Supplementary Material.

\section{Purchasing Managers' Indices: the data and forecasting experiments}\label{sec:experiments}

\subsection{Input data}\label{sec:datasets}

\subsubsection{Purchasing Managers' Indices}

IHS Markit Composite Purchasing Managers' Indices (PMIs) are computed by aggregating the views of senior purchasing executives from around 400 companies on a wide range of economic variables pertinent to the manufacturing and services sectors. Composite PMIs can take values between zero and $100$, with a reading above $50$ indicating economic expansion.
\textcolor{black}{All PMI series have been seasonally adjusted using the X-12 ARIMA method; full}
 details of the survey methodology are available at \url{https://ihsmarkit.com/products/pmi.html}. Currently, monthly composite PMI data are available for thirteen major economies, which are all included in our analyses.
The longest series are from the UK, Italy and Germany, which start from Jan-98,
and the shortest is from Australia, which starts from May-16.

PMI series are diffusion indices, which treat a fixed set of component economic indicators effectively as dummy variables that represent whether the trend in the component is positive or negative. For PMIs, these component series are the survey responses of individual purchasing managers. It is therefore reasonable to assume that diffusion indices are stationary, given the well-documented cyclical nature of macroeconomic time series, see \citet{zarnowitz1991business}. Moreover, Augmented Dickey-Fuller (ADF) unit root tests on the in-sample period between Jan-98
and Dec-17 result in
\mbox{$p$-values} above $0.1$ for all series, indicating stationarity for the time being, against the alternative hypothesis of first-order
non-stationarity.
Broadly, the series also do not reject the null of second-order stationarity according to the wavelet-based tests of \cite{Nason13,Nason:locits}.

\subsubsection{Network construction}\label{sec:netcon}
GNAR and GNARX models require us to specify a network structure to associate with the PMI dataset.
The time series for a particular country's PMI corresponds to a single vertex in the network --- one vertex for each country.
Rather than attempting to learn the network structure indirectly from the time series as in
\cite{leeming:phd} or \cite{knight2019generalised},
we suggest that a construction based on the flow of exports between countries
that provides a reasonable reflection of the association between component series. 
This rests on the assumption that a country's business confidence is partly dependent on the level of economic activity of its major export partners.
Since (i) model orders are chosen by BIC optimisation,
(ii) network parameters are highly significant and
(iii) forecast performance is improved when compared to models that ignore network information, these all suggest that our choice of networks is
reasonable.
Evaluating performance by forecast accuracy is honest and exacting.

\begin{figure}
    \centering
    \includegraphics[width=0.45\textwidth]{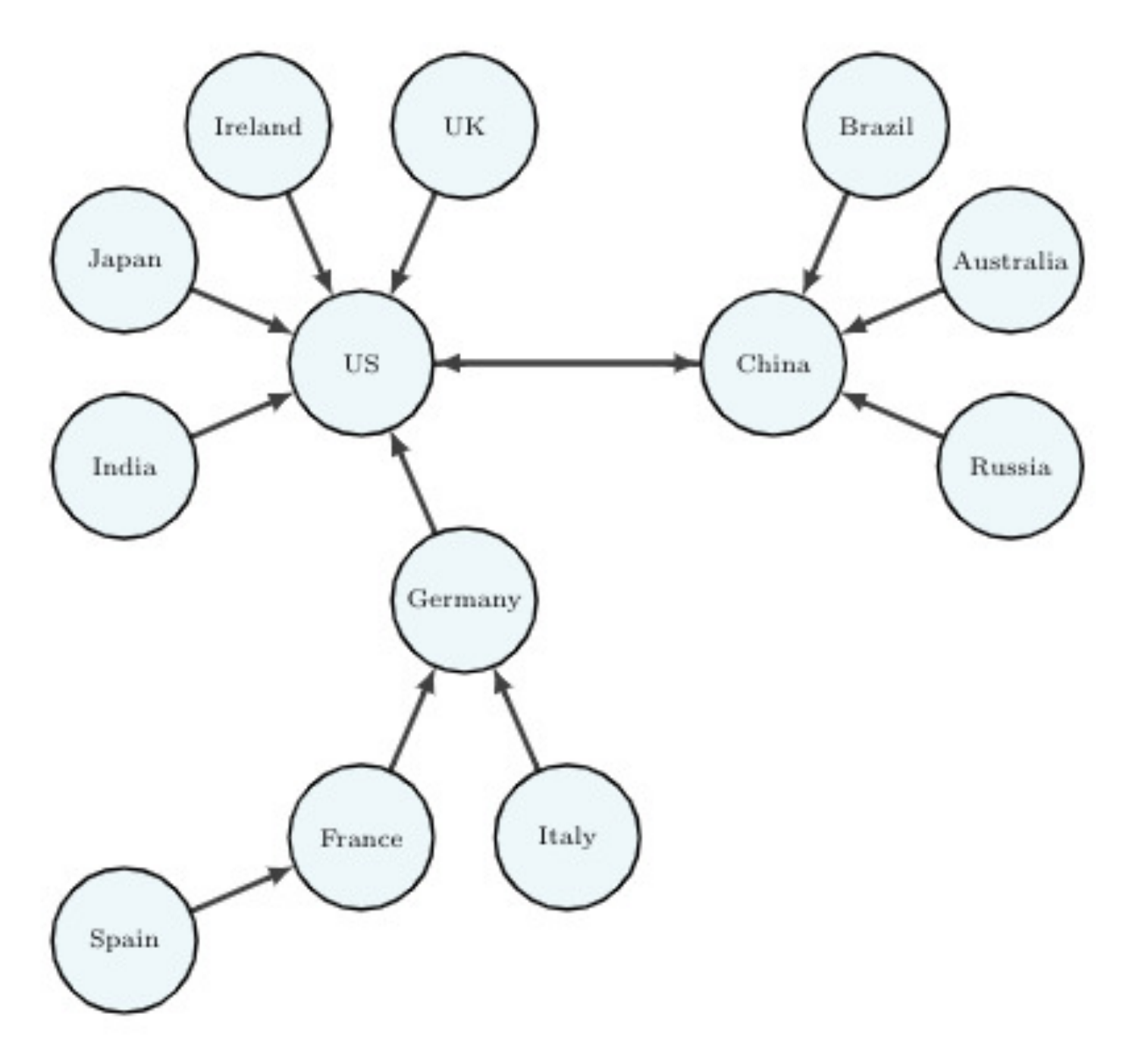}
    \caption{Nearest neighbour network based on 2019 export data}
    \label{fig:GNARgraph}
\end{figure}

Our experiments evaluate two different networks:
\begin{enumerate}
    \item \textbf{Fully-connected trade network}. Each country is connected to every other country.
    Edge weight $w_{i,j}$ is set to be the export quantity between countries $i$ and $j$.
    The weights are then normalized so that $\sum_j w_{i,j}=1$ for all $i$. In this case, $s_j\in\left\{0,1\right\}$ for all $j$.
    
    \item \textbf{Nearest neighbour trade network}. (Figure~\ref{fig:GNARgraph})
    Each country has a single outgoing edge to the country that it exports the most too.
    
    \end{enumerate}
Export data was obtained from the UN Comtrade International Trade Statistics Database (\url{https://comtrade.un.org/data/}), using the latest available observations.

\subsubsection{Exogenous variables}

The GNARX model's advantage is that it permits the multivariate time series $\textbf{Y}_t$ to depend on
vertex-specific external time series regressors $\textbf{X}_t$; two series in our examples.
The first, $X_{1, i, t}$, are differenced country stringency indices \citep{hale2020oxford} for country $i$, which aim to quantify
the severity of COVID-19 mitigation policies. These are composite indices that take values between zero and $100$, with higher values reflecting stricter  measures. The stringency indices are mainly related to containment policies, such as the closing of schools, workplaces and public events. The full list of constituent indicators is available at \url{https://github.com/OxCGRT/covid-policy-tracker/blob/master/documentation}. Values of the stringency indices before the COVID-19 outbreak are automatically set to zero, rather than treated as missing.

The second {\color{black} regressor}, $X_{2, i, t },$ {\color{black} is the first difference} of country $i$ new COVID-19 death rates, which may impact economic conditions via channels beyond the previously discussed interventions. For example, consumers may extrapolate a rise in \textcolor{black}{the} confirmed number of COVID-19 deaths into the future, leading to consumer precautionary saving in anticipation of greater future lockdown measures. \textcolor{black}{As higher death rates would likely also be associated with stricter mitigation policies, omitting death rates would likely lead to  model errors being correlated with the stringency index regressor, resulting in inconsistent parameter estimates}.
Our experiments use the daily number of confirmed COVID-19 deaths per million in a given country, averaged for each month \citep{owidcoronavirus}, available at \url{https://ourworldindata.org/covid-deaths}. \textcolor{black}{The sample correlation between the differenced stringency indices and the differenced COVID-19 death rate of the same country is only $0.46$, suggesting that multicollinearity is not a significant problem.
The correlation only rises modestly to $0.57$ for the non-differenced series.}

\begin{figure}[h]
    \centering
    \resizebox{0.6\textwidth}{!}{\includegraphics{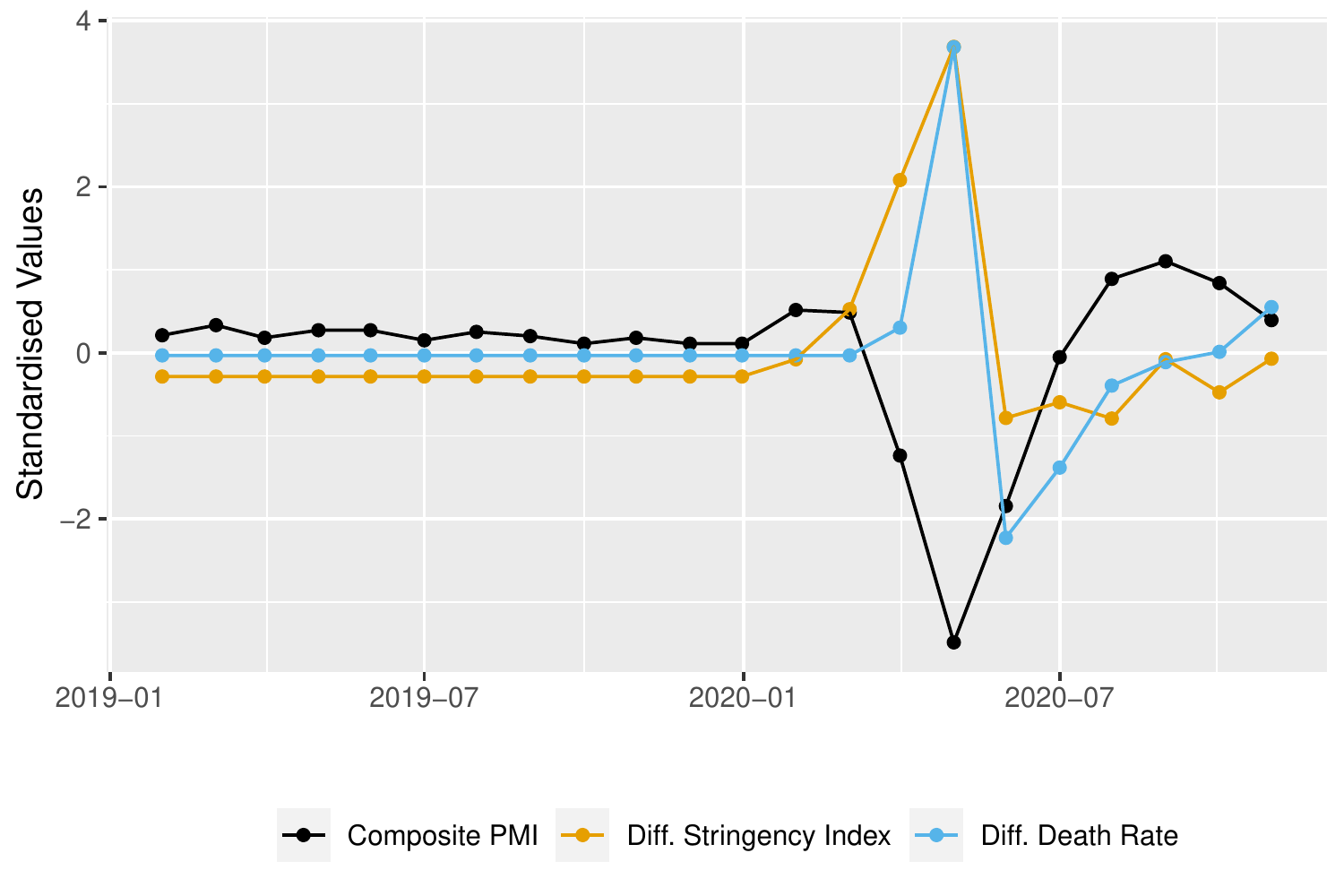}}
    \caption{\textcolor{black}{Plot of UK Composite PMI together with UK stringency index and death rate covariates, where all variables have been standardised to  have zero mean and unit variance.}
    \label{fig:UKcovariates}}
\end{figure}

\textcolor{black}{Figure~\ref{fig:UKcovariates} shows the UK PMI and the two exogenous variables starting from Jan~2019, where all series have been standardised to have zero mean and unit variance so that they can be compared on the same axes. The most prominent feature of the UK PMI series is the large six and eight standard deviation declines during Mar and Apr 2020.
This fall in purchasing manager confidence coincides with the first UK lockdown (rise in stringency index), but also reflects broader worsening of global macroeconomic conditions caused by the COVID-19 pandemic, see, e.g.,~\cite{long2020world}.}

\subsection{GNAR forecasting performance}\label{sec:experiment1}
Here, model order is selected to minimise the Bayesian Information Criterion (BIC) subject to an order limit of \textcolor{black}{$p=12$. For computational feasibility, we adopt a stagewise approach, where $p$ is first selected to minimise BIC, followed by $s_j$ for $j=1,...,p$}.
\textcolor{black}{This allows us to capture network effects lasting up to a \textcolor{black}{year} in duration}\textcolor{black}{. For the exogenous regressors in later experiments, we instead impose a lower order limit of three months, due to the lack of nonzero observations. In that case, $p_h'$ for $h=1,...,H$ are optimised at the same time as $s_j$.} All PMI observations up to Dec-17 are treated as the in-sample period for model order selection and estimation of both GNAR and VAR models. \textcolor{black}{We refer readers to
the Supplementary Material Section~7 for the results if model orders were instead selected to minimise forecasting errors in a subset of the in-sample period. Finally, we provide simulation evidence for the consistency of the model selection procedure using stagewise BIC optimisation for GNARX in the Supplementary Material Section~8,} {\color{black} which shows similar performance to `global search' BIC.}

Unlike GNAR, VAR models do not easily handle missing values. Specifically, GNAR models deal with missingness
by modifying connection weights \citep[Section~2.6]{knight2019generalised} and
because of this advantage our GNAR model can make use of the {\em entire} unbalanced panel starting from Jan-98.
If Australia were included, then
the in-sample period for the VAR model would contain only 20 months and, for this reason, we exclude Australia from this analysis.

Table~\ref{tab:experiment1a} shows the composite PMI forecasting performance of four different
{\color{black} stagewise-BIC selected} GNAR models, a VAR model,
\textcolor{black}{a naive forecast
(predicting next from current) and univariate AR models. The latter involve a separate AR model estimated for each country, with order determined by BIC optimisation. For all but three countries, AR$(1)$ was selected. However, for fair comparison with the GNAR and VAR models where autoregressive order is the same for every series, AR$(1)$ was used for all countries. It is worth noting that AR$(1)$ is identical to GNAR$(1,[0])$
with \mbox{local-$\alpha$}.}
{\color{black} We have used mean-squared forecast error (MSFE) in our article, which computes the empirical
average of the squared differences between our point forecasts and the eventual out-turn.}
The lowest forecast error is attained by the local-$\alpha$ GNAR model with the fully-connected network.

\begin{table}
\caption{\label{tab:experiment1a}Out-of-sample composite PMI forecast performance between Jan-18 and Oct-20 using GNAR,
VAR and naive forecasting. MSFE=Out-of-sample mean-squared forecasting error; SE = standard error}
\centering
\framebox{%
\begin{tabular}{lrrr}
\hline
Model & Model order & Parameters & MSFE (SE) \\ 
\hline
Global-$\mathbf{\alpha}$ GNAR, fully-connected net & \textcolor{black}{2, [1, 1]} & \textcolor{black}{4} & \textcolor{black}{41.9 (8.2)}\\
Global-$\mathbf{\alpha}$ GNAR, nearest-neighbour net & \textcolor{black}{2, [1, 1]} &  \textcolor{black}{4} & \textcolor{black}{40.7 (7.4)}\\
Local-$\mathbf{\alpha}$ GNAR, fully-connected net & 1, [1] & 13 & {\color{black} 38.4} (6.8)\\
Local-$\mathbf{\alpha}$ GNAR, nearest-neighbour net & 1, [1] & 13 & 39.3 (7.1)\\
\\
VAR & 2 & 288 & 57.3 (9.3)\\
{\color{black} AR} & {\color{black} 1} & {\color{black} 12} & {\color{black} 39.1 (7.0)}\\
Naive forecast & - & -  & 40.6 (7.2)\\
\hline
\end{tabular} }
\end{table}
\textcolor{black}{Furthermore, the} mean-squared forecast error of all four GNAR models are substantially lower than that of the VAR model, indicating
overfitting in the latter. The local-$\alpha$ GNAR\textcolor{black}{$(1, [1])$} model contains only thirteen parameters, whilst the VAR($2$) model contains $2N^2=288$ parameters.  Our results indicate that when a  simple model is appropriate, BIC optimisation for VAR may still lead to overfitting, but the same does not hold for GNAR models. Indeed, only three of the four GNAR models outperformed the naive forecast \textcolor{black}{and one of the two \mbox{local-$\alpha$}
GNAR models outperformed the univariate AR models}, reflecting the difficulty in forecasting
economic indicators  using only  past values.
{\color{black} On the other hand, the best GNAR model {\em does}
improve the forecast performance and a few percent improvement can translate into substantial
economic value when one is dealing with GDP-like magnitudes, see Section~\ref{sec:scenario}.}


Table~\ref{tab:exp1Estimates} shows estimates for the local-$\alpha$ GNAR\textcolor{black}{$(1, [1])$} model, \textcolor{black}{with} $p$-values computed \textcolor{black}{using} HC2 robust standard errors, \textcolor{black}{which are heteroskedasticity-consistent estimators first proposed by \cite{mackinnon1985some} that adjust residuals using the projection matrix leverage values. These were calculated with} the \textbf{sandwich} package of~\citet{zeileis2004econometric}. \textcolor{black}{In this experiment, the use of robust standard errors produces more conservative $p$-values.}

\begin{figure}[h]
    \centering
    \resizebox{0.6\textwidth}{!}{\includegraphics{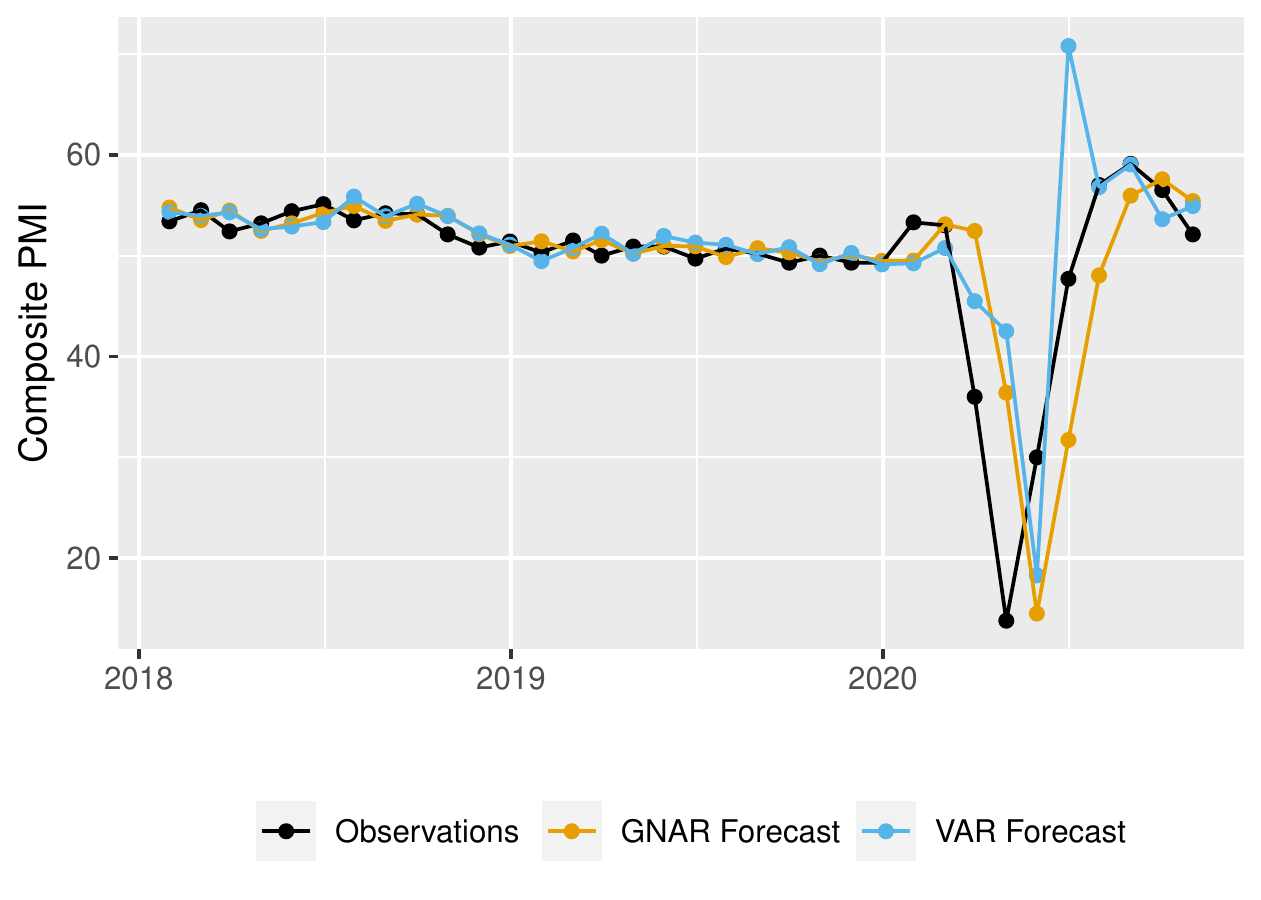}}
    \caption{Out-of-sample one-step-ahead rolling composite UK PMI forecasts using
    the local-$\alpha$ GNAR$(1, [1])$ and the VAR$(2)$ models.
    \label{fig:UK}}
\end{figure}

Figure~\ref{fig:UK} plots the UK out-of-sample one-step-ahead rolling forecasts for this best-performing model, along with those
from the VAR$(2)$ model (see  Supplementary Material, Section~6, for other countries).
The VAR forecasts appear to underestimate the initial PMI decline at the start of the crisis period, but
substantially overestimate the recovery.
This leads to substantially higher mean-squared forecast error during the COVID-19 crisis period starting from Feb-20, relative to  GNAR.
\begin{table}
\caption{\label{tab:exp1Estimates}Local-$\alpha$ GNAR$(1, [1])$ parameter estimates (to two decimal places) using the fully-connected network. 
Parameter \mbox{$p$-values} were all $< 0.001\%$.}
\centering
\fbox{%
\begin{tabular}{lr|lr|lr}
\hline
Parameter & Estimate & Parameter & Estimate & Parameter & Estimate \\ 
\hline
$\alpha_{{\rm USA},1}$ & 0.93 &  $\alpha_{{\rm Japan},1}$ & 0.87 & $\alpha_{{\rm Germany},1}$ & 0.90 \\
$\alpha_{{\rm UK},1}$ & 0.90 &  $\alpha_{{\rm Italy},1}$ & 0.89 & $\alpha_{{\rm China},1}$ & 0.88 \\
$\alpha_{{\rm Spain},1}$ & 0.92 &  $\alpha_{{\rm Ireland},1}$ & 0.92 & $\alpha_{{\rm Brazil},1}$ & 0.88 \\
$\alpha_{{\rm Russia},1}$ & 0.92 &  $\alpha_{{\rm India},1}$ & 0.91 & $\alpha_{{\rm France},1}$ & 0.92 \\
$\beta_{1,1}$ & 0.07 && && \\
\hline
\end{tabular}}
\end{table}

Overall, BIC choice selects for higher autoregressive order in the global-$\alpha$ GNAR models compared to the local-$\alpha$ variant,
which also results in higher out-of-sample mean-squared forecast error. This suggests that the PMIs' autoregressive dynamics
differs between countries, although the autoregressive parameters in Table~\ref{tab:exp1Estimates} are  similar.
Moreover, BIC choice of the local-$\alpha$ models also favours exploiting the network structure implied by export flows, whether by using the fully connected or  the nearest neighbour network: $\mathbf{s} \ne \mathbf{0}_p$ for both of the best-performing GNAR models.

\subsection{GNARX forecasting performance with external regressors}\label{sec:experiment2}

Given the shortness of the exogenous regressors, it is not feasible now to evaluate out-of-sample forecasting performance for the GNARX models, so we switch to in-sample forecast evaluation. For example, predicting `October'
based on exogenous variables
that only existed since early Spring, i.e.\ few observations, is not sensible. Here, we estimate model parameters on the whole
series and then, given a time point,  we use the estimates, but use
observations only from previous time points. In future, with more data, we will be
able to switch to out-of-sample forecasting performance evaluation.

We can now reintroduce the shorter Australian PMI series, using the same method for handling missing observations
as  in Section~\ref{sec:experiment1}.
On our relatively short time series with missing data, vector autoregressions with exogenous variables (VARX) as described in, for example, \citet{tsay2013multivariate},
which contain $N^2(p+p'+1)$ parameters, are infeasible.
Recall that some VAR models may be estimated as separate linear regressions for each country. A simple VARX(1,0) model would involve the estimation of 13 coefficients corresponding to the stringency indices and 13  corresponding to the death rates in each country. However, our dataset contains fewer than 13 nonzero observations for each exogenous variable of each country, clearly leading to a parameter identification problem. 

Table~\ref{tab:experiment2} shows in-sample mean-squared forecasting errors for our
{\color{black} stagewise-BIC selected} GNARX models using the fully-connected network (results for the nearest neighbour \textcolor{black}{network and no network} were always worse and reported in the Supplementary Material, Section~11).
The inclusion of exogenous variables results in substantially lower BIC and an almost halving of
the mean-squared forecast error across \textcolor{black}{all settings} compared to GNAR without external regressors. The lowest BIC model is the global-$\alpha$ \textcolor{black}{GNARX$(5, [1, 0, 1, 0, 1], 3, 2)$}, with estimates shown in Table~\ref{tab:experiment2fit2} \textcolor{black}{and}
\mbox{$p$-values} calculated as in Table~\ref{tab:experiment1a}.

\begin{table}
\caption{\label{tab:experiment2}In-sample composite PMI forecast performance between Jan-98 and Oct-20 using GNARX models
	with the fully-connected network.
	MSFE=In-sample mean-squared forecasting error.}
\centering
\fbox{%
\begin{tabular}{lrrr}
\hline
Model & Model order & BIC & MSFE (SE) \\ 
\hline
\multicolumn{4}{|l|}{\textbf{Without external regressors}} \tabularnewline
\hline
Global-$\mathbf{\alpha}$ GNAR & \textcolor{black}{4, [1, 1, 1, 0]} & \textcolor{black}{12653} & \textcolor{black}{7.1 (0.9)}\\
Local-$\mathbf{\alpha}$ GNAR & \textcolor{black}{3, [1, 1, 1]} & \textcolor{black}{12806} & \textcolor{black}{6.7 (0.9)}\\
\hline
 & & & \\
\hline
\multicolumn{4}{|l|}{\textbf{Stringency indices only}} \tabularnewline
\hline
Global-$\mathbf{\alpha}$ GNARX & \textcolor{black}{5, [1, 0, 1, 0, 1], 3} & \textcolor{black}{11323} & \textcolor{black}{4.2 (0.3)} \\
Local-$\mathbf{\alpha}$ GNARX & \textcolor{black}{2, [1, 1], 3} & \textcolor{black}{11331} & \textcolor{black}{4.0 (0.3)} \\
\hline
 & & & \\
\hline
\multicolumn{4}{|l|}{\textbf{COVID-19 death rates only}} \tabularnewline
\hline
Global-$\mathbf{\alpha}$ GNARX & \textcolor{black}{4, [1, 1, 1, 0], 3} & \textcolor{black}{12020} & \textcolor{black}{5.5 (0.8)} \\
Local-$\mathbf{\alpha}$ GNARX & \textcolor{black}{2, [1, 1], 3} & \textcolor{black}{12212} & \textcolor{black}{5.5 (0.8)} \\
\hline
 & & & \\
\hline
\multicolumn{4}{|l|}{\textbf{With both external regressors}} \tabularnewline
\hline
Global-$\mathbf{\alpha}$ GNARX & \textcolor{black}{5, [1, 0, 1, 0, 1], 3, 2} & \textcolor{black}{11175} & \textcolor{black}{4.0 (0.3)} \\
Local-$\mathbf{\alpha}$ GNARX & \textcolor{black}{2, [1, 1], 3, 1} & \textcolor{black}{11199} & \textcolor{black}{3.8 (0.3)} \\
\hline
\end{tabular}}
\end{table}

Table~\ref{tab:experiment2fit2} shows \textcolor{black}{that  the 1st and 5th positive autoregressive coefficients are significant}. As before, current-period PMIs are also significantly positively dependent on previous-period PMIs of neighbours weighted by export volumes. On the other hand, the negative \textcolor{black}{$\beta_{3,1}$ and $\beta_{5,1}$} parameter estimates are \textcolor{black}{ smaller in magnitude}.
\begin{table}
\caption{\label{tab:experiment2fit2}Global-$\alpha$ \textcolor{black}{GNARX$(5, [1, 0, 1, 0, 1], 3, 2)$} model parameter estimates using the fully-connected network and both exogenous regressors. $\dagger$ means \mbox{$p$-value} is \textcolor{black}{$< 0.001$}. Values significant at the 5\% level in bold.}
\centering
\fbox{%
\begin{tabular}{lrr|lrr|lrr}
\hline
Param.\ & Est.\ & \mbox{$p$-value} \% & Param.\ & Est.\ & \mbox{$p$-value} \% & Param.\ & Est.\ & \mbox{$p$-value} \% \\  \hline

$\boldsymbol{\alpha}_1$ & \textcolor{black}{$0.83$} & $\dagger$  &  $\boldsymbol{\beta}_{1,1}$ & \textcolor{black}{$0.28$} & \textcolor{black}{$\dagger$} & \textcolor{black}{$\boldsymbol{\lambda}_{1,2}$} & \textcolor{black}{$0.15$} & \textcolor{black}{$0.8$} \\

$\textcolor{black}{\alpha_2}$ & \textcolor{black}{$0.00$} & \textcolor{black}{$97.5$} & \textcolor{black}{$\boldsymbol{\beta}_{3,1}$} & \textcolor{black}{$-0.15$} & \textcolor{black}{$4.2$} & \textcolor{black}{$\boldsymbol{\lambda}_{1,3}$}& \textcolor{black}{$0.10$} & \textcolor{black}{$3.3$} \\

\textcolor{black}{$\alpha_3$} & \textcolor{black}{$0.03$} & \textcolor{black}{$59.2$} & \textcolor{black}{$\boldsymbol{\beta}_{5,1}$} & \textcolor{black}{$-0.08$} & \textcolor{black}{$3.5$} &
\textcolor{black}{$\boldsymbol{\lambda}_{2,0}$}& \textcolor{black}{$-0.91$} & \textcolor{black}{$0.2$} \\

$\textcolor{black}{\alpha_4}$ & \textcolor{black}{$-0.05$} & \textcolor{black}{$34.0$} & \textcolor{black}{$\boldsymbol{\lambda}_{1,0}$} & \textcolor{black}{$-0.31$} & \textcolor{black}{$\dagger$} & 
\textcolor{black}{$\lambda_{2,1}$}& \textcolor{black}{$0.42$} & \textcolor{black}{$21.3$} \\

$\boldsymbol{\alpha}_5$ & \textcolor{black}{$0.09$} & \textcolor{black}{$0.6$} & \textcolor{black}{$\lambda_{1,1}$} & \textcolor{black}{$-0.09$} & \textcolor{black}{$24.8$} &
\textcolor{black}{$\lambda_{2,2}$}& \textcolor{black}{$0.41$} & \textcolor{black}{$11.3$}\\

\hline
\end{tabular}}
\end{table}

The zero-lag coefficient for COVID-19 intervention stringency indices is highly significant, matching our expectation of a strong coincident negative impact of containment policies on business confidence. A country's COVID-19 death rate also has
a significant negative contemporaneous impact on the PMI, indicating that the pandemic might be
influencing business confidence through channels beyond policy interventions.

\begin{figure}
    \centering
        \resizebox{0.6\textwidth}{!}{\includegraphics{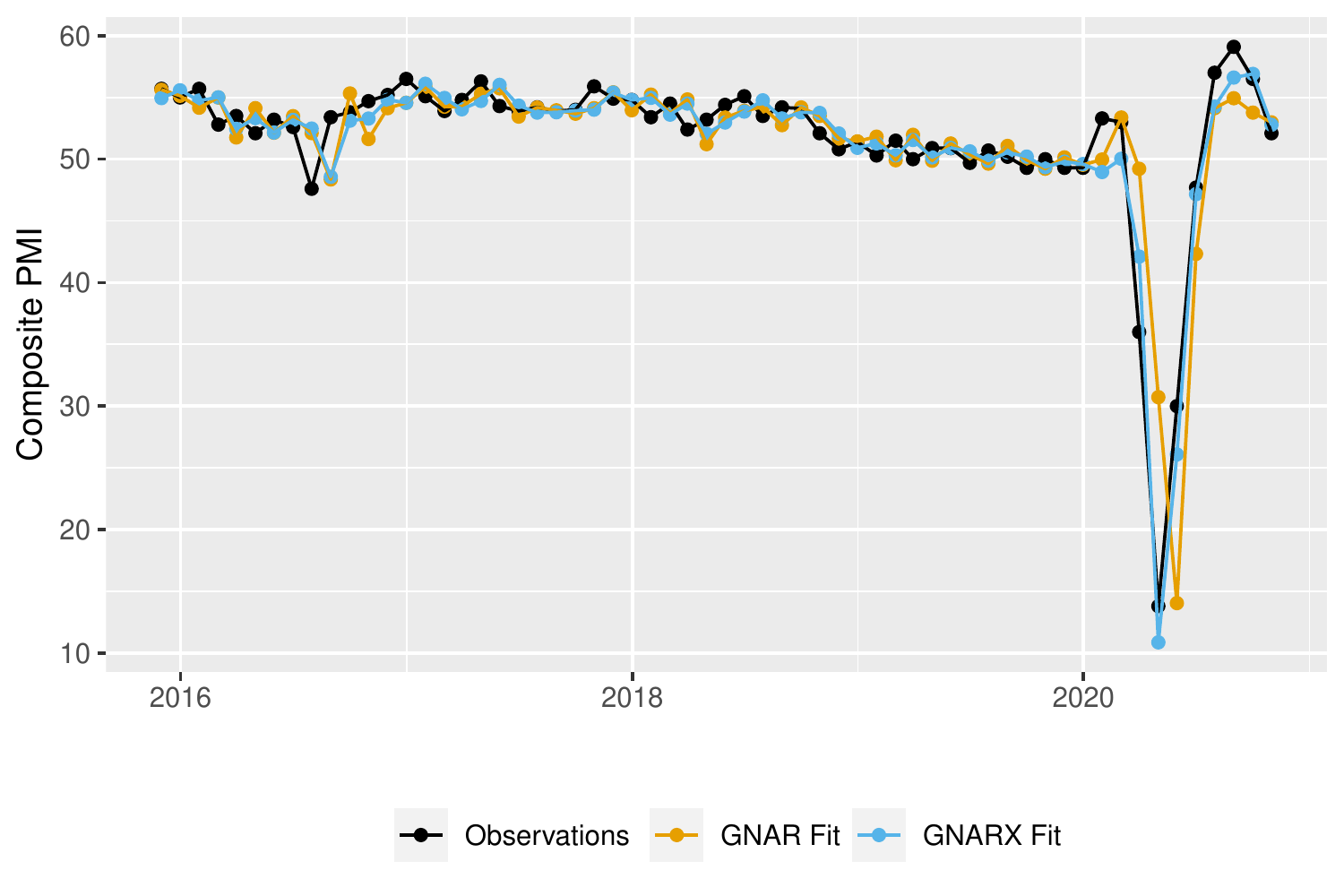}}
	 \caption{UK: Comparison of \textcolor{black}{GNARX$(5, [1, 0, 1, 0, 1])$} and \textcolor{black}{GNARX$(5, [1, 0, 1, 0, 1], 3, 2)$} fits. }
    \label{fig:UKFit}
\end{figure}

Figure~\ref{fig:UKFit} shows the global-$\alpha$ \textcolor{black}{GNARX$(5, [1, 0, 1, 0, 1], 3, 2)$} UK fits over the most recent five-year period, to  illustrate the quality of fit during the COVID-19 period
(other countries' charts are given in the Supplementary Material, Section~10). The inclusion of stringency indices and COVID-19 death rates allows the GNARX model to anticipate both the initial composite PMI decline at the start of the crisis and the subsequent recovery,
leading to a much lower mean-squared forecasting error than the corresponding GNAR model.

\section{UK scenario analysis}\label{sec:scenario}

\begin{figure}
    \centering
    \includegraphics[width=0.6\textwidth]{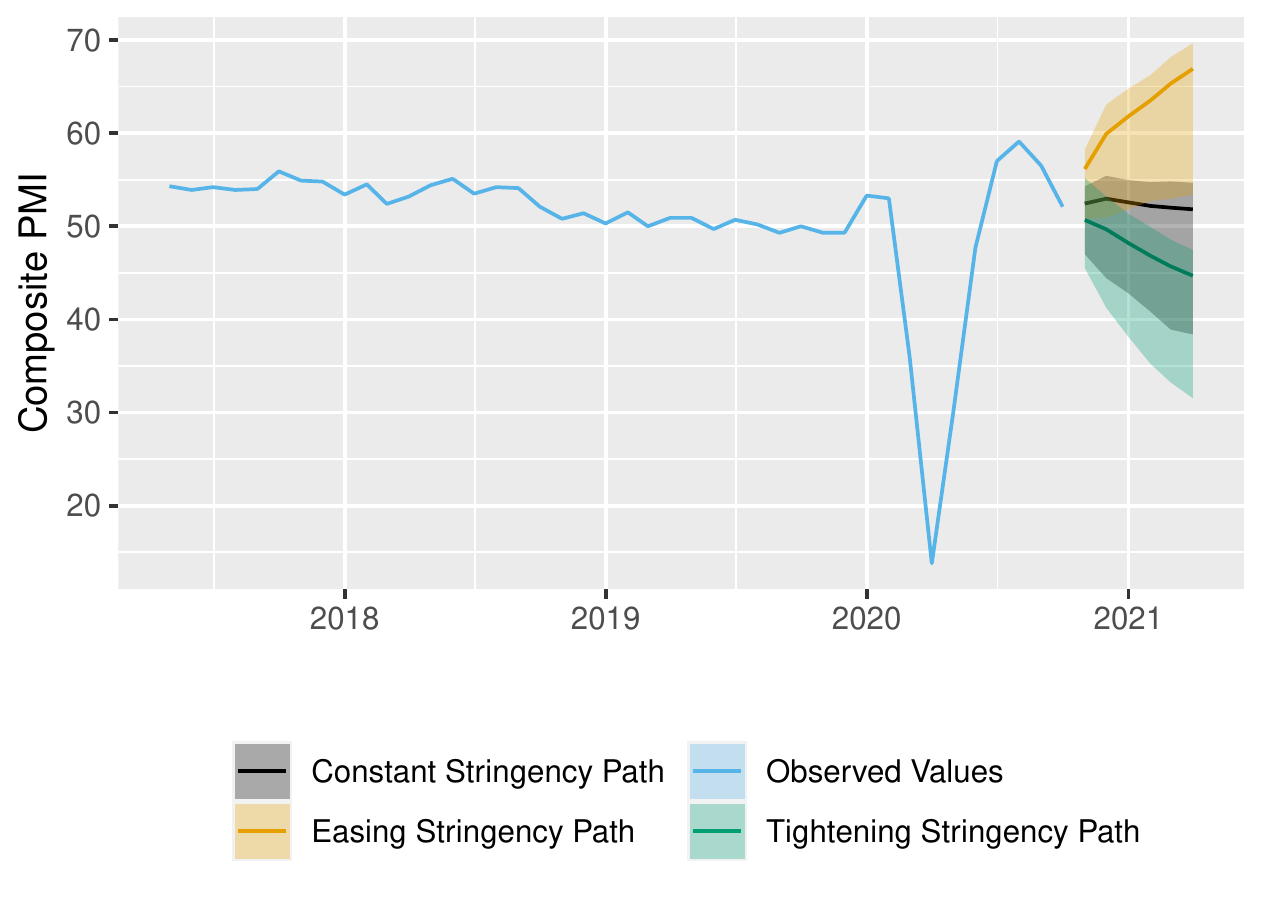}
    \caption{Forecasts of UK composite PMI under three different scenarios for policy interventions using the \textcolor{black}{GNARX$(5, [1, 0, 1, 0, 1], 3, 2)$} model. 
    	\mbox{$95$\%-coverage} prediction intervals estimated using $1000$ bootstrap samples.}
    \label{fig:UKforecast}
\end{figure}

We now use the fitted the global-$\alpha$ \textcolor{black}{GNARX$(5, [1, 0, 1, 0, 1], 3, 2)$} model from the previous section to produce six-month forecasts of the UK composite PMI series under different assumptions on the evolution of the UK stringency index.
\textcolor{black}{For other examples of conditional forecasting on multivariate economic time series, we refer readers to the seminal paper by \cite{waggoner1999conditional} and the more recent summary provided by \cite{banbura2015conditional}, which describe conditional forecasting techniques and applications for VAR models. We would note that these describe more difficult problem than ours, as we condition on exogenous variables.}
Figure~\ref{fig:UKforecast} provides forecasts of the UK composite PMI under three different policy intervention scenarios for the six month period from
Nov-20 to Apr-21, assuming no change in the stringency indices of other countries from Aug-20 or in COVID-19 death rates:
\begin{enumerate}
    \item \textbf{Easing COVID-19 intervention measures}. The stringency index decreases linearly from the Oct-20 value of $67.9$ to $0.0$ between Nov-20 and Apr-21.
    \item \textbf{No change in COVID-19 intervention measures}. The stringency index is constant at the Oct-20 value of $67.9$ between Nov-20 and
    Apr-21.
    \item \textbf{Tightening in COVID-19 intervention measures}. The stringency index increases linearly from the Oct-20 value of $67.9$ to $100.0$ between Nov-20 and Apr-21.
\end{enumerate}
As might be expected, the GNARX model predicts significantly higher UK composite PMI under assumption of the UK stringency index falling to zero, holding death rates constant. Using bootstrap prediction intervals, computed using the algorithm introduced in the Supplementary Material Section~12, the expected PMI under the easing scenario exceeds the $97.5$th percentile of the PMI under the tightening and constant stringency scenarios after only one month.  The Apr-21 prediction of PMI is \textcolor{black}{$62.1$} in the easing scenario (which would be a record high) and \textcolor{black}{$47.6$} in the tightening scenario.
In all cases, the estimated sampling distribution of the PMI is negatively skewed.

We now estimate how these composite PMI paths would translate into headline quarter-on-quarter real GDP growth, using data sourced from \url{https://www.ons.gov.uk/economy/grossdomesticproductgdp/timeseries/ihyq/qna}. At the time of writing, 2020 Q3 is the latest available observation of UK GDP. Given that the GDP series is sampled quarterly, while the PMI series is sampled monthly, we assume a mixed-frequency linear relationship between the GDP growth and PMI series, using the MIDAS regression model \citep{ghysels2004midas}. Analysis of both the unrestricted MIDAS model and the restricted MIDAS model, using exponential Almon lag polynomial weights, suggest that quarterly GDP is primarily dependent on the ``$2/3$'' lag of composite PMI. E.g., 2020 Q3 GDP depends only on the Jul-20 PMI observation  in our  simplified model.

Table~\ref{tab:midasforecast} shows
conditional forecasts for 2020 Q4, 2021 Q1 and 2021 Q2 UK GDP growth using the PMI predictions produced by  GNARX  for each of the three COVID-19 intervention scenarios, assuming only the $2/3$ lag of composite PMI is relevant. Our results suggest a \textcolor{black}{$4.5$\%} difference in
2021 Q1 GDP growth and a \textcolor{black}{$4.8$\%} difference in 2021 Q2 GDP growth between the easing and tightening scenarios.
\begin{table}
\caption{\label{tab:midasforecast}Forecasts for UK quarterly real GDP growth \textcolor{black}{(\%)} based on PMI forecasts for the tightening, constant and easing intervention measure scenarios. 95\% prediction intervals in brackets.}
\centering
\fbox{%
\begin{tabular}{lrrr}
\hline
Quarter & Tightening & Constant & Easing \\
\hline
2020 Q4 & -0.1 (-4.1, 3.8) & -0.1 (-4.1, 3.8) & -0.1 (-4.1, 3.8) \\
2021 Q1 & \textcolor{black}{-1.2 (-5.8, 2.9)} & \textcolor{black}{0.3 (-4.1, 4.4)} & \textcolor{black}{3.3 (-1.3, 7.4)} \\
2021 Q2 & \textcolor{black}{-1.6 (-7.1, 2.2)} & \textcolor{black}{-0.1 (-5.7, 3.2)} & \textcolor{black}{3.2 (-2.0, 7.1)} \\
\hline
\end{tabular}}
\end{table}
Table~\ref{tab:midasforecast}'s bootstrap prediction intervals 
take into account both the uncertainties in the GDP regression model and the underlying PMI forecasts.

\textcolor{black}{Lending support to the strongly positive GDP growth path in our easing restriction scenario, the most recent (at the time of writing) \cite{monetary} Monetary Policy Report forecasts a material recovery in UK GDP in the event of an easing in Covid-related restrictions. The authors suggest that the economic impact of such easing would be driven by a substantial increase in consumer spending, and a rise in business investment to a lesser extent as sales rise and uncertainty diminishes.}


\section{Discussion}\label{sec:conclusion}

We showed how to apply network time series models to multivariate PMI time series for many countries and
how forecast performance could be  improved by incorporating network information.
We introduced the GNARX model, which permits exogenous variables to be incorporated into GNAR models for the first time and
\textcolor{black}{improves} forecasting results further still \textcolor{black}{by} using COVID-19 intervention stringency indices and  COVID-19 death rates.
Parameter estimates suggest a highly significant negative impact of harsher intervention measures and COVID-19 deaths on composite PMI and suggests that this effect can be transmitted to neighbours through the export channel. 

GNAR(X) model order choice using BIC selects  highly parsimonious models that result in excellent forecasting performance,
considering the inherent difficulties in forecasting economic indicators. On the other hand, the BIC-chosen VAR model appears to overfit, leading to relatively high forecast errors. Future work might compare the GNAR forecasting performance to that of BVAR models \citep{giannone2015prior}.

Our network was constructed by considering what looked `sensible' rather than optimality with respect to  forecasting performance and
further development work is required. Some preliminary
ideas have been suggested relating to network construction in different
fields, see  \cite{leeming:phd,knight2019generalised} and references therein.

Our three different scenarios earlier all operated under the simplifying assumption that policy interventions would be modified independently of the COVID-19 death rate. Future work may relax this unrealistic assumption, and incorporate other variables to capture attempts by
governments to mitigate the negative impact of containment measures through fiscal stimulus. 

{\color{black} We have used mean-squared forecasting error throughout our work as it is a valid measure,
well-understood and meaningful.
There are, of course, many other forecast quality metrics that could be used. For example,
the actual proportion of forecasts whose out-turn performed well with respect to (or within) previously
established forecasting intervals, see~\cite{ehm16} for more examples and discussion.}

Network time series modelling is young and much remains unexplored.
For  GNARX
models, an obvious extension might permit exogenous regressor coefficients  to vary by node.
Regarding the COVID-19 experiments discussed above, this would allow economies to react differently to a given stringency level of policy interventions. Intuitively, economies where a larger proportion of employees are able to work remotely should be less vulnerable to strict lockdown measures \citep{papanikolaou2020working}.

We are also conducting research
where more than one network time series interacts. So, for example, one can already see suggestions in the media on how countries' lockdown
policies influence each other. For example, the second UK lockdown on 5th November was made politically easier because of recent lockdowns
in major European neighbours. Hence, the stringency index time series (and the COVID-19 deaths) could also be represented by network
time series and there seems \textcolor{black}{to be} no reason why the PMI  and  stringency index networks would have to be the same. \textcolor{black}{Finally, further} extensions
to network time series might be considered, such as time-varying parameters that can be used to create a locally stationary network time series
process, or permitting edge-distances to change and be modelled (e.g.\ as exports vary between countries over time).

\section*{Acknowledgements}
This article was prepared under the auspices of the Imperial College COVID-19 Response Team
(Economics Team). We are grateful to Dr Katharina Hauck, Deputy Director of the Abdul Latif Jameel
Institute for Disease and Emergency Analytics (J-IDEA), School of Public Health, Imperial College,
London for enthusiastic encouragement and helpful comments on an earlier version of this article.

\bibliographystyle{guy3}
\bibliography{bibliography}
\end{document}


%
%

{\bf \Large \color{black} \em Part 1: Models and Theory}

{\color{black}
\section{The models of Zhu {\em et al.} (2017) and Knight {\em et al.} (2020)}
For completeness the model from~\citet{zhu2017network} is as follows. Let $Y_{i ,t}$ be
the response collected from the $i$th subject at time $t$, where $i= 1, \ldots, N$ and
$t$ is thought of as time. Each subject can be thought of as a node in a network
and each node may have associated a \mbox{$p$-dimensional} node-specific random (covariate)
vector $Z_i = ( Z_{i, 1}, \ldots, Z_{i, p} )^\intercal$. Their network vector autoregression model
(NAR) is given by
\begin{equation}
Y_{i, t} = \beta_{0} + Z_i^\intercal \gamma + \beta_1 n_i^{-1} \sum_{j=1}^N a_{i, j} Y_{j, (t-1)}
	+ \beta_2 Y_{i, (t-1)} + \epsilon_{i, t},
\end{equation}
where $n_i = \sum_{j\neq i} a_{i, j}$ is the total number of nodes that $i$ follows, where
$a_{i, j} = 1$ is there is a relationship (edge) between $i$ and $j$ and $a_{i, j} = 0$, otherwise,
$\beta_0$ is a level effect parameter, $\beta_1$ is a network effect parameter, $\beta_2$
is the standard AR$(1)$ parameter and $\gamma$ is a \mbox{$p$-dimensional} covariate parameter.

The model of~\cite{knight2019generalised} is
\begin{equation}\label{eqn:GNARX}
    X_{i,t}=\sum_{j=1}^{p}\left(\alpha_{i,j}X_{i,t-j} + \sum_{c=1}^C \sum_{r=1}^{s_j}  \beta_{j, r, c} \sum_{q\in\mathcal{N}^{(r)}(i)}\omega_{i,q, c}X_{q,t-j}\right) + u_{i,t},
\end{equation}
where the quantities here are the same as in the main paper except here there is an
extra index $c$ in the network part,
which represents one of  $C$ possible different edge types (for example,
in epidemics, $C=2$ might arise because viruses might spread by wind (edge type $c=1$)
or by physical communication (edge type $c=2$).
This model was developed from the NARIMA model from~\cite{knight2016}, which does not
have edges types, but introduces differencing and a moving average component as is found in
standard ARIMA modelling for time series, see, e.g.\
\cite{priestley:spectral}, \cite{brockwell:time},  \cite{hamilton:time} or \cite{chatfield:the}.
}

%
%

\textcolor{black}{\section{Stationarity conditions for the GNARX model}}

\textcolor{black}{\begin{theorem}\label{thm:stationarity}
The GNARX model, as defined in~(1) of the main report, is stationary under the conditions that the external regressor time series $\mathbf{X}_t$ are stationary and
\begin{equation}\label{eqn:station-cond}
\sum_{j=1}^p\left(|\alpha_{i,j}|+\sum_{c=1}^C\sum_{r=1}^{s_j}|\beta_{j,r,c}|\right)<1\;\;\; \forall i\in 1,...,N.
\end{equation}
\end{theorem}}

\textcolor{black}{The GNARX model can be expressed in VARX form as
\begin{equation}\label{eqn:VAR-form2}
    \mathbf{Y}_t=\phi_1\mathbf{Y}_{t-1}+ \cdots +\phi_p\mathbf{Y}_{t-p}+\Lambda_0\mathbf{X}_{t}+\Lambda_1\mathbf{X}_{t-1}+ \cdots +\Lambda_{p'}\mathbf{X}_{t-p'}+\mathbf{u}_t,
\end{equation}
where the coefficient matrices are restricted such that $\phi_k=\text{diag}\left\{\alpha_{i,k}\right\}+\sum_{r=1}^{s_k}\beta_{k,r}W^{(r)}$, $[W^{(r)}]_{l,m}=w_{l,m}\mathbb{I}\left\{m\in\mathcal{N}^{(r)}(l)\right\}$, and $\Lambda_{k}=\lambda_{k}I_N$.
Equation \ref{eqn:VAR-form2} can be expressed compactly in terms of lag polynomials as
\begin{equation}\label{eqn:compact-VAR-form}
    \phi(B)\mathbf{Y}_t = \Lambda(B)\mathbf{X}_{t},
\end{equation}
where $\phi(z) = I_n-\phi_1z-...-\phi_pz^p$ and $\Lambda(z) = I_n-\Lambda_1z-...-\Lambda_pz^p$.
\cite{knight2019generalised} demonstrate that under Equation \ref{eqn:station-cond}, the causality criterion $\text{det}\, \phi(z)\ne 0$ for all $|z|\le 1$, where $z\in\mathbb{C}$, is satisfied.
By \cite{brockwell1991time}, we can thus obtain the representation 
\begin{equation}
    \mathbf{Y}_t=\sum_{j=0}^{\infty}\psi_j\mathbf{X}_{t-j},
\end{equation}
where the coefficient matrices $\psi_j$ correspond to the lag polynomial $\psi(z)=\phi^{-1}(z)\Lambda(z)$. As $\mathbf{X}_t$ are stationary by assumption, $\mathbf{Y}_t$ is therefore stationary.
}

%
%

%
%

\section{Consistency and asymptotic normality of GNARX parameter estimates}
A straightforward application of~\citet{lutkepohl2005new} Proposition~5.1, with an additional assumption that the external regressor time series are stationary, results in the following asymptotic properties of ${\hat\gamma}^{FGLS}$, which we provide without proof.
\begin{theorem}\label{thm:asymptotics}
Let ${\hat\gamma}^{FGLS}$ and ${\hat\Sigma}_u$ be defined according to~\textcolor{black}{(5)}
and~\textcolor{black}{(6) from the main paper,}
respectively, for stationary multivariate time series $\left\{\textbf{Y}_t\right\}_{t=1, \ldots,T}$ and $\left\{\textbf{X}_t\right\}_{t=1, \ldots, T}$ as in~\textcolor{black}{(1)}. Then
 ${\hat\gamma}^{FGLS}\overset{P}{\rightarrow}\gamma$
and
\begin{equation}\label{eqn:distribution}
    \sqrt{T}\left({\hat\gamma}^{FGLS}-\gamma\right)\overset{d}{\rightarrow}N\left(0,\left[R^\intercal \left\{\plim (T^{-1}ZZ^\intercal )\otimes{\hat\Sigma}_u^{-1}\right\}R\right]^{-1}\right),
\end{equation}
where $\overset{P}{\rightarrow}$ and $\overset{d}{\rightarrow}$ denote convergence in probability and distribution, respectively, and
$\plim$ is the probability limit.
\end{theorem}

%
%

\section{Definition of the Model Matrix, $R$}
Please refer to the notation in the main paper.
From equation~\eqref{eqn:VAR-form2} we have
  \begin{align}\label{eqn:vec-phi}
    \underset{N^2\times1}{\vecJ (\phi_k)} & =\begin{bmatrix}
\alpha_{1,k} &
\mathbf{0}_N &
\alpha_{2,k} & 
\mathbf{0}_N & 
\cdots & \alpha_{N,k}
\end{bmatrix}^\intercal
+\sum_{r=1}^{s_k}\beta_{k,r}\vecJ \left\{W^{(r)}\right\},\\
    \text{vec}(\Lambda_{k}) & =\lambda_{k} \vecJ (I_N), \label{eqn:vec-Lambda}
\end{align}
where $[\textbf{x}]_i$ denotes the $i$th element of  $\textbf{x}$ and $\mathbf{0}_a$ denotes the length $a$-vector of zeros.

By vertically stacking the $\vecJ (\phi_k)$ and $\vecJ (\Lambda_k)$ vectors,  the $P\times M$ matrix $R$ is
\begin{multline}
\underset{N^2(p+p'+1)\times M}{R} = \left[
  \begin{matrix}
    A & R_1 & 0_{N^2\times N} & 0_{N^2\times s_2} & \cdots\\ 
    0_{N^2\times N} & 0_{N^2\times s_1} & A & R_2 & \cdots \\
    \vdots  & \vdots & \vdots & \vdots & \\ 
    0_{N^2\times N} & 0_{N^2\times N} & 0_{N^2\times s_1} & 0_{N^2\times s_2} & \cdots \\
    0_{N^2\times N} & 0_{N^2\times N} & 0_{N^2\times s_1} & 0_{N^2\times s_2} & \cdots \\
    0_{N^2\times N} & 0_{N^2\times N} & 0_{N^2\times s_1} & 0_{N^2\times s_2} & \cdots \\
    \vdots  & \vdots & \vdots & \vdots & \\
    0_{N^2\times N} & 0_{N^2\times N} & 0_{N^2\times s_1} & 0_{N^2\times s_2} & \cdots \\
  \end{matrix}\right.                
\\
  \left.
  \begin{matrix}
    \cdots & 0_{N^2\times N} & 0_{N^2\times s_p} & 0 & 0 & \cdots & 0 \\ 
    \cdots & 0_{N^2\times N} & 0_{N^2\times s_p} & 0 & 0 & \cdots & 0 \\ 
    & \vdots & \vdots & \vdots & \vdots  & & \vdots\\ 
    \cdots & A & R_p & 0 & 0 & \cdots & 0 \\
    \cdots & 0_{N^2\times N} & 0_{N^2\times s_p} & L & \textbf{0}_{N^2} & \cdots & \textbf{0}_{N^2} \\
    \cdots & 0_{N^2\times N} & 0_{N^2\times s_p} & \textbf{0}_{N^2} & L & \cdots & \textbf{0}_{N^2} \\
    & \vdots & \vdots & \vdots & \vdots  & \ddots & \vdots \\
    \cdots & 0_{N^2\times N} & 0_{N^2\times s_p} & \textbf{0}_{N^2} & \textbf{0}_{N^2} & \cdots & L \\
  \end{matrix}\right],
\end{multline}
where $0_{a\times b}$ denotes the $a\times b$ matrix of zeroes,
$\underset{N^2\times s_k}{R_k}  =\left[\vecJ \left\{W^{(1)}\right\},  \ldots , \vecJ \left\{W^{(s_k)}\right\}\right]$,
$\underset{N^2\times 1}{L}  =\left[1, \mathbf{0}_N^\intercal,1, \mathbf{0}_N^\intercal ,1, \ldots,1\right]^\intercal$ and
$\underset{N^2\times N}{A} =\left[\textbf{e}_1,0_{N\times N},\textbf{e}_2, 0_{N\times N},\textbf{e}_3, \ldots,\textbf{e}_N\right]^\intercal$.

\vspace{0.5cm}

{\bf \Large \color{black} \em Part 2: Exploratory Data Analysis}

%
%
\section{Second-order stationary tests}
\begin{figure}[p]
\centering
\resizebox{\tw\textwidth}{!}{\includegraphics{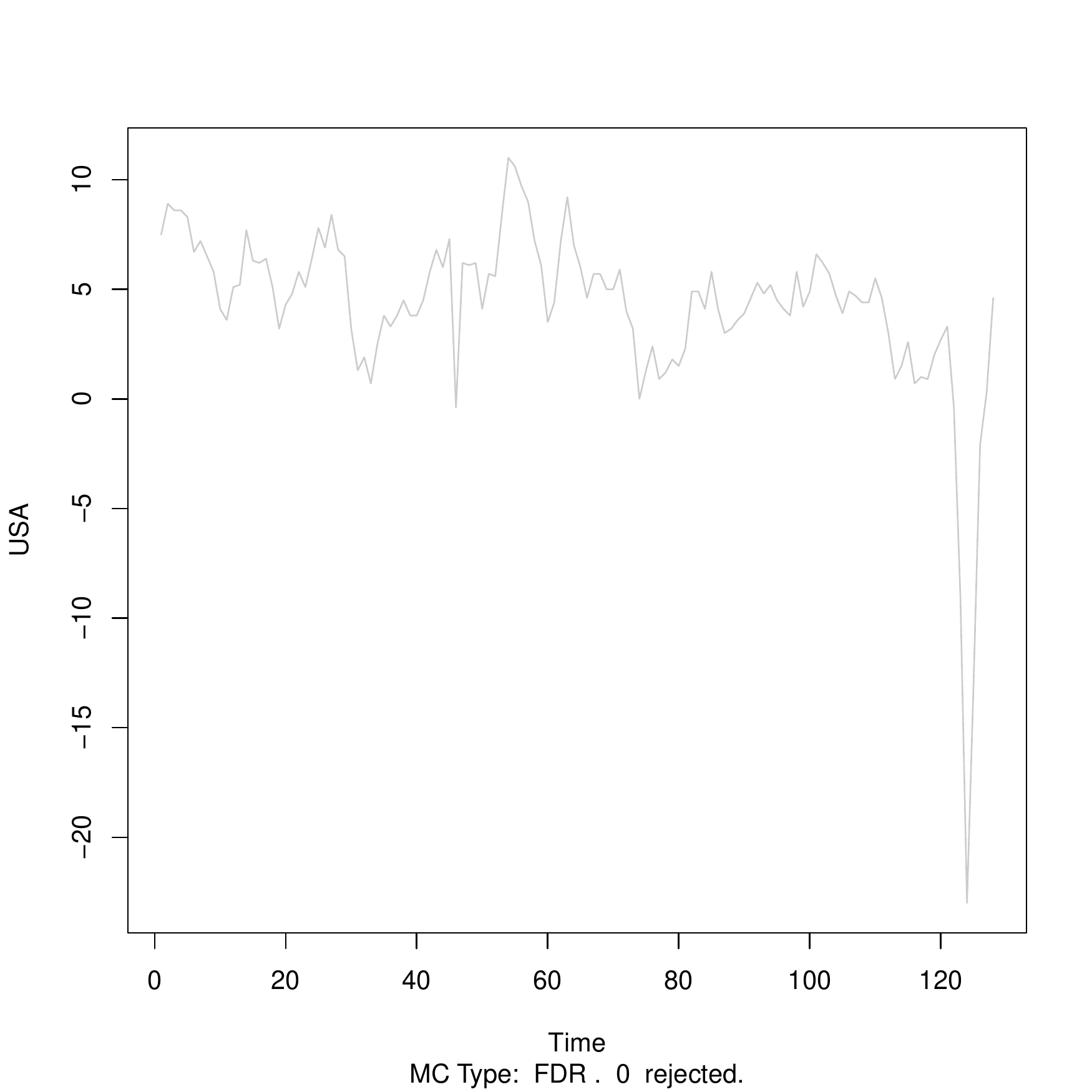}}
\caption{\label{fig:usa} Stationarity test plot for USA PMIs. No formal non-stationarities discovered}
\end{figure}

\begin{figure}[p]
\centering
\resizebox{\tw\textwidth}{!}{\includegraphics{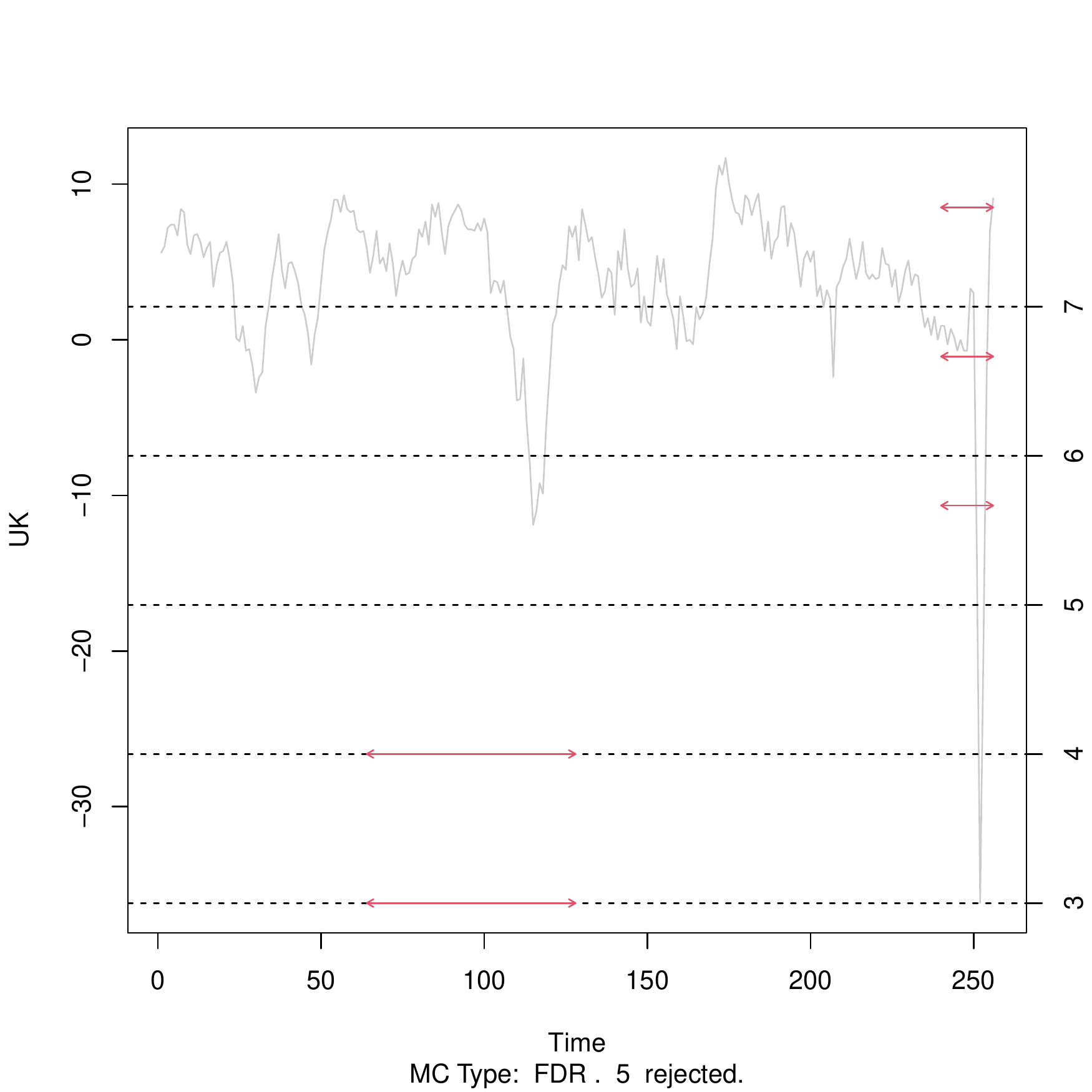}}
\caption{\label{fig:uk} Stationarity test plot for UK PMIs. Non-stationarities discovered during the great financial crisis and COVID-19.}
\end{figure}

\begin{figure}[p]
\centering
\resizebox{\tw\textwidth}{!}{\includegraphics{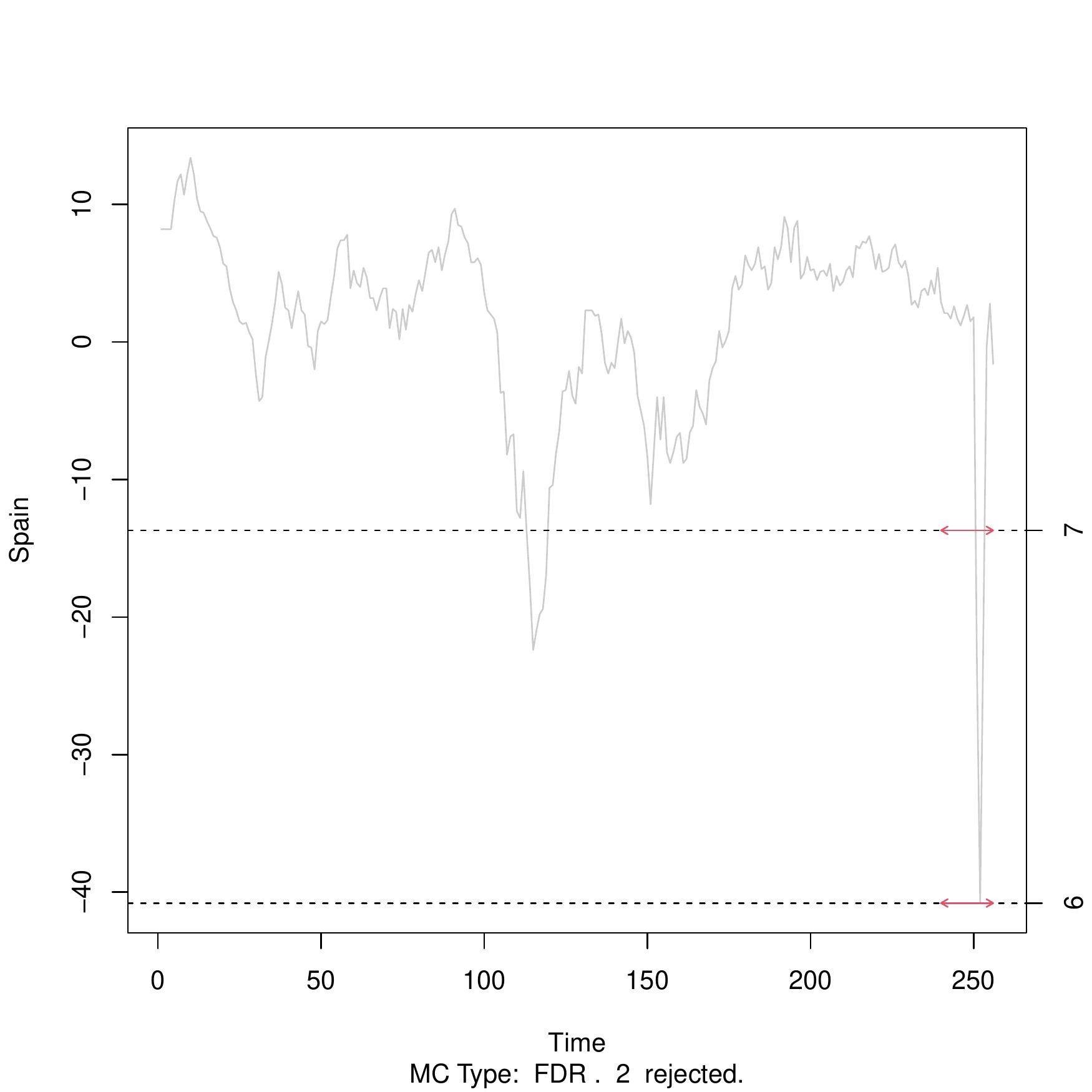}}
\caption{\label{fig:spain} Stationarity test plot for Spanish PMIs. Non-stationarities discovered during COVID-19.}
\end{figure}

\begin{figure}[p]
\centering
\resizebox{\tw\textwidth}{!}{\includegraphics{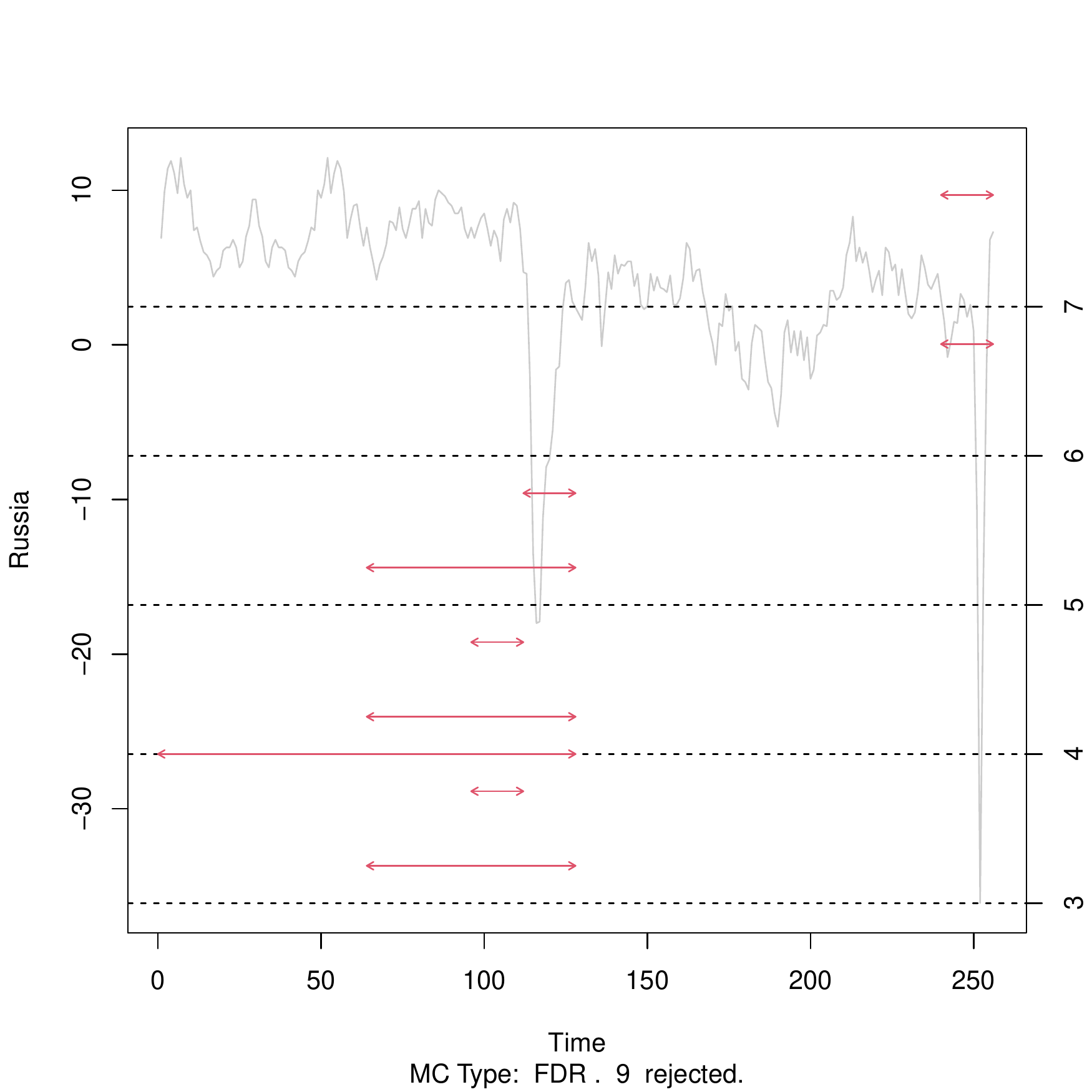}}
\caption{\label{fig:russia} Stationarity test plot for Russian PMIs. Non-stationarities discovered during the great financial crisis and COVID-19.}
\end{figure}

\begin{figure}[p][p]
\centering
\resizebox{\tw\textwidth}{!}{\includegraphics{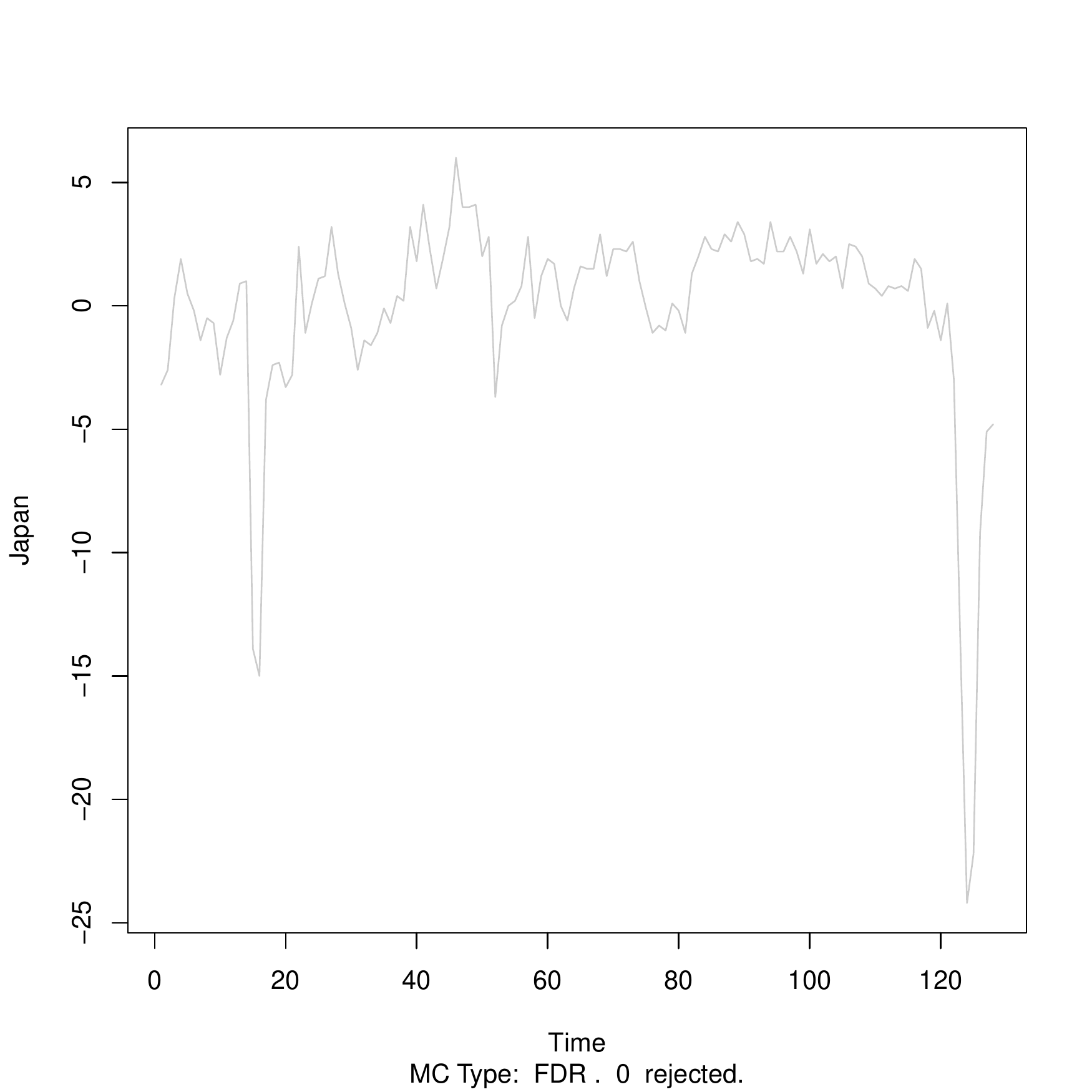}}
\caption{\label{fig:japan} Stationarity test plot for Japanese PMIs. No formal non-stationarities discovered}
\end{figure}

\begin{figure}[p][p]
\centering
\resizebox{\tw\textwidth}{!}{\includegraphics{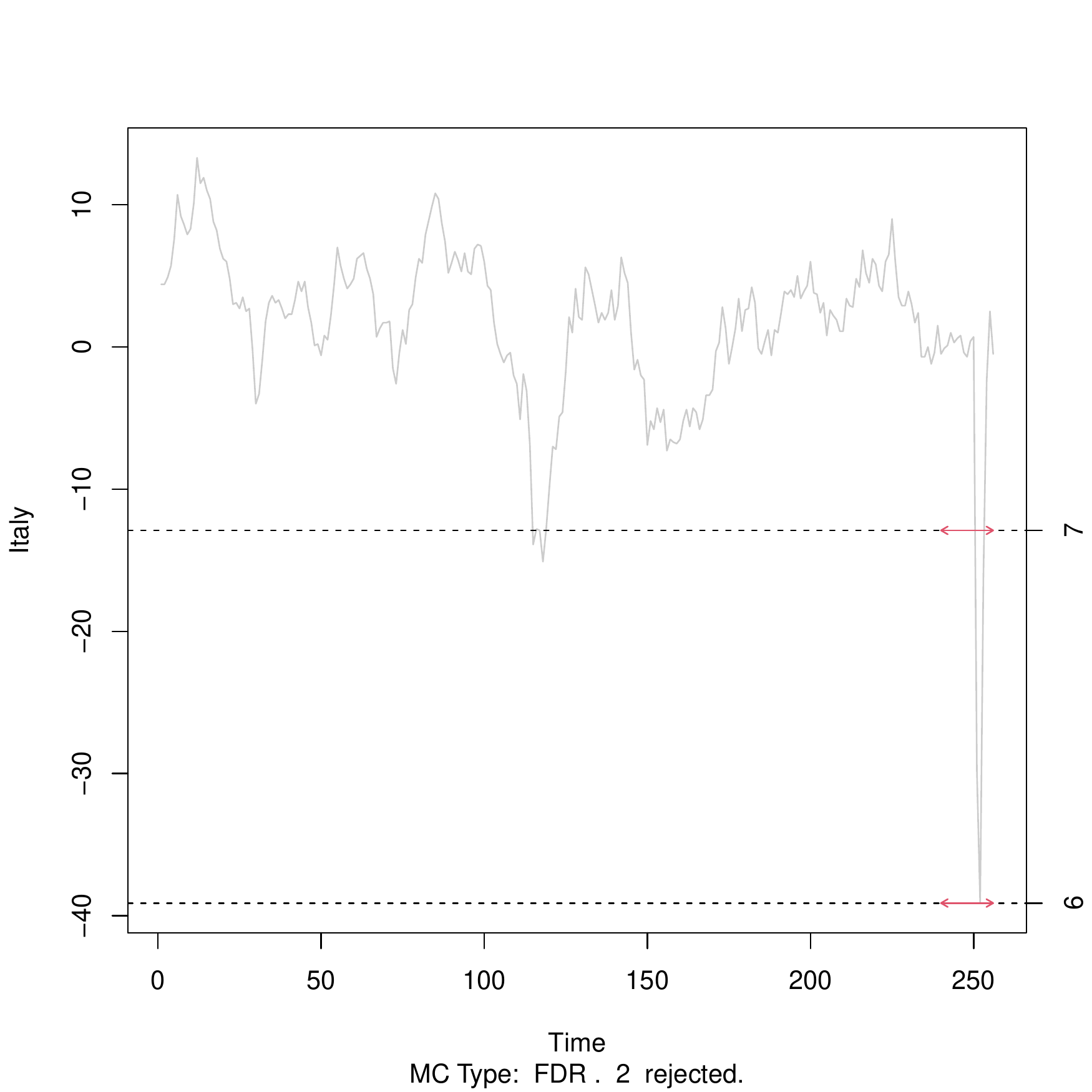}}
\caption{\label{fig:italy} Stationarity test plot for Italian PMIs. Non-stationarities discovered during COVID-19.}
\end{figure}

\begin{figure}[p]
\centering
\resizebox{\tw\textwidth}{!}{\includegraphics{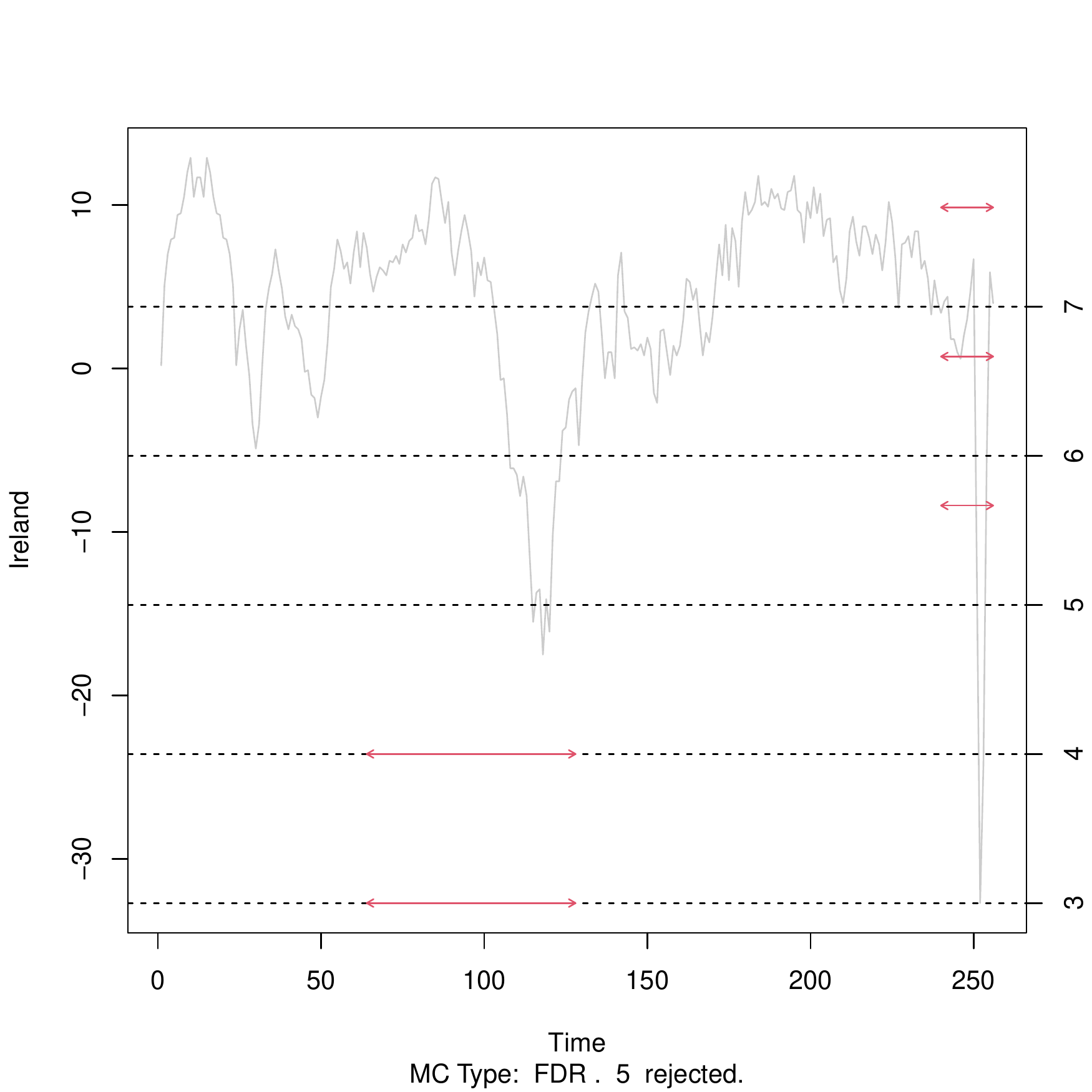}}
\caption{\label{fig:ireland} Stationarity test plot for Irish PMIs. Non-stationarities discovered during the great financial crisis and COVID-19.}
\end{figure}

\begin{figure}[p]
\centering
\resizebox{\tw\textwidth}{!}{\includegraphics{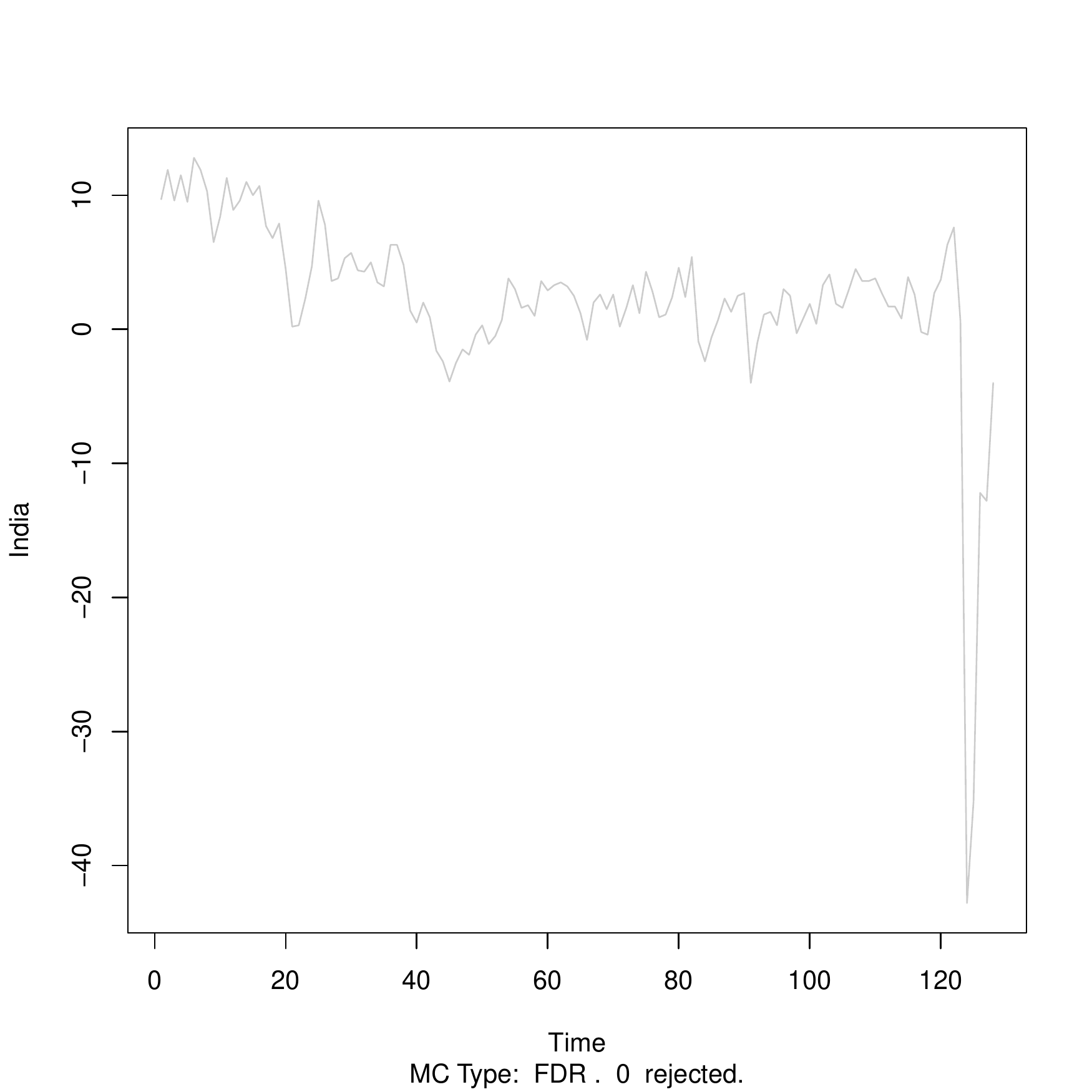}}
\caption{\label{fig:india} Stationarity test plot for Indian PMIs. No formal non-stationarities discovered}
\end{figure}

\begin{figure}[p]
\centering
\resizebox{\tw\textwidth}{!}{\includegraphics{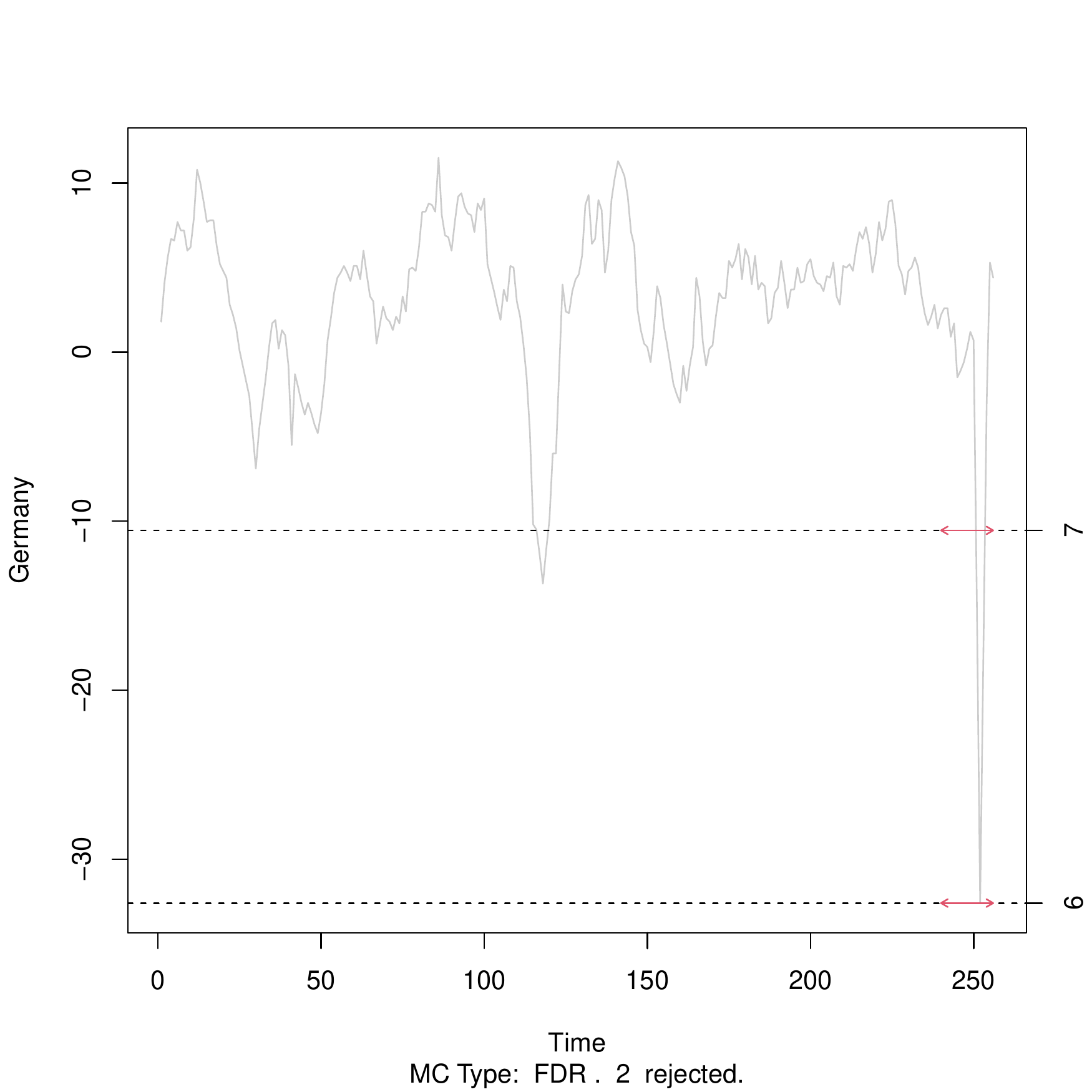}}
\caption{\label{fig:germany} Stationarity test plot for German PMIs. Non-stationarities discovered during COVID-19.}
\end{figure}

\begin{figure}[p]
\centering
\resizebox{\tw\textwidth}{!}{\includegraphics{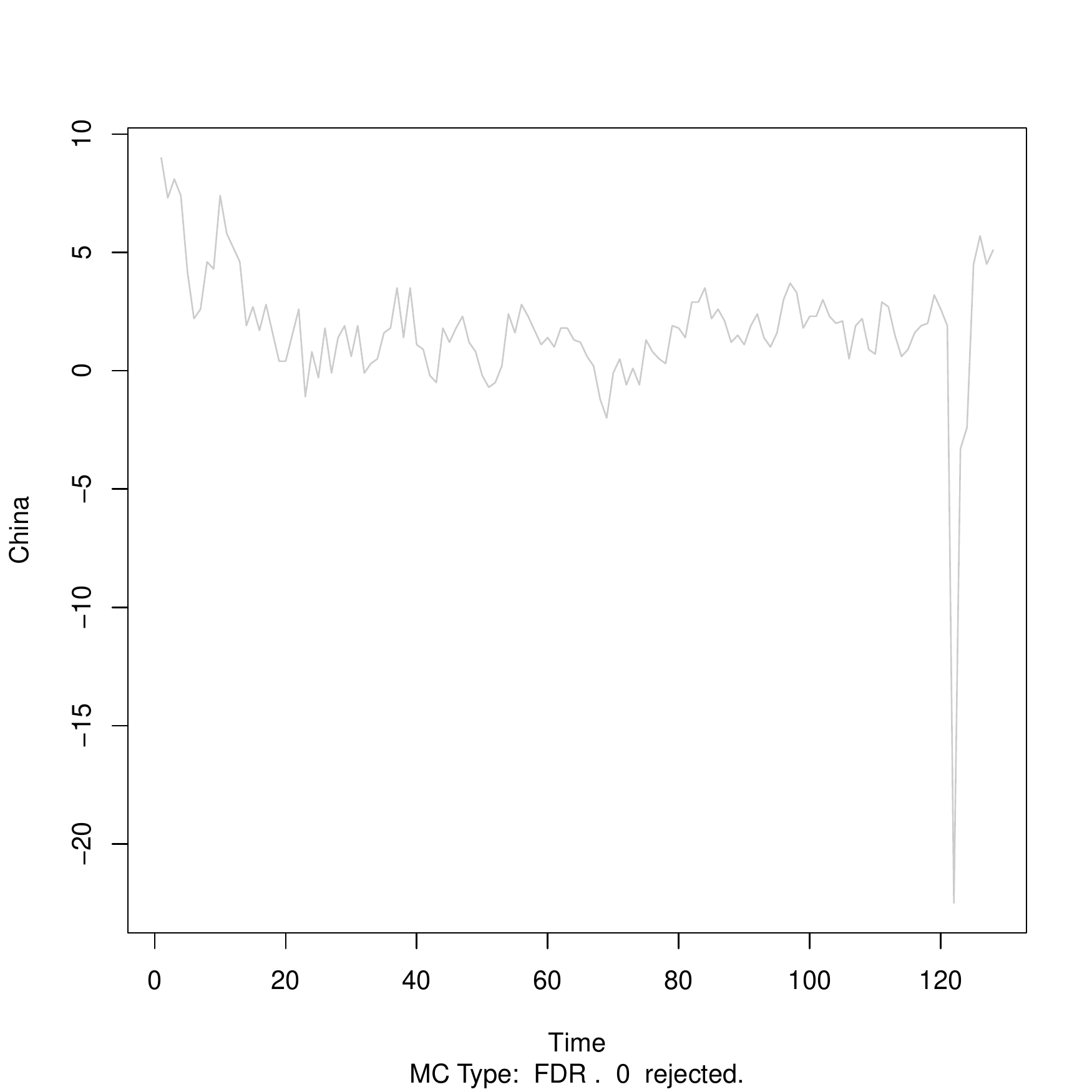}}
\caption{\label{fig:china} Stationarity test plot for Chinese PMIs. No formal non-stationarities discovered}
\end{figure}

\begin{figure}[p]
\centering
\resizebox{\tw\textwidth}{!}{\includegraphics{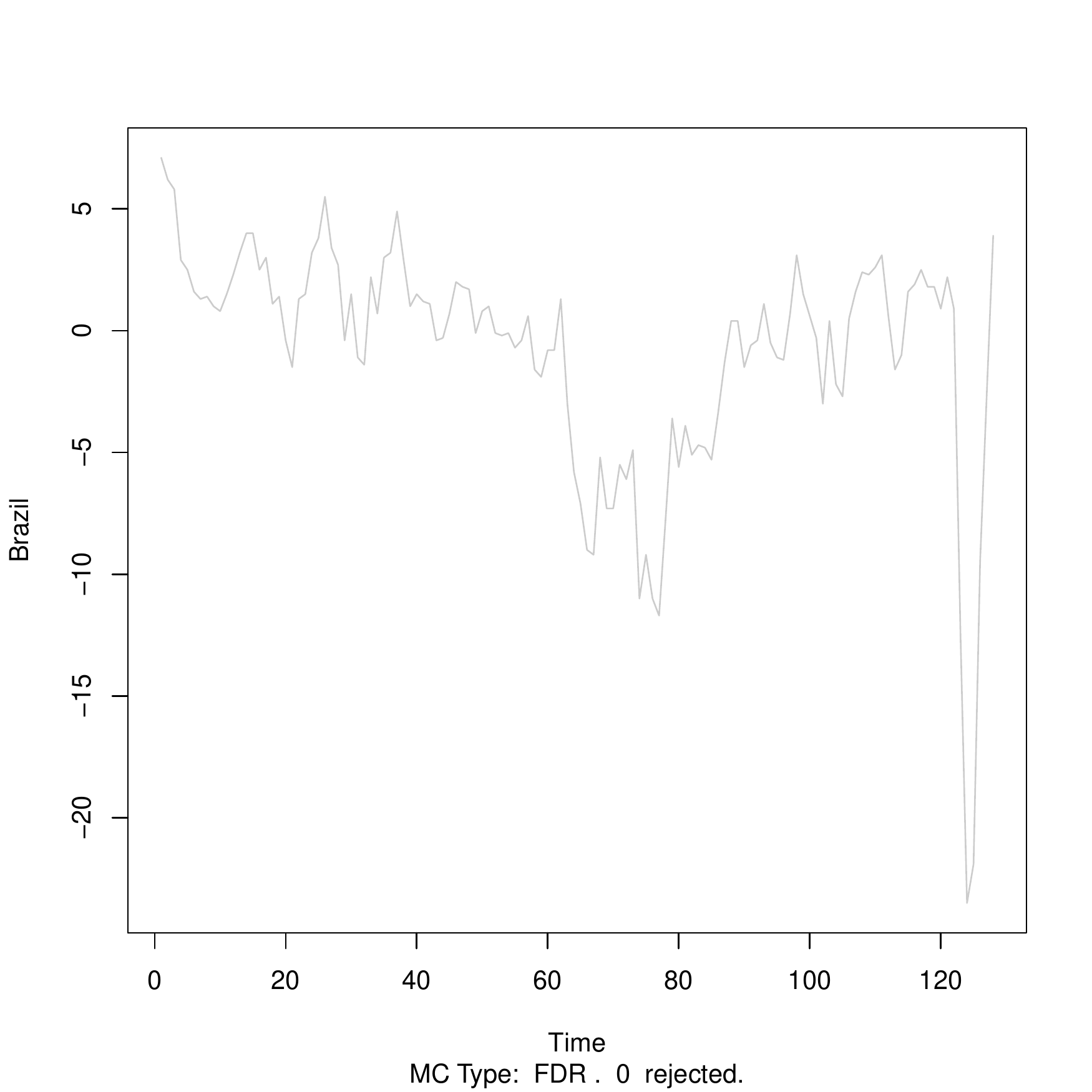}}
\caption{\label{fig:brazil} Stationarity test plot for Brazilian PMIs. No formal non-stationarities discovered}
\end{figure}

\begin{figure}[p]
\centering
\resizebox{\tw\textwidth}{!}{\includegraphics{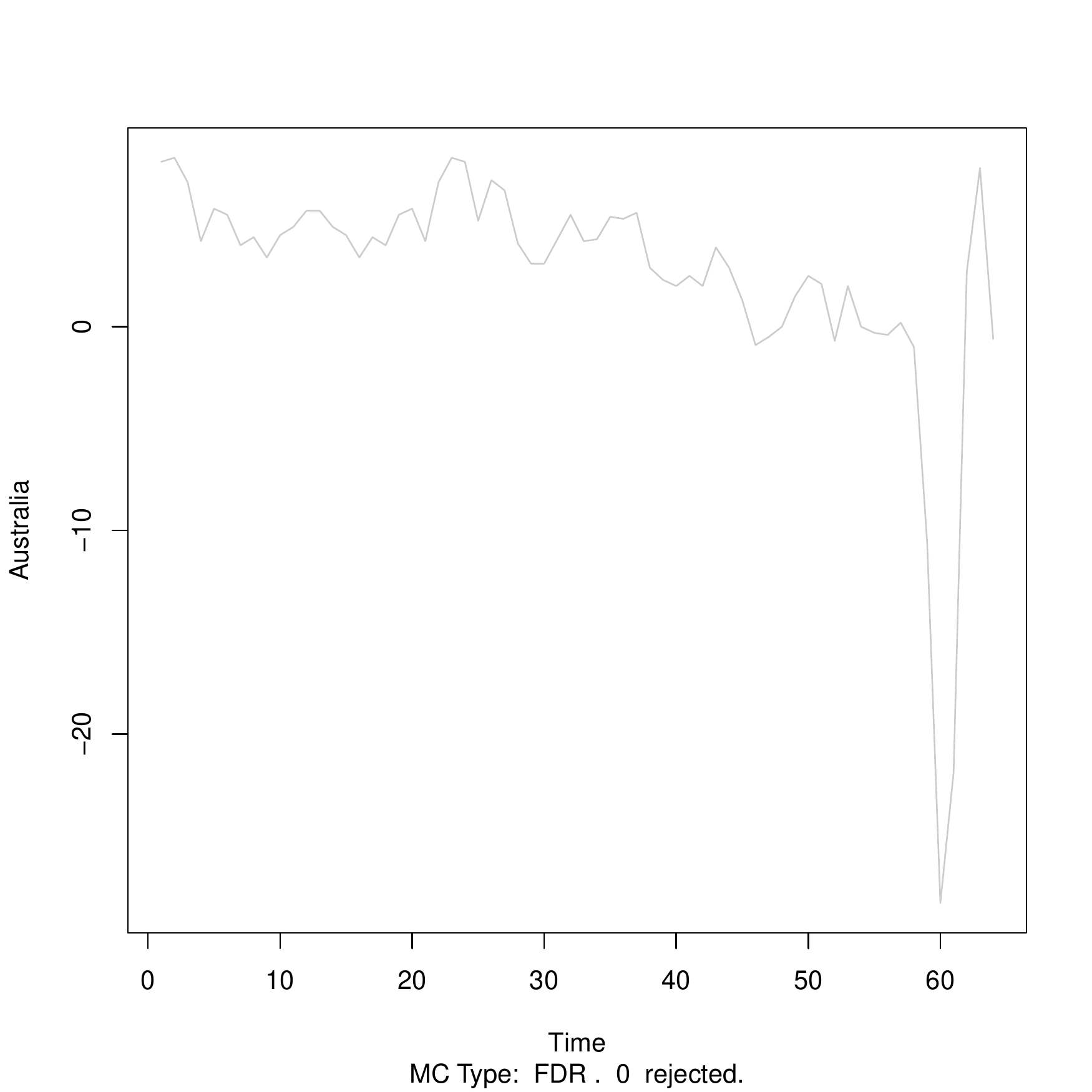}}
\caption{\label{fig:aus} Stationarity test plot for Australian PMIs. No formal non-stationarities discovered}

\end{figure}

\begin{figure}[p]
\centering
\resizebox{\tw\textwidth}{!}{\includegraphics{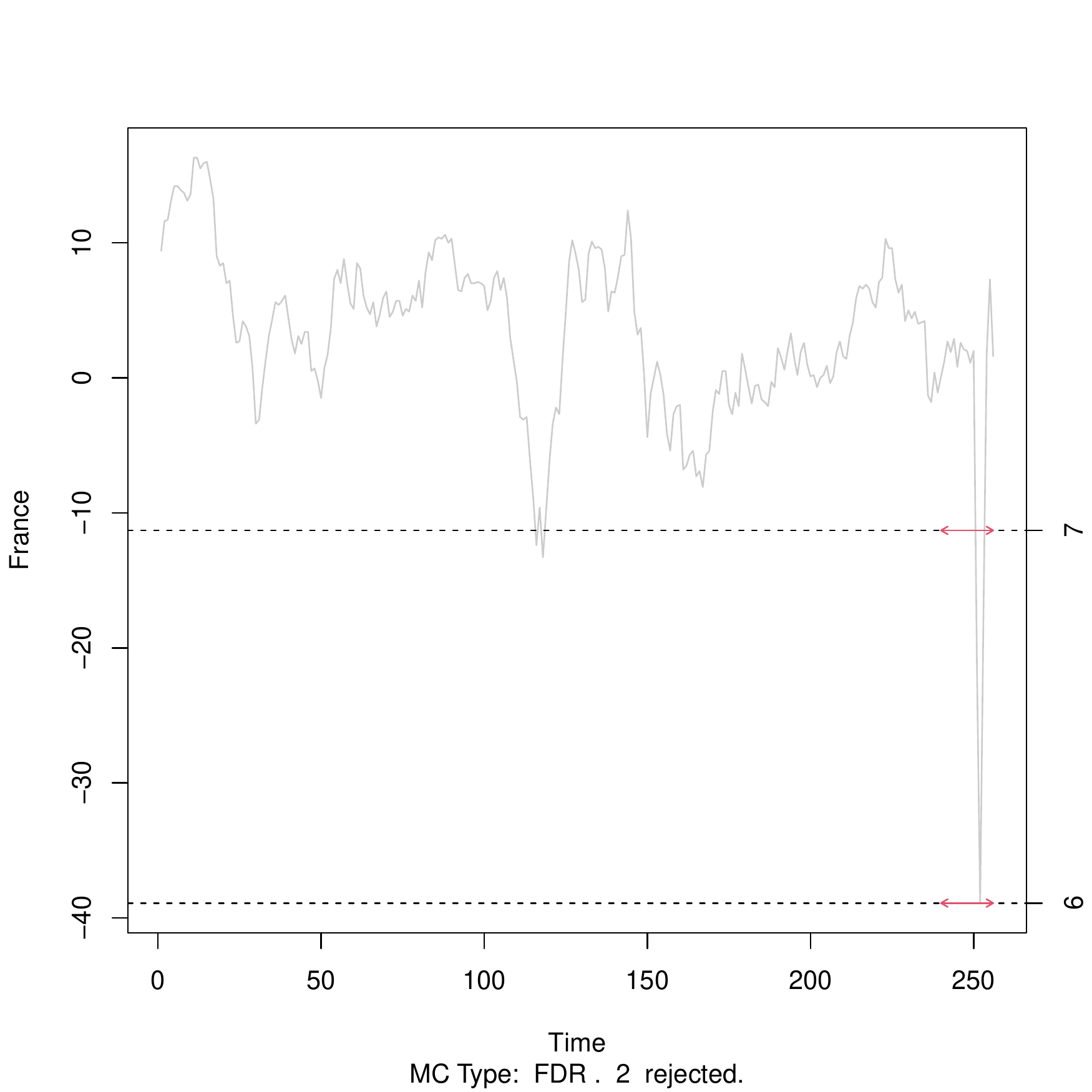}}
\caption{\label{fig:france} Stationarity test plot for French PMIs. Non-stationarities discovered during COVID-19.}
\end{figure}

We used the second-order stationary test from \cite{Nason13,Nason:locits} on the demeaned purchasing managers' indices from each country.
The results are summarised by Figures~\ref{fig:usa} to~\ref{fig:france}. Outside of COVID and, in a few cases, the great financial crisis, there is
little evidence to reject the null hypothesis of second-order stationarity. Table~\ref{tab:countriesnonstat} shows which countries had
no non-stationarities, those that exhibited non-stationarities during the COVID-19 pandemic and those that exhibited
them during the COVID-19 pandemic and great financial crisis.

Although some of these time series do exhibit non-stationarities, for forecasting the \textcolor{black}{2007--08 global} financial crisis was about 160 time periods ago, was relatively short-lived (with respect to the extent of the rest of the data) and will not unduly influence the forecasts (the methods we are using place
greater weight on more recent observations). Secondly, the non-stationarities due to the COVID-19 are precisely what we expect the exogenous
variables to model --- they, in effect, act as a kind of intervention variables.
\begin{table}
\caption{Countries' PMI non-stationarity status. \label{tab:countriesnonstat}}
\centering
\fbox{%
\begin{tabular}{c|c}
Non-stationarity type & Countries\\\hline
None & USA, Japan, India, China, Brazil, Australia  \\
COVID only & Spain, Italy, Germany, France\\
COVID + GFC & UK, Russia, Ireland \\ \hline
\end{tabular}}
\end{table}

%
%

\section{\textcolor{black}{Covariate plots}}
\begin{figure}[p]
\centering
\resizebox{\tw\textwidth}{!}{\includegraphics{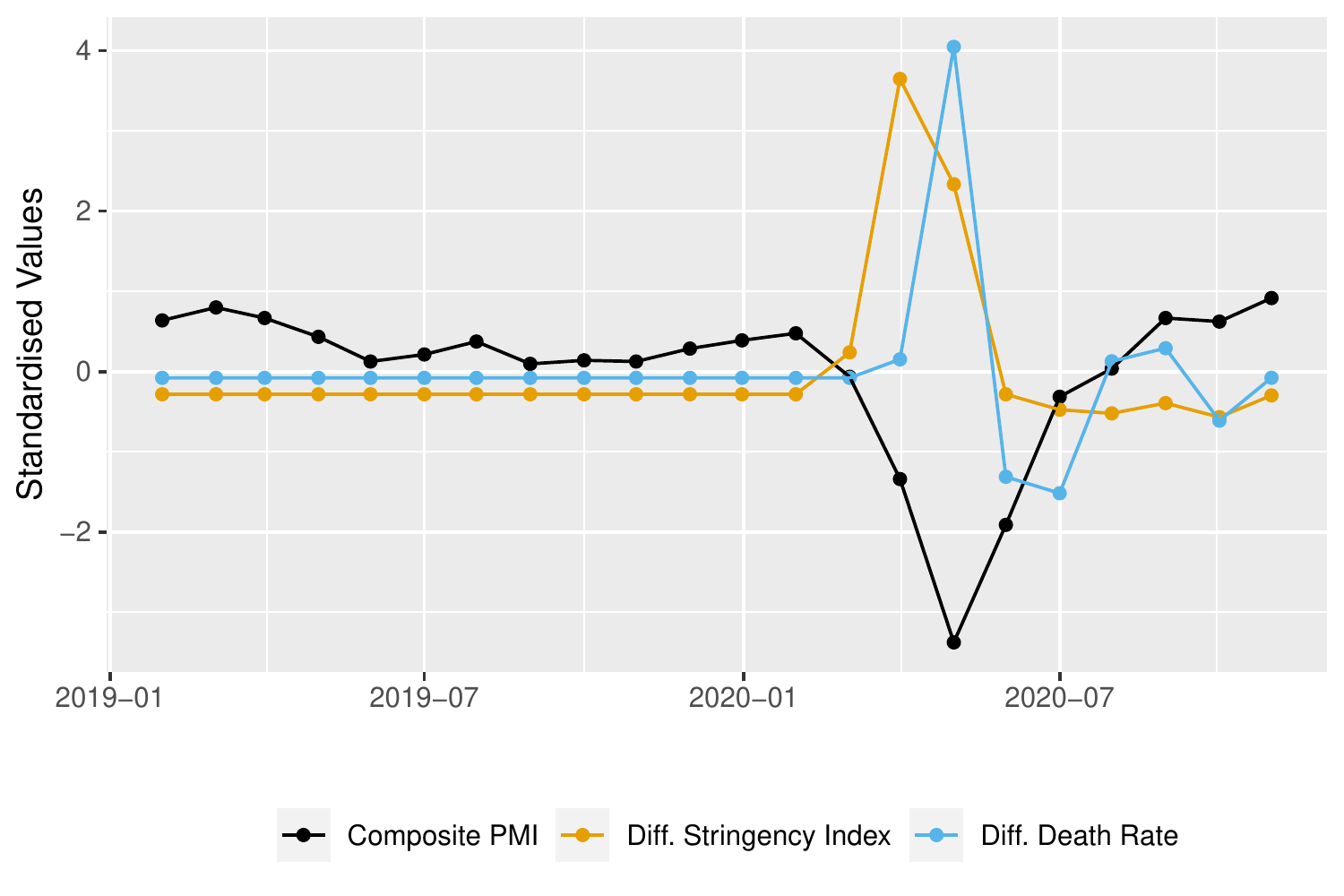}}
\caption{\textcolor{black}{Plot of US Composite PMI together with US stringency index and US death rate covariates, where all variables have been standardised to zero mean and unit variance.}\label{fig:covus}}
\end{figure}

\textcolor{black}{Below, Figures~\ref{fig:covus} to~\ref{fig:covfrance} show the plots of covariates for all country nodes, excluding the UK (which can be found in the main article as Figure~2). All series have been standardised to zero mean and unit variance so that they may be plotted on the same axes.}

\begin{figure}[p]
\centering
\resizebox{\tw\textwidth}{!}{\includegraphics{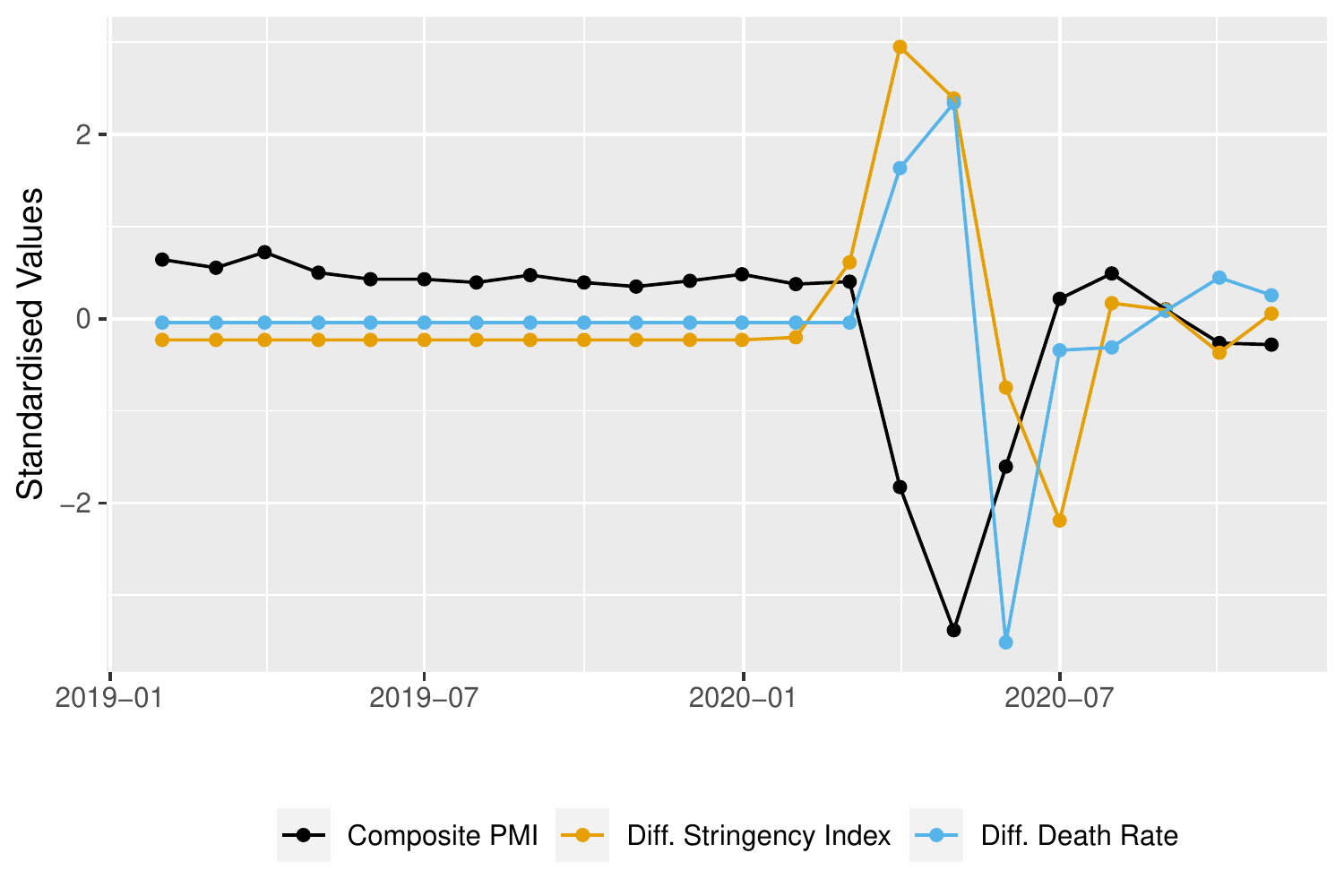}}
\caption{\textcolor{black}{Plot of Spain Composite PMI together with Spain stringency index and Spain death rate covariates, where all variables have been standardised to zero mean and unit variance.}}
\end{figure}

\begin{figure}[p]
\centering
\resizebox{\tw\textwidth}{!}{\includegraphics{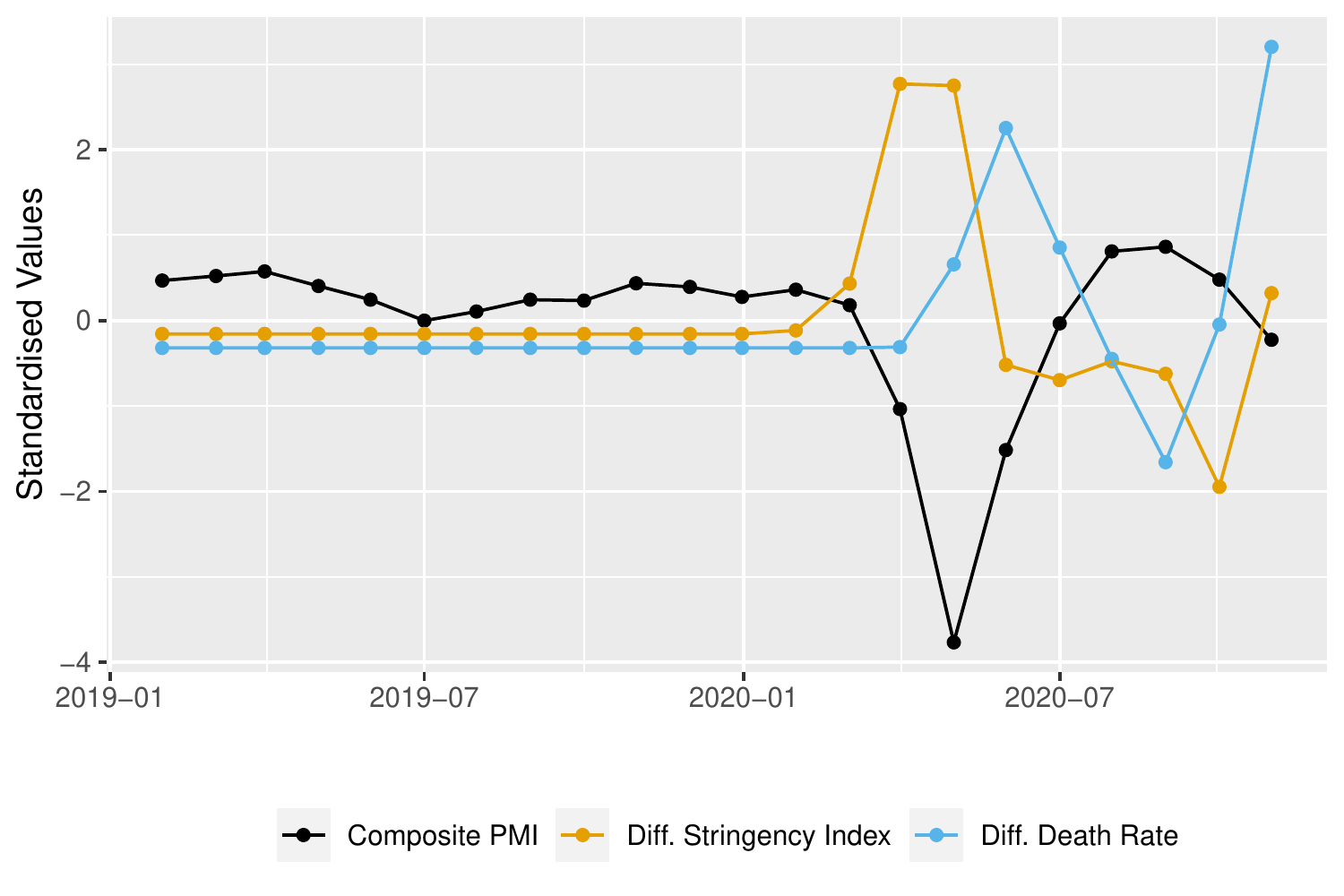}}
\caption{\textcolor{black}{Plot of Russia Composite PMI together with Russia stringency index and Russia death rate covariates, where all variables have been standardised to zero mean and unit variance.}}
\end{figure}

\begin{figure}[p]
\centering
\resizebox{\tw\textwidth}{!}{\includegraphics{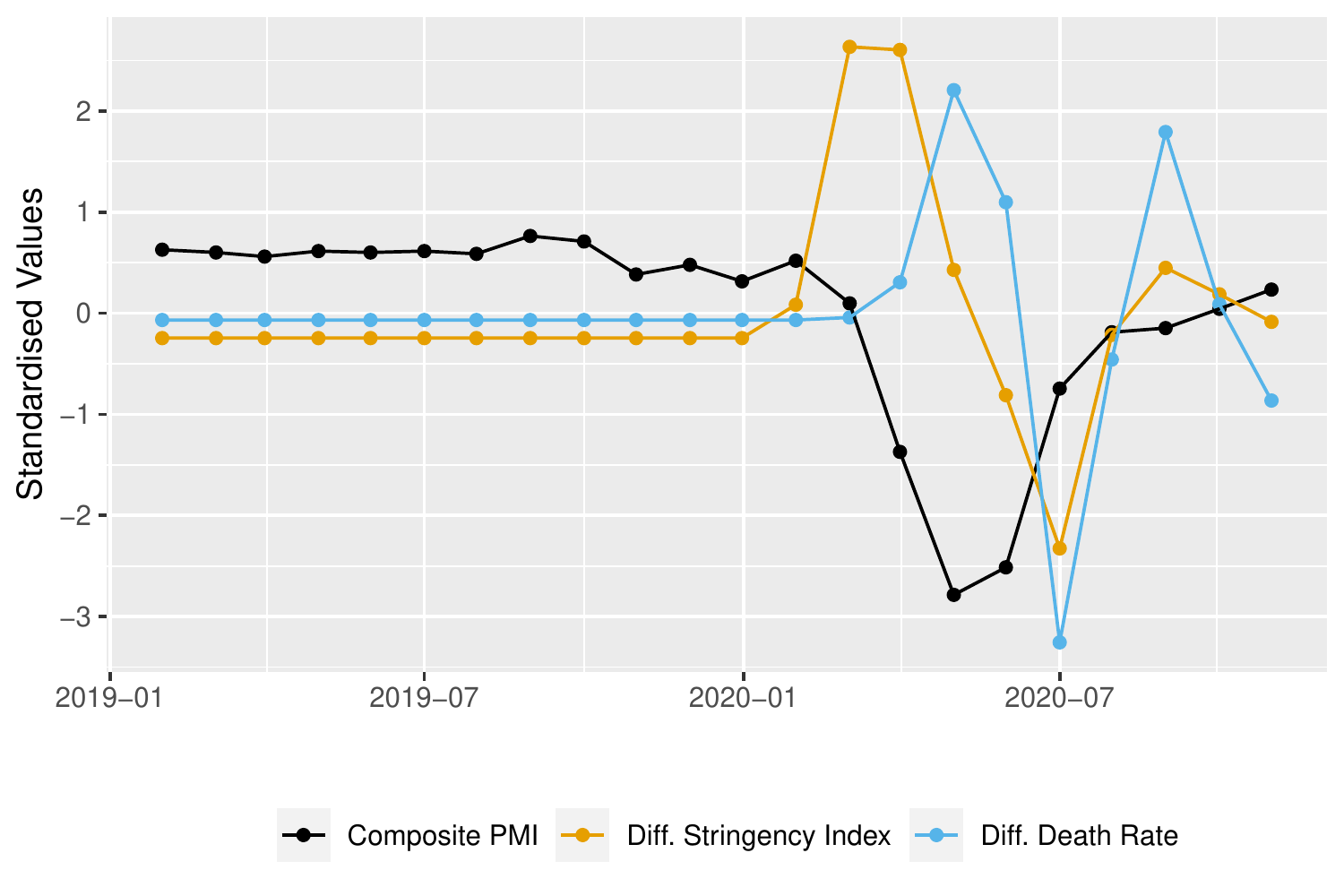}}
\caption{\textcolor{black}{Plot of Japan Composite PMI together with Japan stringency index and Japan death rate covariates, where all variables have been standardised to zero mean and unit variance.}}
\end{figure}

\begin{figure}[p]
\centering
\resizebox{\tw\textwidth}{!}{\includegraphics{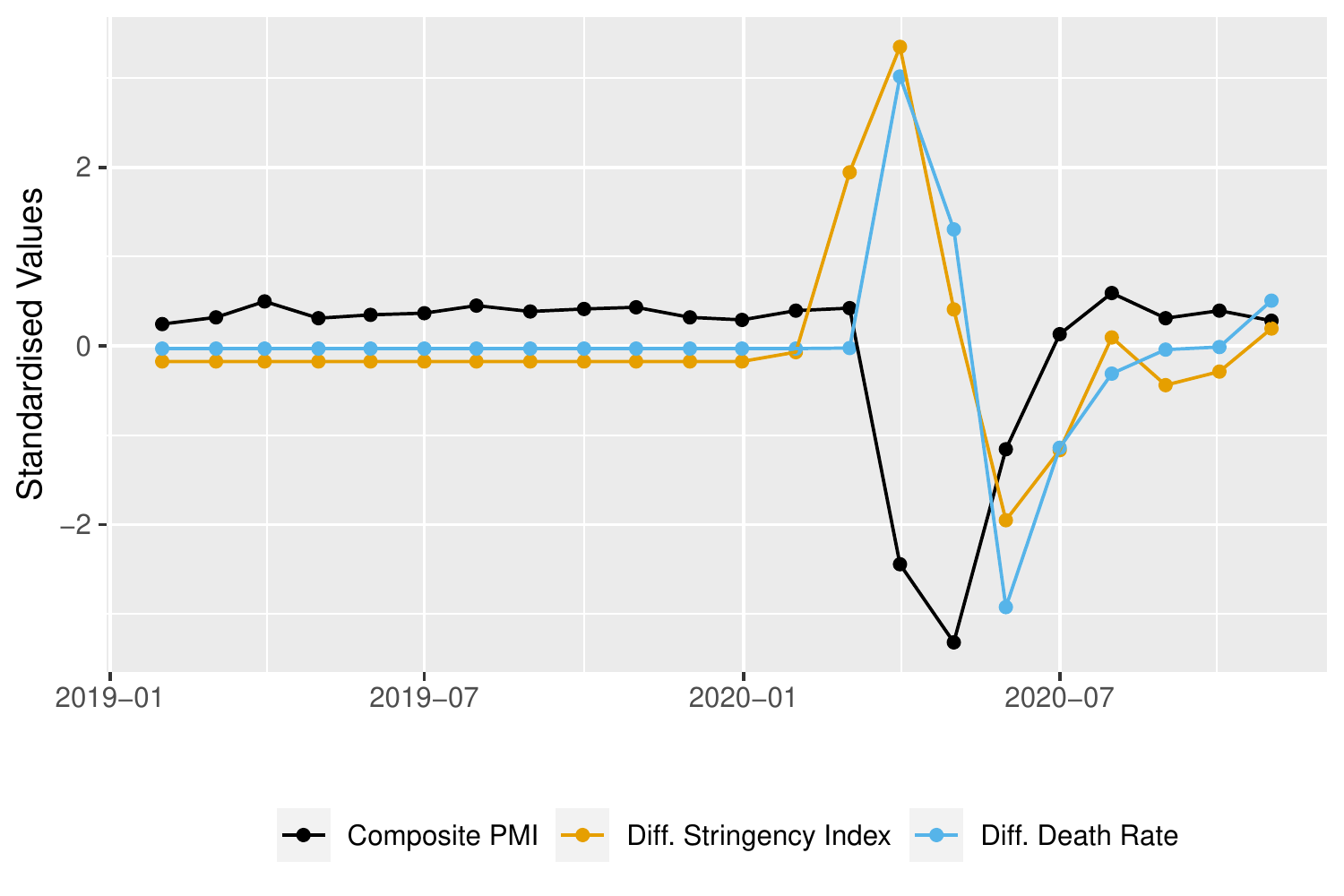}}
\caption{\textcolor{black}{Plot of Italy Composite PMI together with Italy stringency index and Italy death rate covariates, where all variables have been standardised to zero mean and unit variance.}}
\end{figure}

\begin{figure}[p]
\centering
\resizebox{\tw\textwidth}{!}{\includegraphics{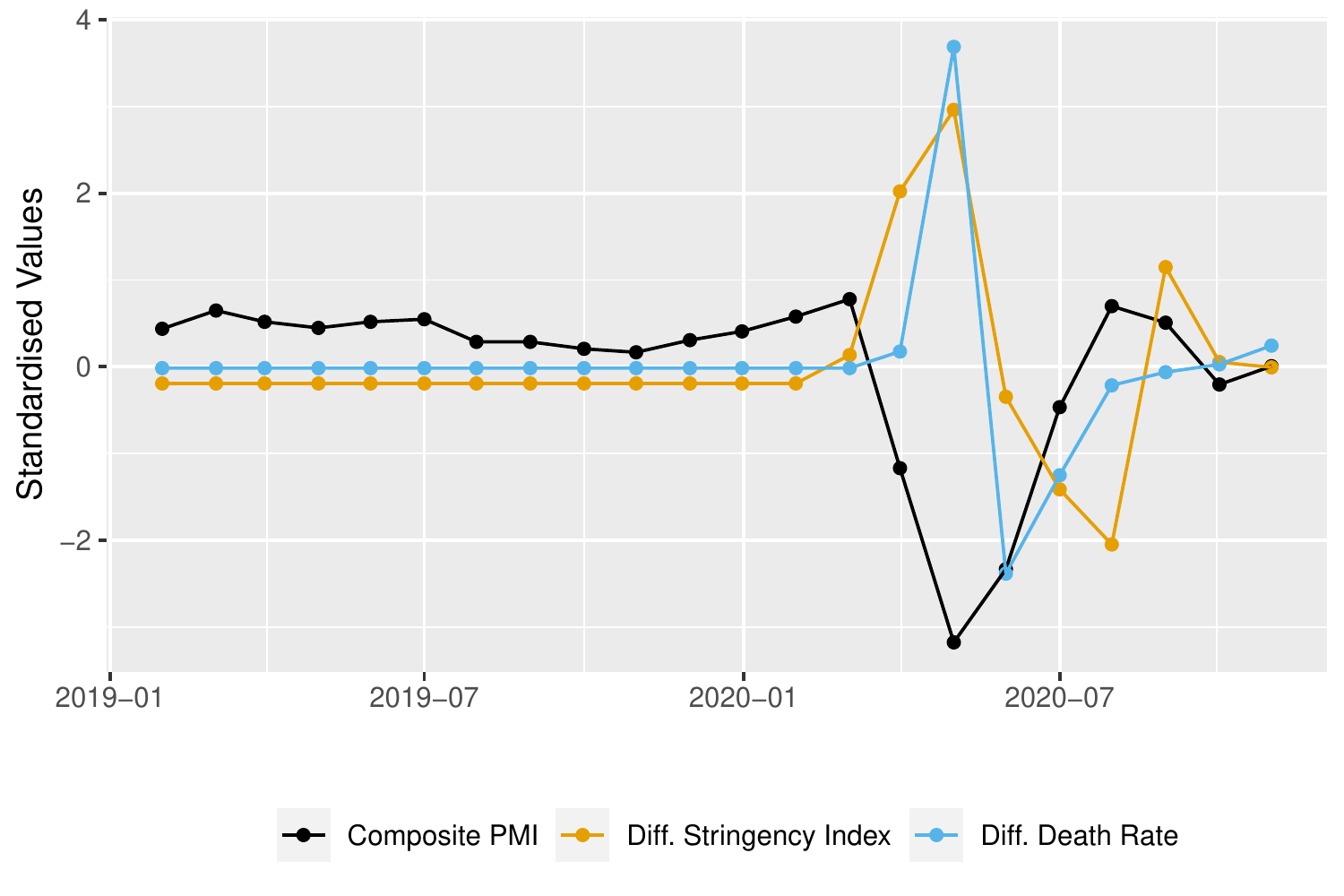}}
\caption{\textcolor{black}{Plot of Ireland Composite PMI together with Ireland stringency index and Ireland death rate covariates, where all variables have been standardised to zero mean and unit variance.}}
\end{figure}

\begin{figure}[p]
\centering
\resizebox{\tw\textwidth}{!}{\includegraphics{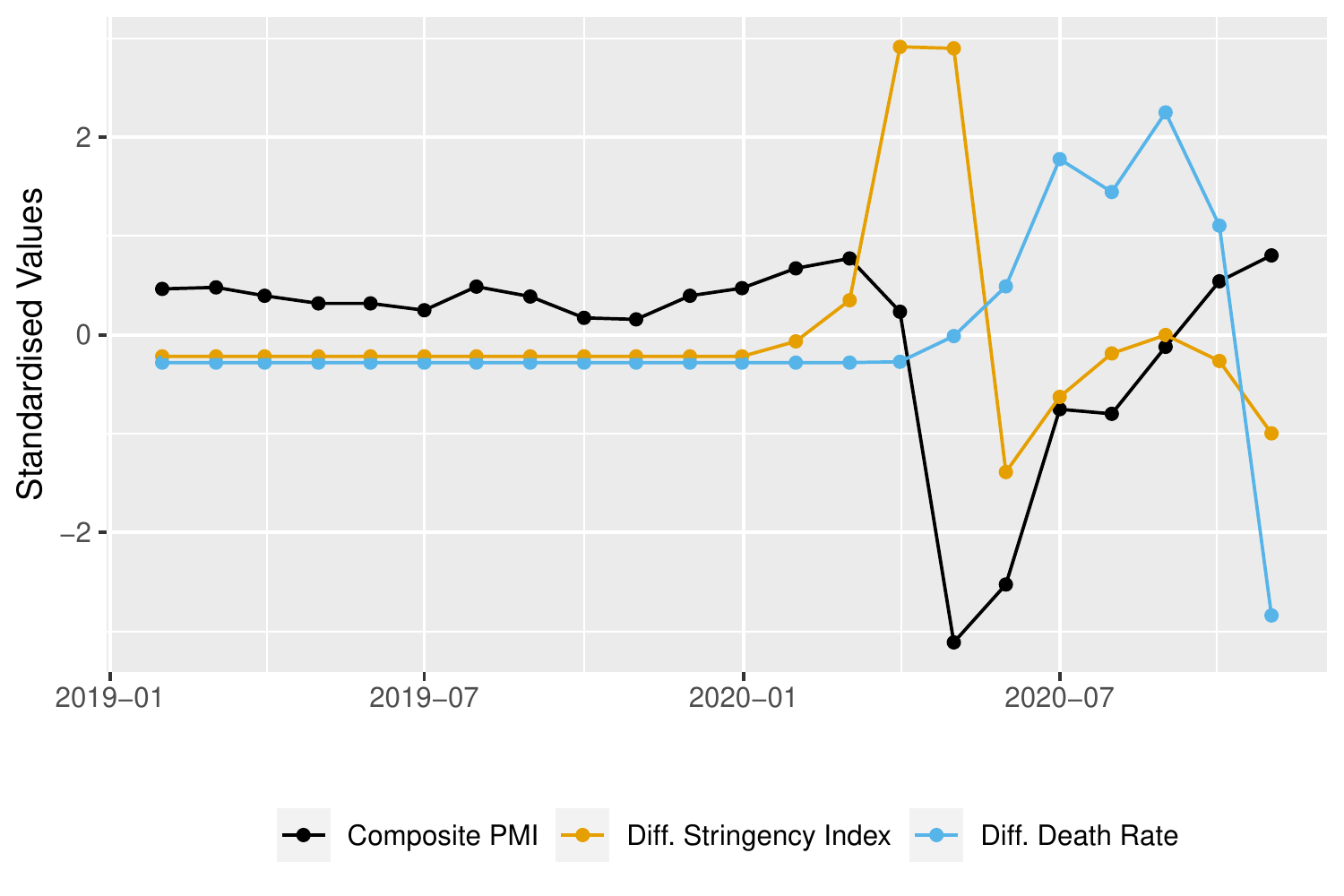}}
\caption{\textcolor{black}{Plot of India Composite PMI together with India stringency index and India death rate covariates, where all variables have been standardised to zero mean and unit variance.}}
\end{figure}

\begin{figure}[p]
\centering
\resizebox{\tw\textwidth}{!}{\includegraphics{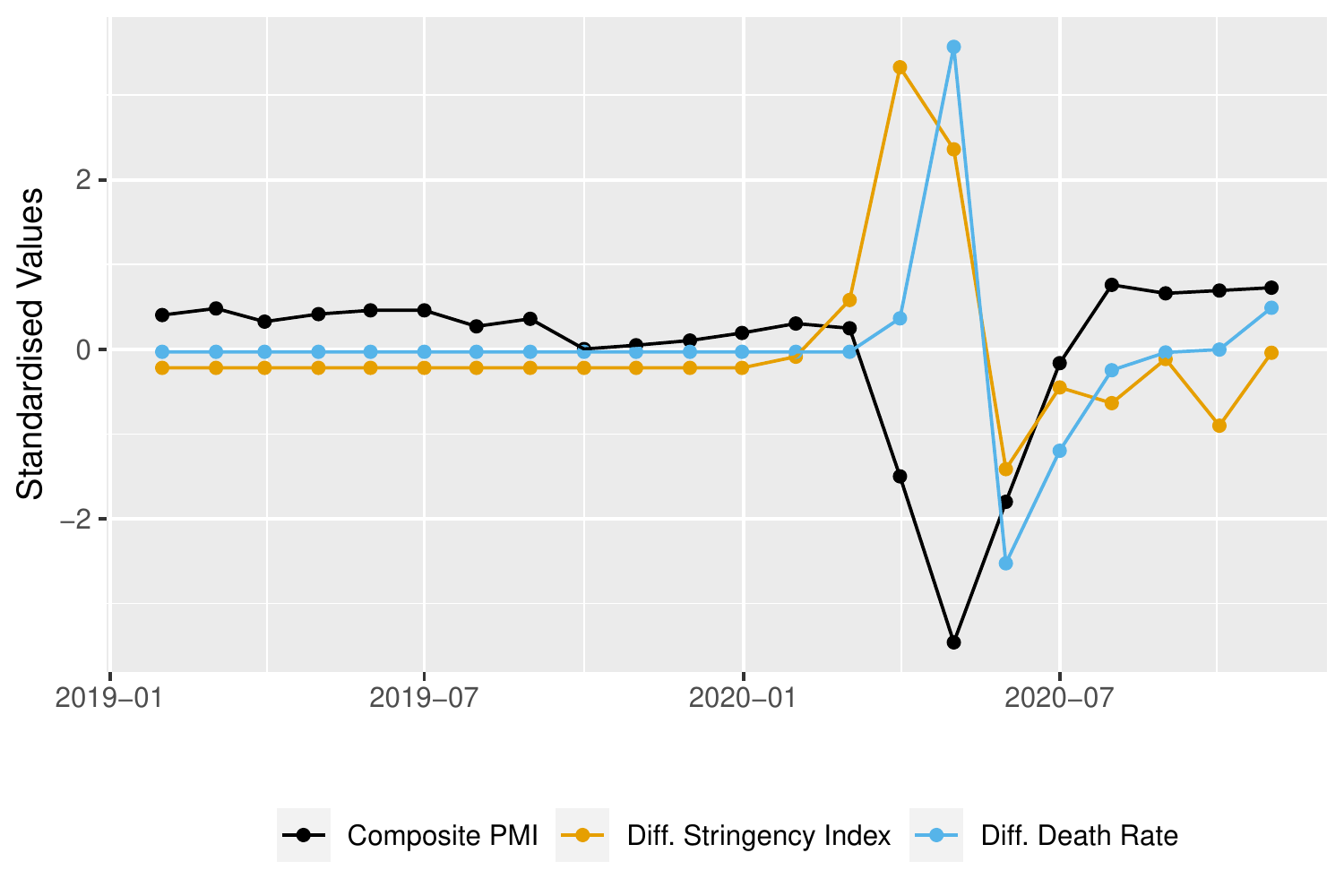}}
\caption{\textcolor{black}{Plot of Germany Composite PMI together with Germany stringency index and Germany death rate covariates, where all variables have been standardised to zero mean and unit variance.}}
\end{figure}

\begin{figure}[p]
\centering
\resizebox{\tw\textwidth}{!}{\includegraphics{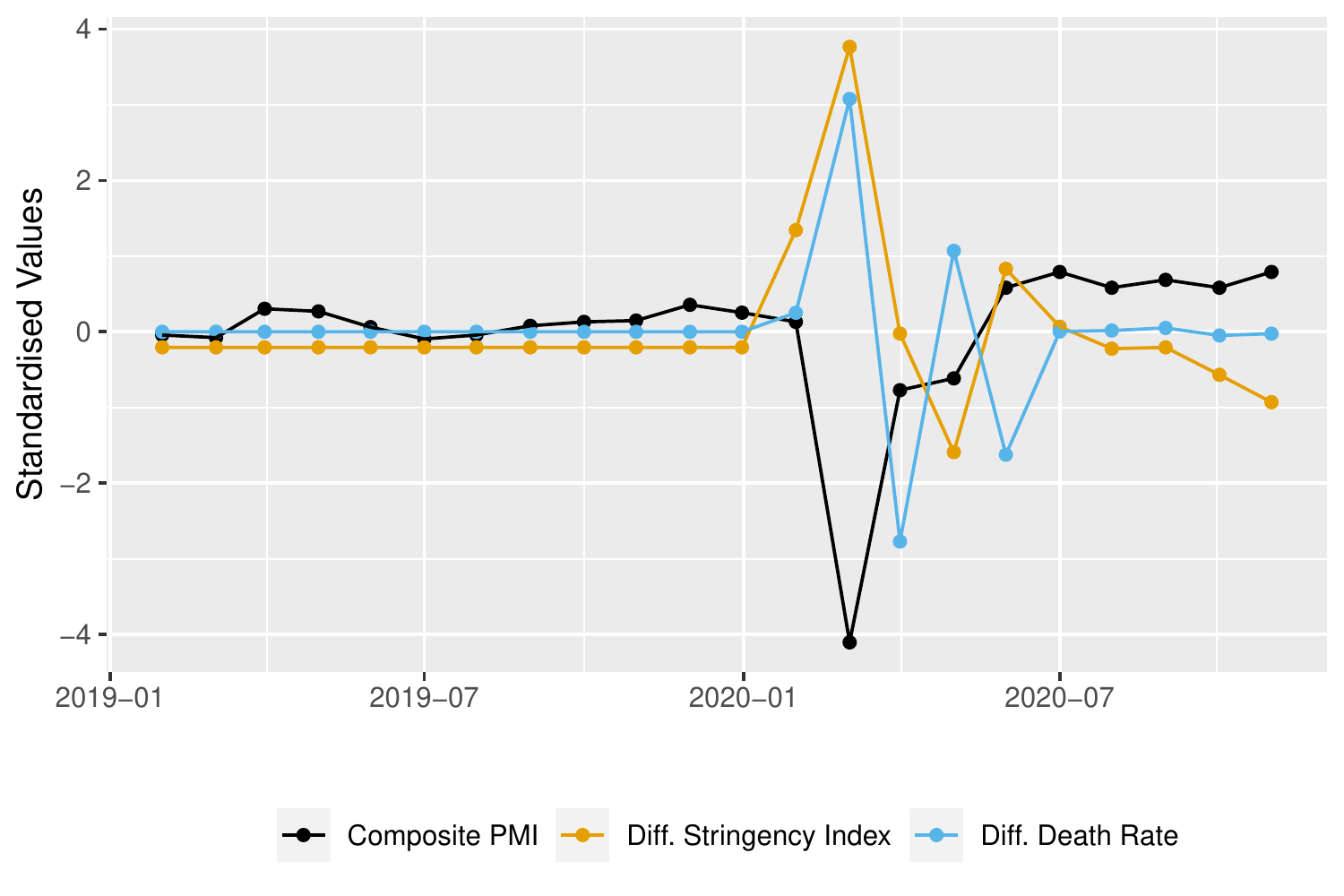}}
\caption{\textcolor{black}{Plot of China Composite PMI together with China stringency index and China death rate covariates, where all variables have been standardised to zero mean and unit variance.}}
\end{figure}

\begin{figure}[p]
\centering
\resizebox{\tw\textwidth}{!}{\includegraphics{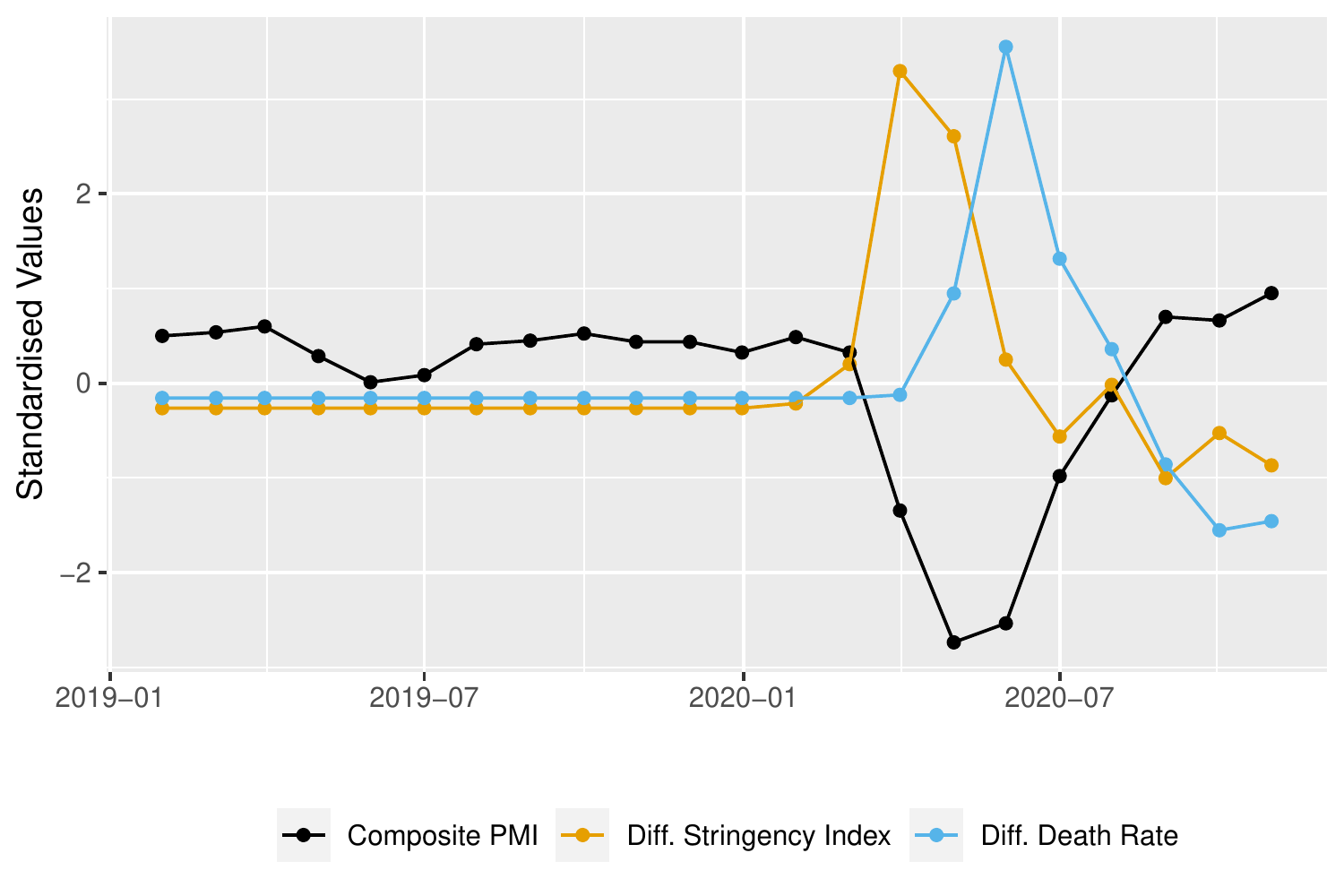}}
\caption{\textcolor{black}{Plot of Brazil Composite PMI together with Brazil stringency index and Brazil death rate covariates, where all variables have been standardised to zero mean and unit variance.}}
\end{figure}
\textbf{}

\begin{figure}[p]
\centering
\resizebox{\tw\textwidth}{!}{\includegraphics{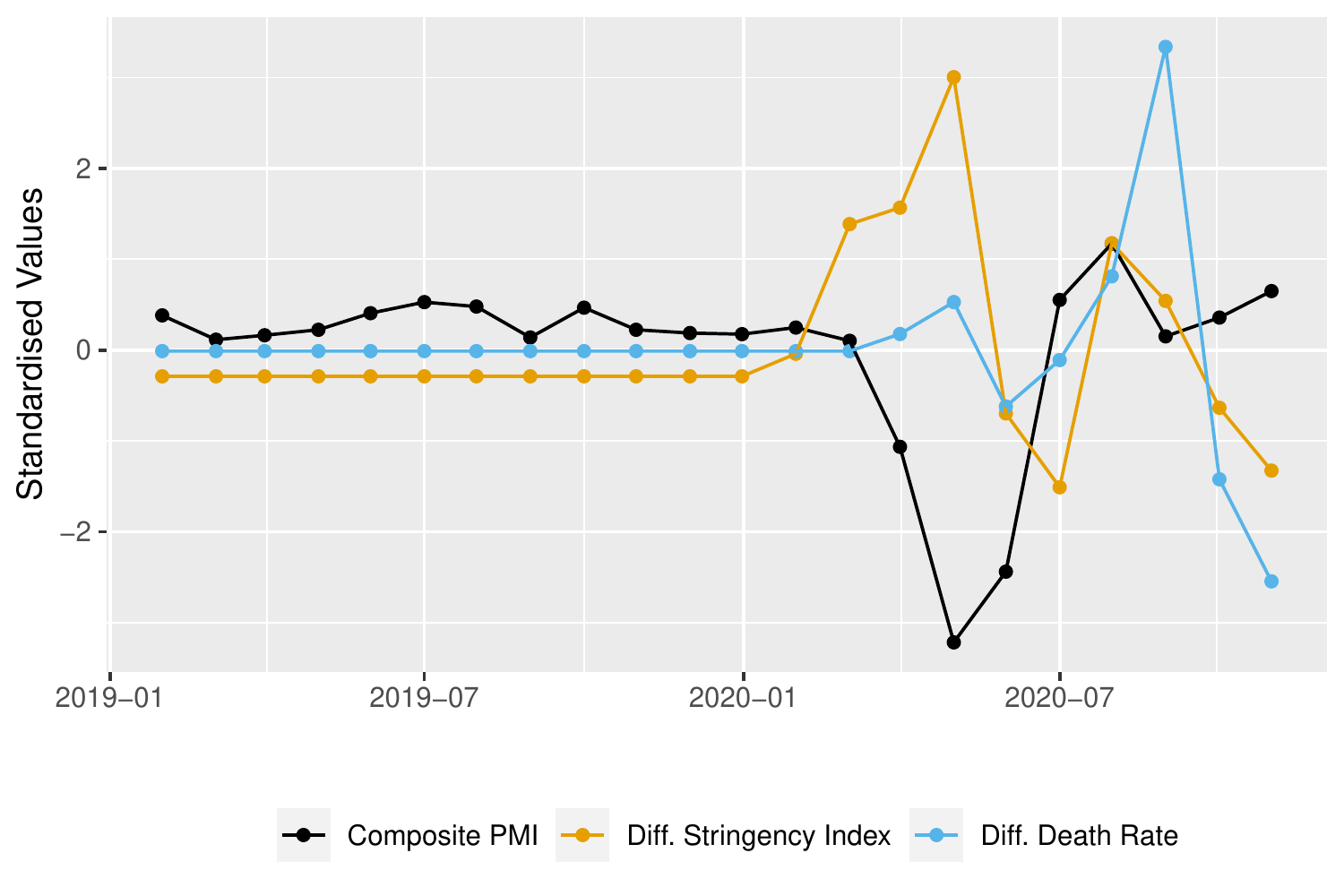}}
\caption{\textcolor{black}{Plot of Australia Composite PMI together with Australia stringency index and Australia death rate covariates, where all variables have been standardised to zero mean and unit variance.}}
\end{figure}
\textbf{}

\begin{figure}[p]
\centering
\resizebox{\tw\textwidth}{!}{\includegraphics{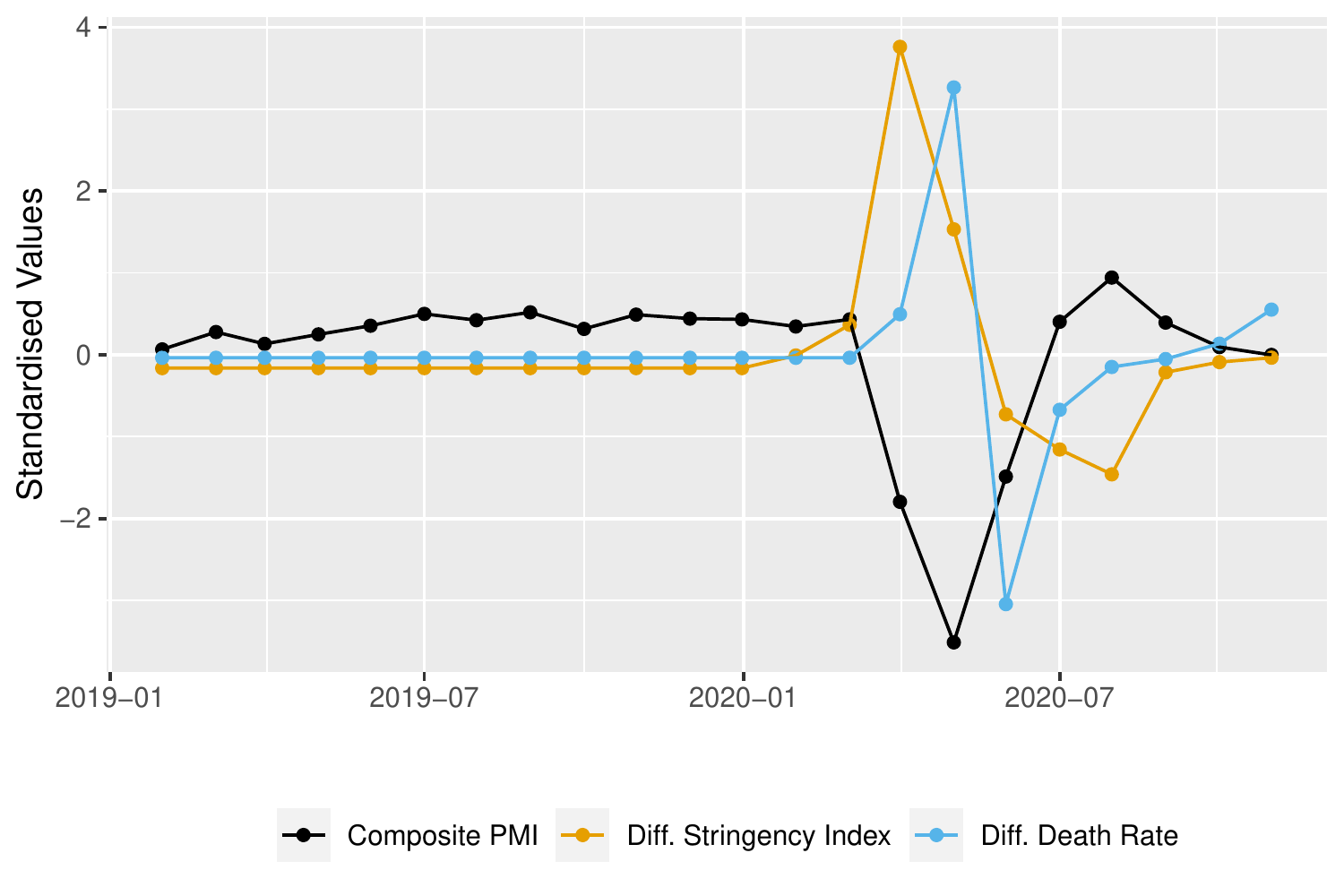}}
\caption{\textcolor{black}{Plot of France Composite PMI together with France stringency index and France death rate covariates, where all variables have been standardised to zero mean and unit variance.\label{fig:covfrance}}}
\end{figure}
\textbf{}

\clearpage

{\bf \Large \color{black} \em Part 3: GNAR and GNARX experiments}

%
%

\section{Additional GNAR \textcolor{black}{experimental results}}

\begin{figure}[p]
    \centering
    \includegraphics[width=0.5\textwidth]{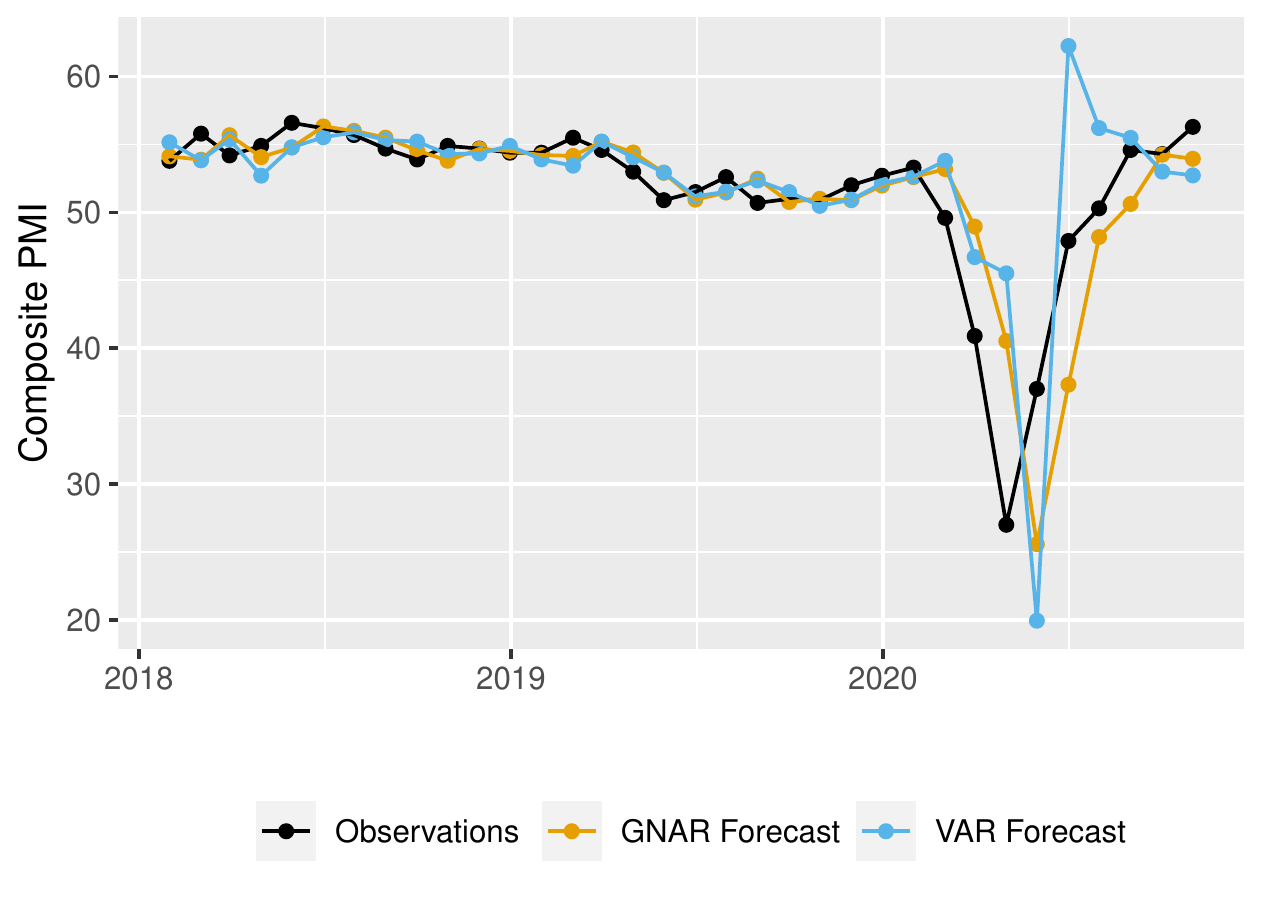}
    \caption{United States: Out-of-sample one-step-ahead rolling composite PMI forecasts of the local-$\alpha$ GNAR($1, 1$) model and the VAR($2$) model.}
    \label{fig:USA}
\end{figure}

\begin{figure}[p]
    \centering
    \includegraphics[width=0.5\textwidth]{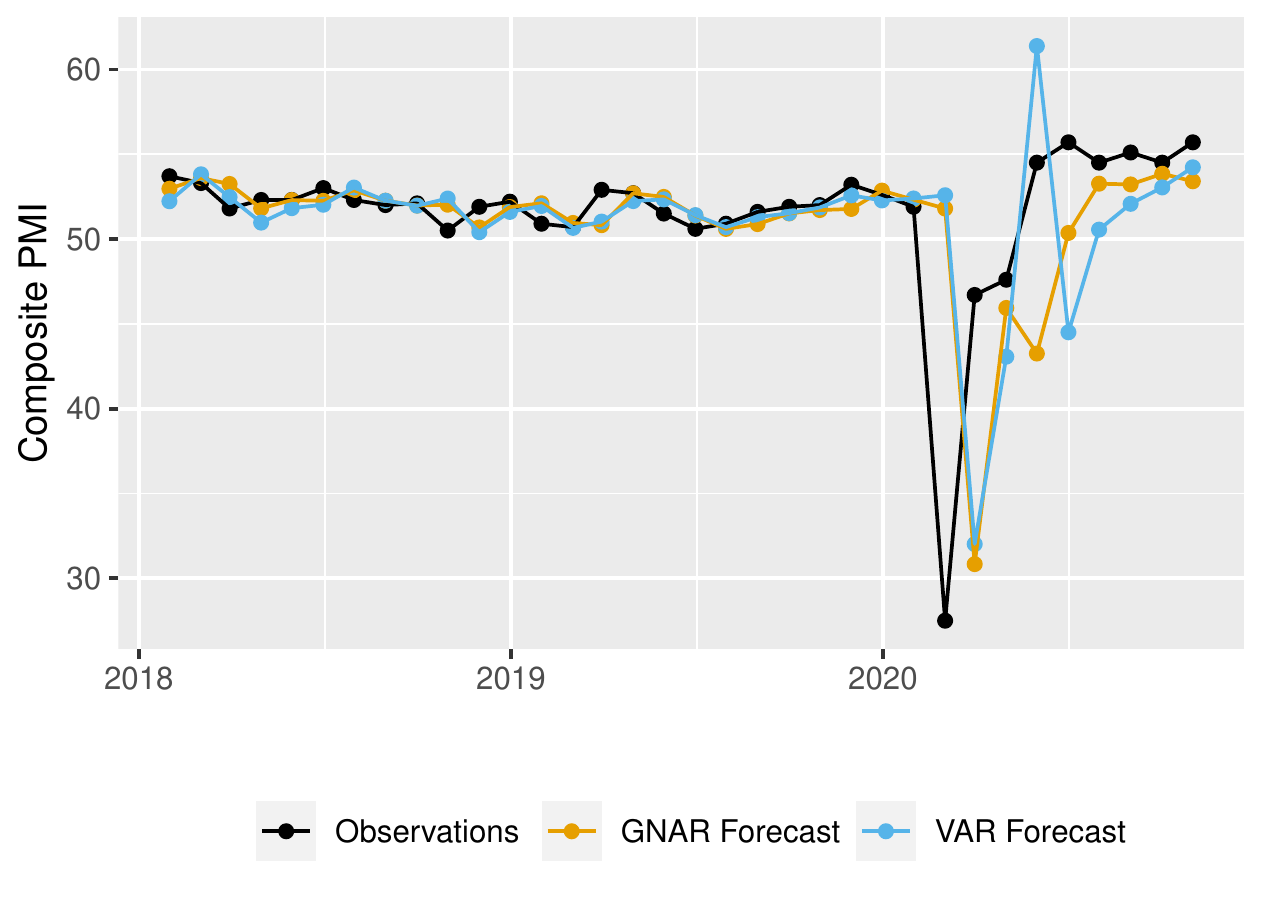}
    \caption{China: Out-of-sample one-step-ahead rolling composite PMI forecasts of the local-$\alpha$ GNAR($1, 1$) model and the VAR($2$) model.}
    \label{fig:China}
\end{figure}

\begin{figure}[p]
    \centering
    \includegraphics[width=0.5\textwidth]{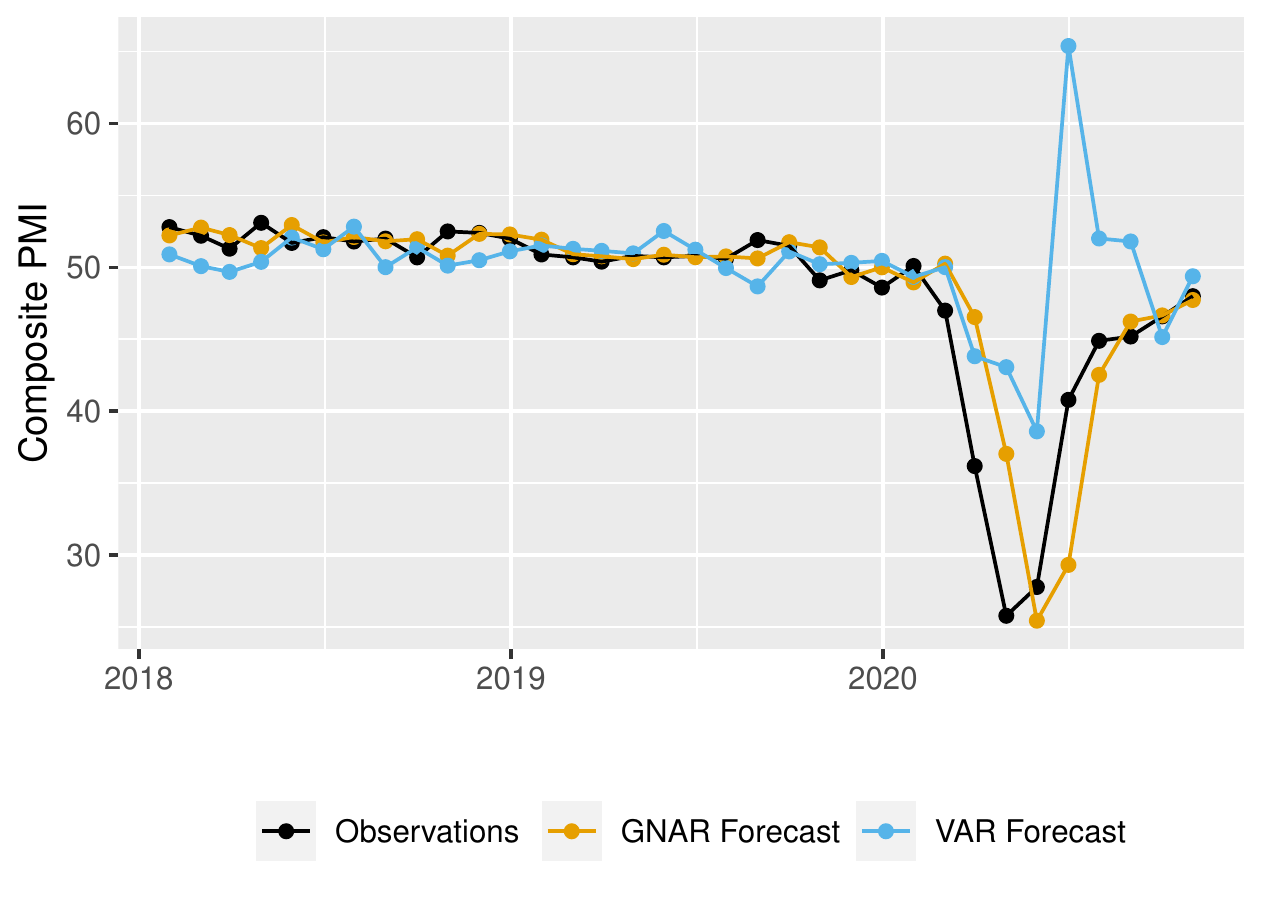}
    \caption{Japan: Out-of-sample one-step-ahead rolling composite PMI forecasts of the local-$\alpha$ GNAR($1, 1$) model and the VAR($2$) model.}
    \label{fig:Japan}
\end{figure}

\begin{figure}[p]
    \centering
    \includegraphics[width=0.5\textwidth]{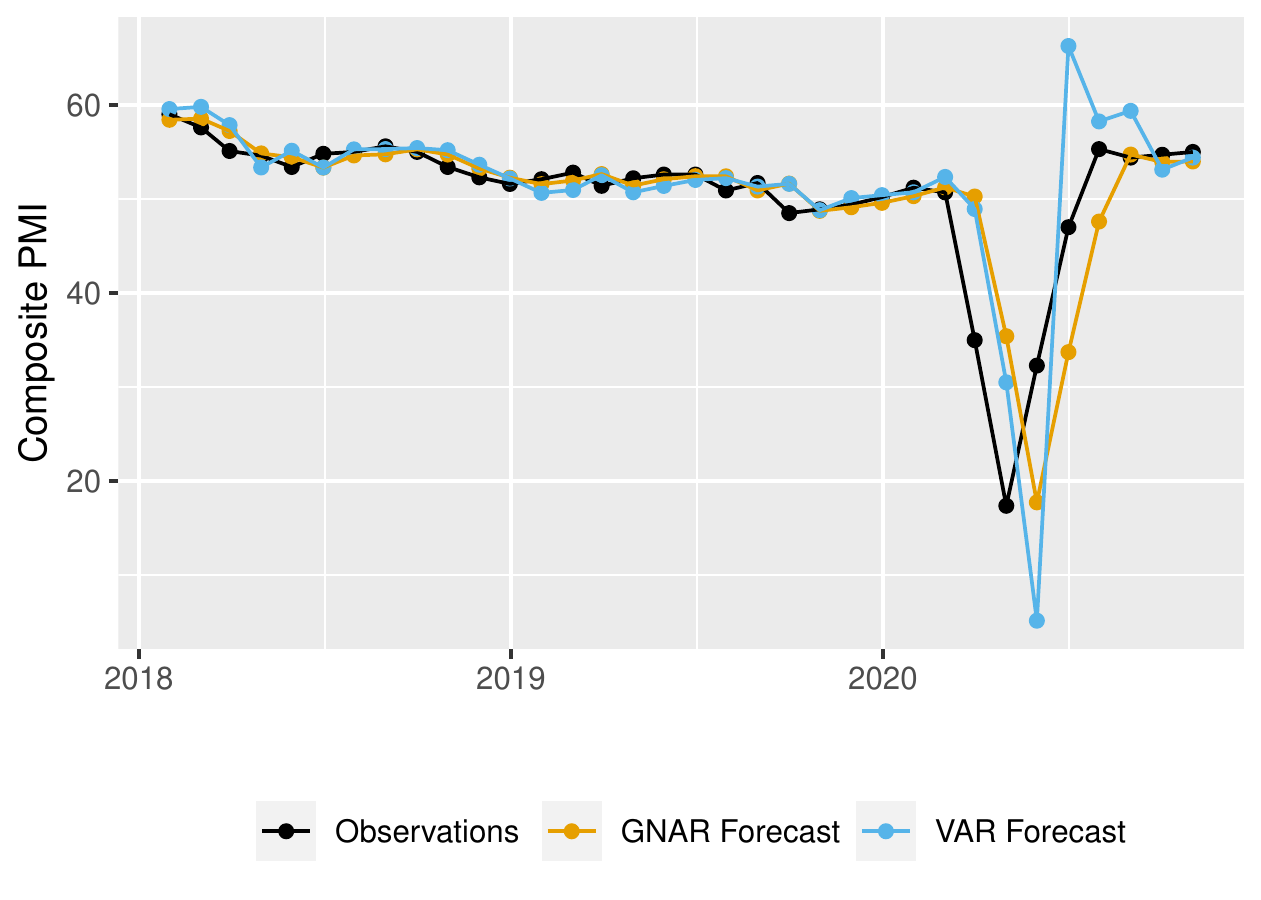}
    \caption{Germany: Out-of-sample one-step-ahead rolling composite PMI forecasts of the local-$\alpha$ GNAR($1, 1$) model and the VAR($2$) model.}
    \label{fig:Germany}
\end{figure}

\begin{figure}[p]
    \centering
    \includegraphics[width=0.5\textwidth]{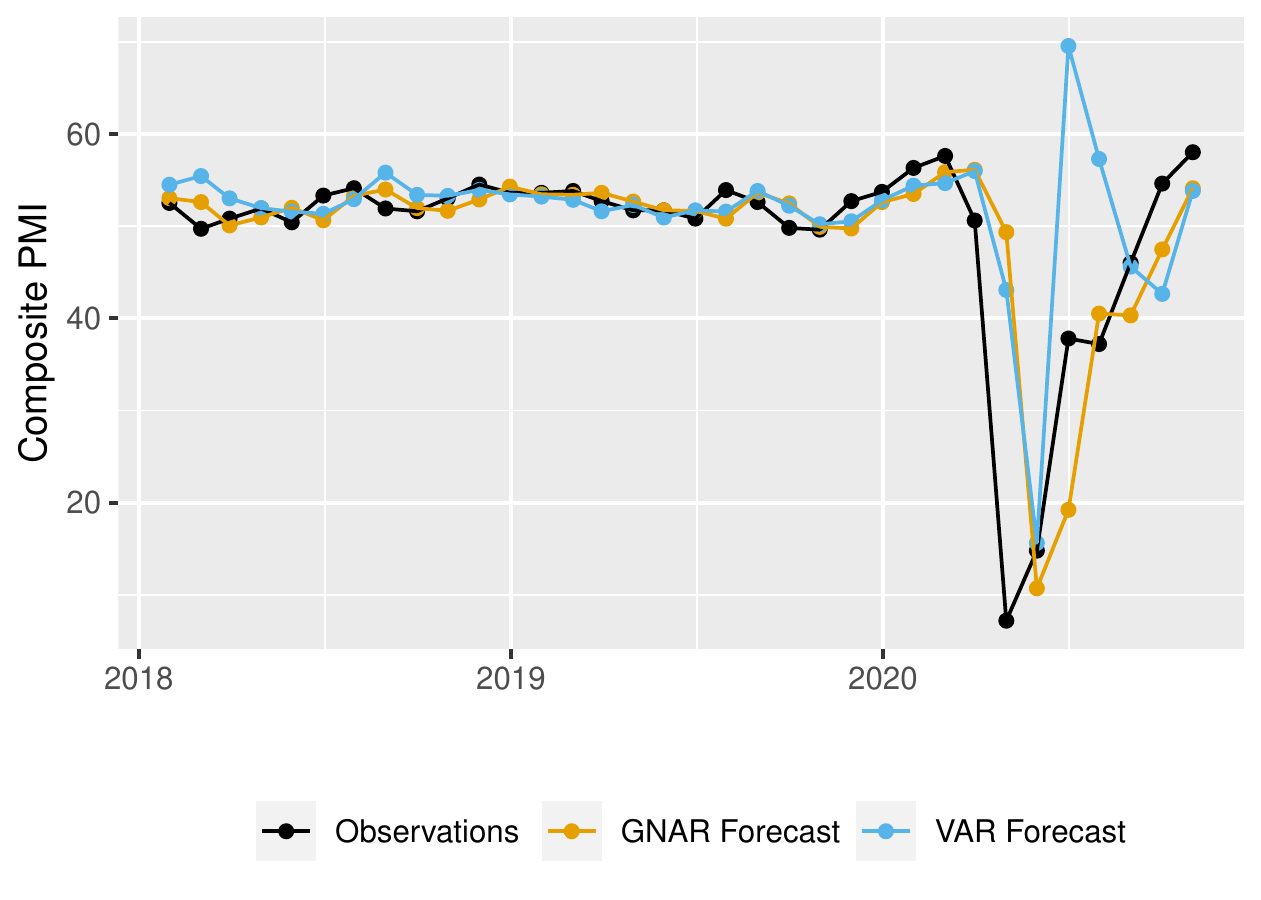}
    \caption{India: Out-of-sample one-step-ahead rolling composite PMI forecasts of the local-$\alpha$ GNAR($1, 1$) model and the VAR($2$) model.}
    \label{fig:India}
\end{figure}

\begin{figure}[p]
    \centering
    \includegraphics[width=0.5\textwidth]{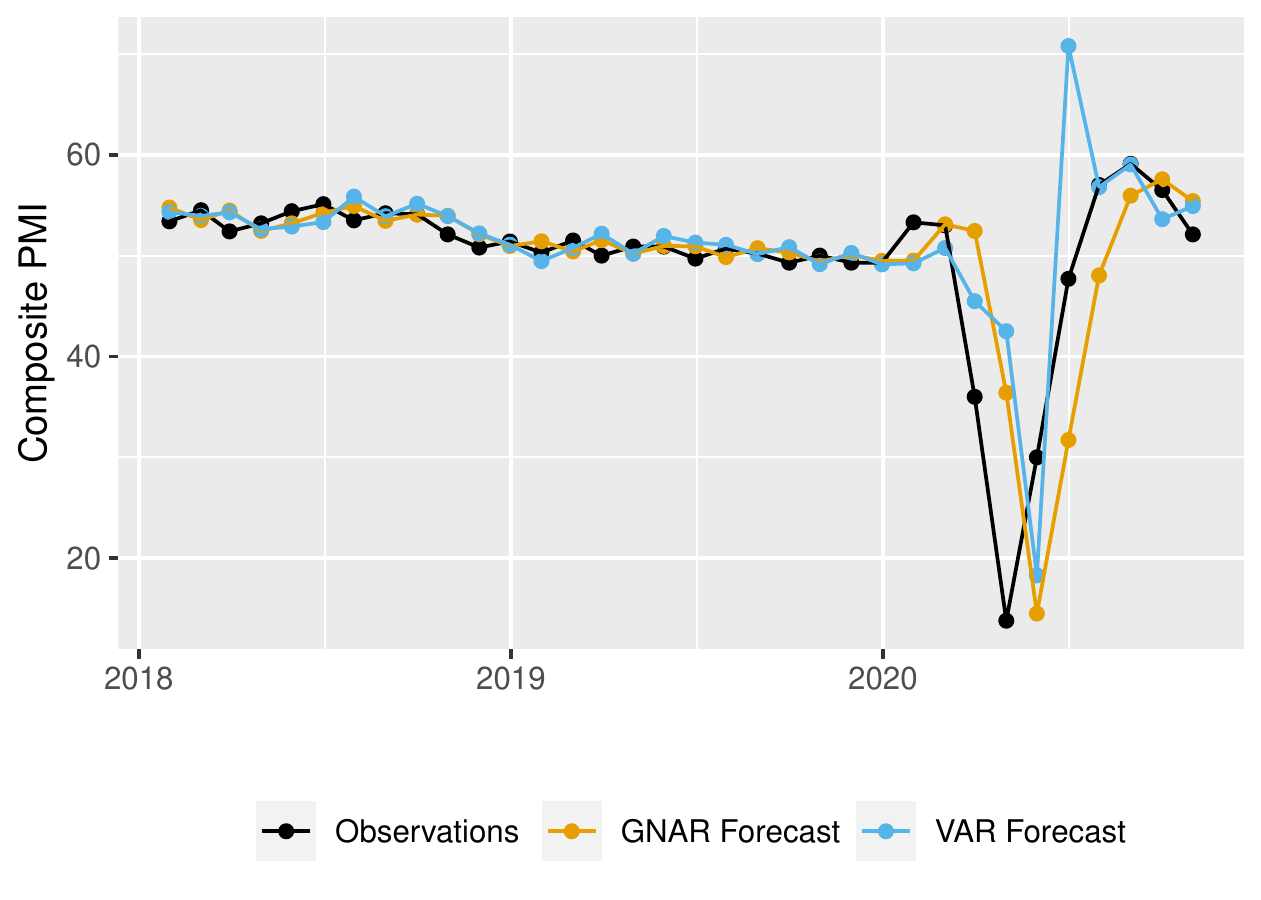}
    \caption{United Kingdom: Out-of-sample one-step-ahead rolling composite PMI forecasts of the local-$\alpha$ GNAR($1, 1$) model and the VAR($2$) model.}
    \label{fig:UK}
\end{figure}

\begin{figure}[p]
    \centering
    \includegraphics[width=0.5\textwidth]{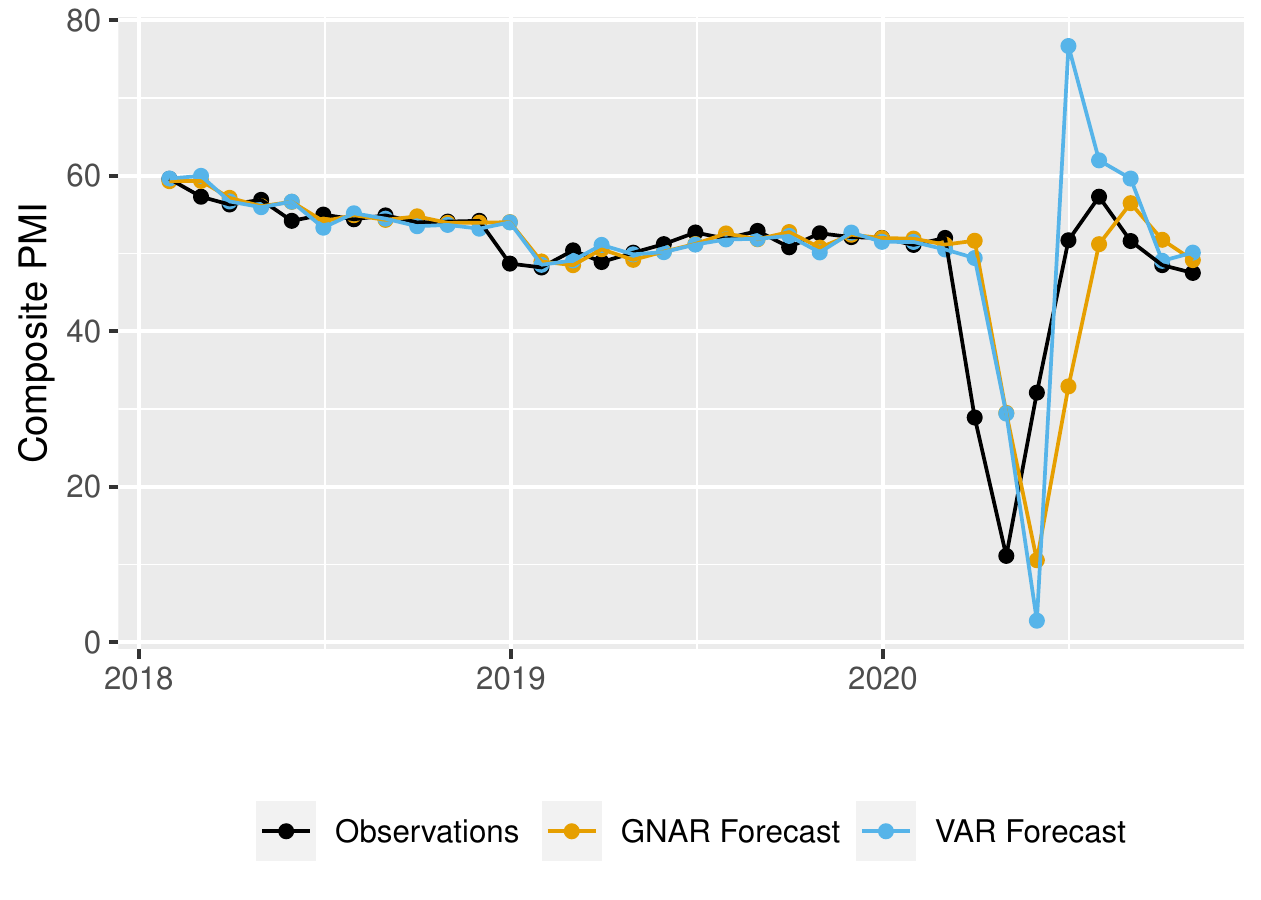}
    \caption{France: Out-of-sample one-step-ahead rolling composite PMI forecasts of the local-$\alpha$ GNAR($1, 1$) model and the VAR($2$) model.}
    \label{fig:France}
\end{figure}

\begin{figure}[p]
    \centering
    \includegraphics[width=0.5\textwidth]{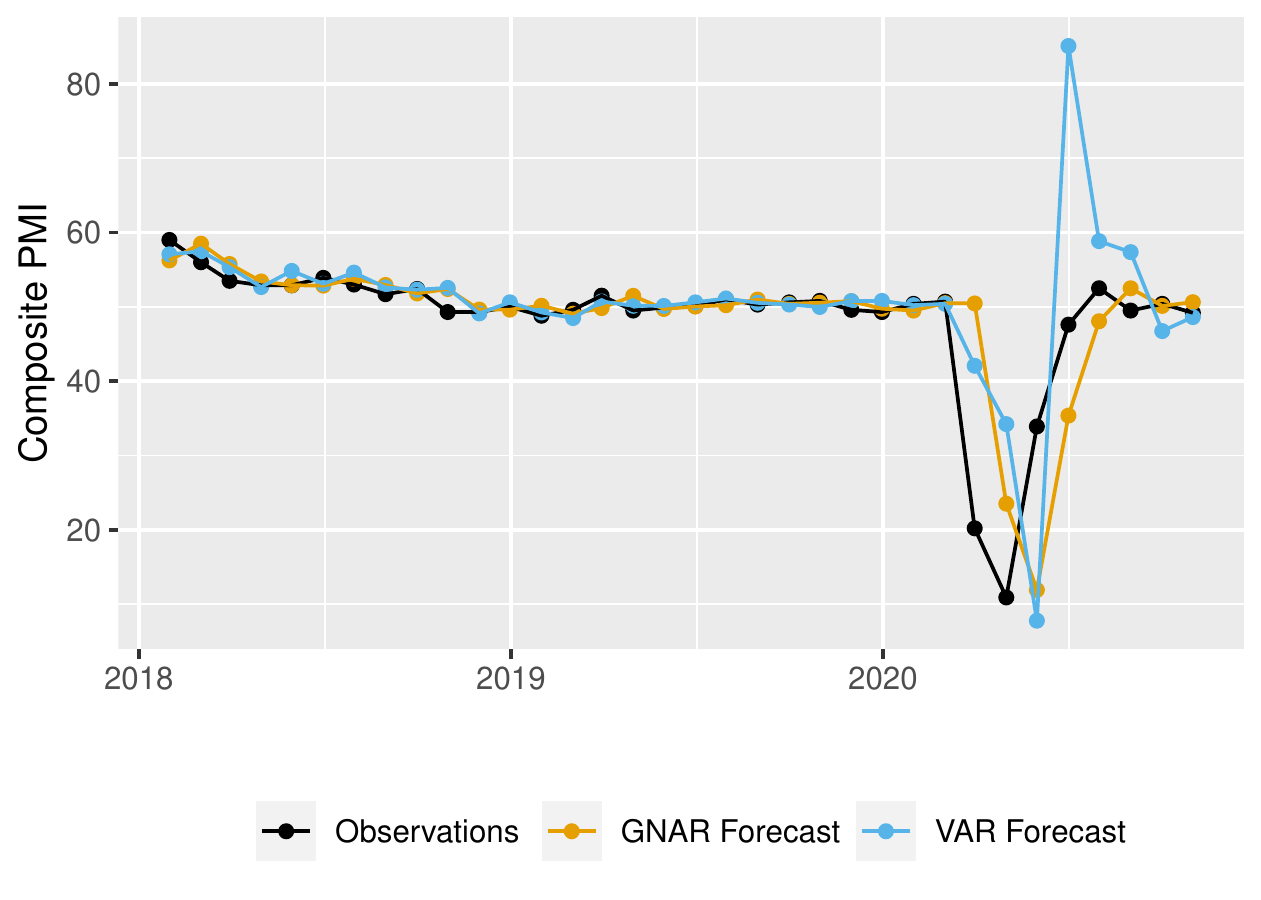}
    \caption{Italy: Out-of-sample one-step-ahead rolling composite PMI forecasts of the local-$\alpha$ GNAR($1, 1$) model and the VAR($2$) model.}
    \label{fig:Italy}
\end{figure}

\begin{figure}[p]
    \centering
    \includegraphics[width=0.5\textwidth]{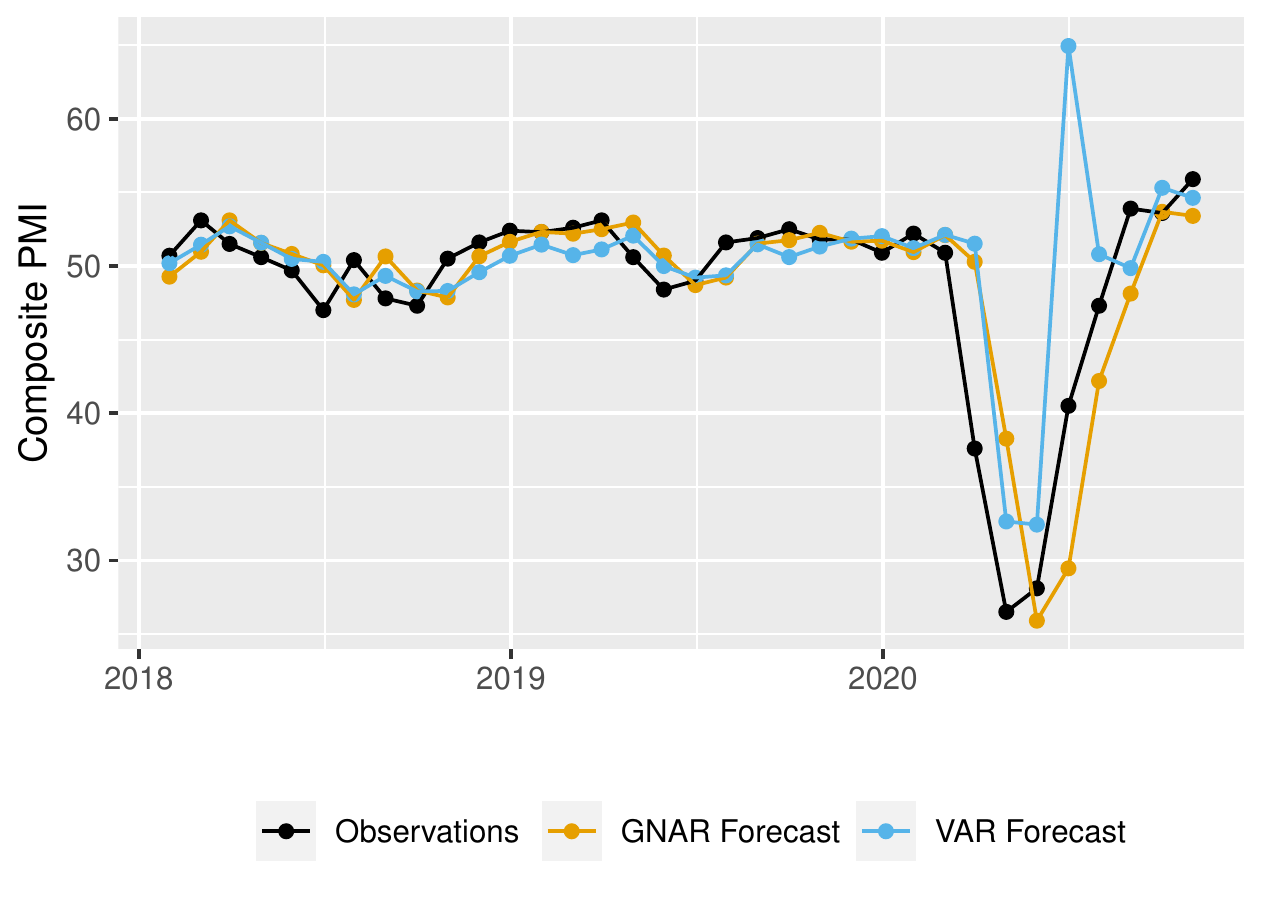}
    \caption{Brazil: Out-of-sample one-step-ahead rolling composite PMI forecasts of the local-$\alpha$ GNAR($1, 1$) model and the VAR($2$) model.}
    \label{fig:Brazil}
\end{figure}

\begin{figure}[p]
    \centering
    \includegraphics[width=0.5\textwidth]{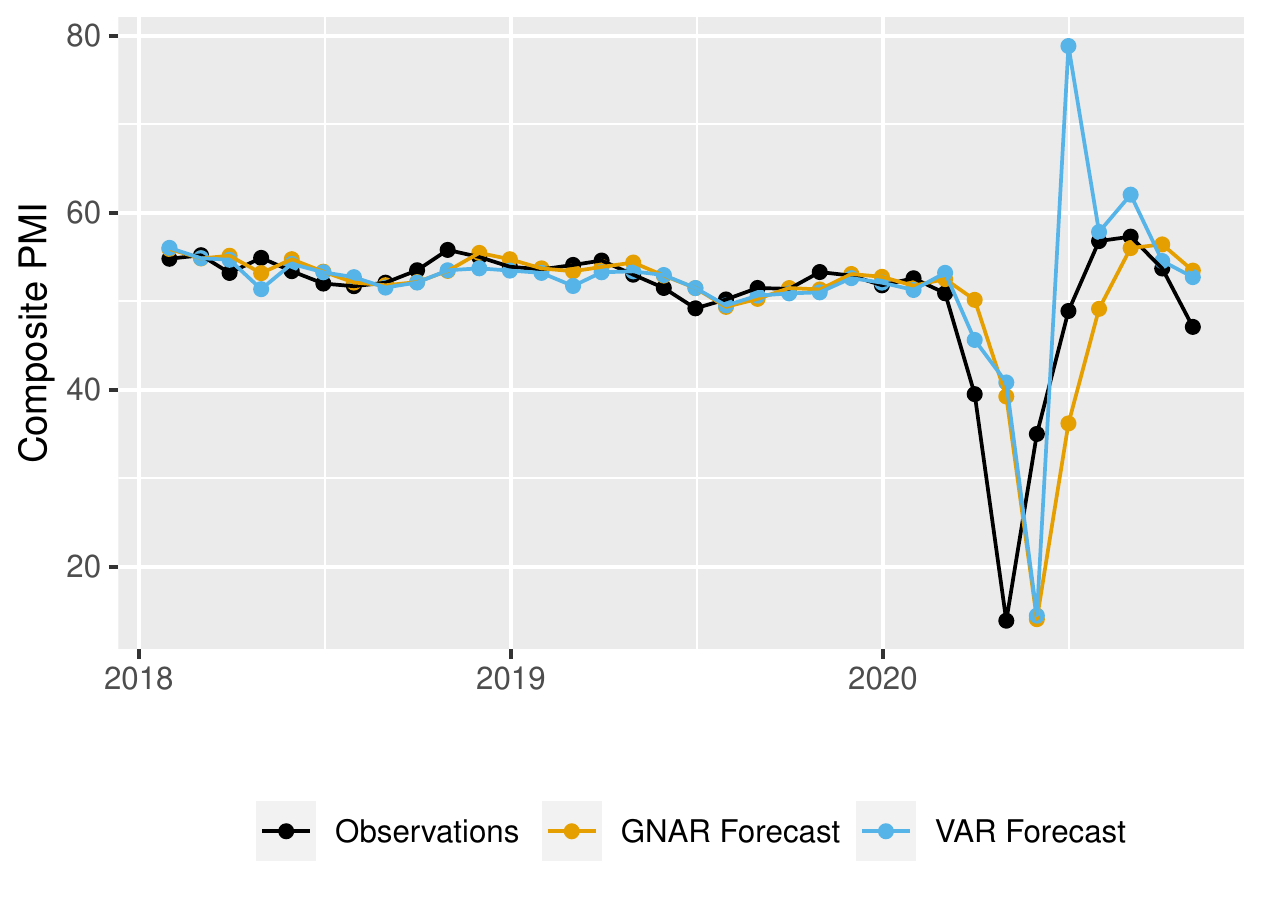}
    \caption{Russia: Out-of-sample one-step-ahead rolling composite PMI forecasts of the local-$\alpha$ GNAR($1, 1$) model and the VAR($2$) model.}
    \label{fig:Russia}
\end{figure}

\begin{figure}[p]
    \centering
    \includegraphics[width=0.5\textwidth]{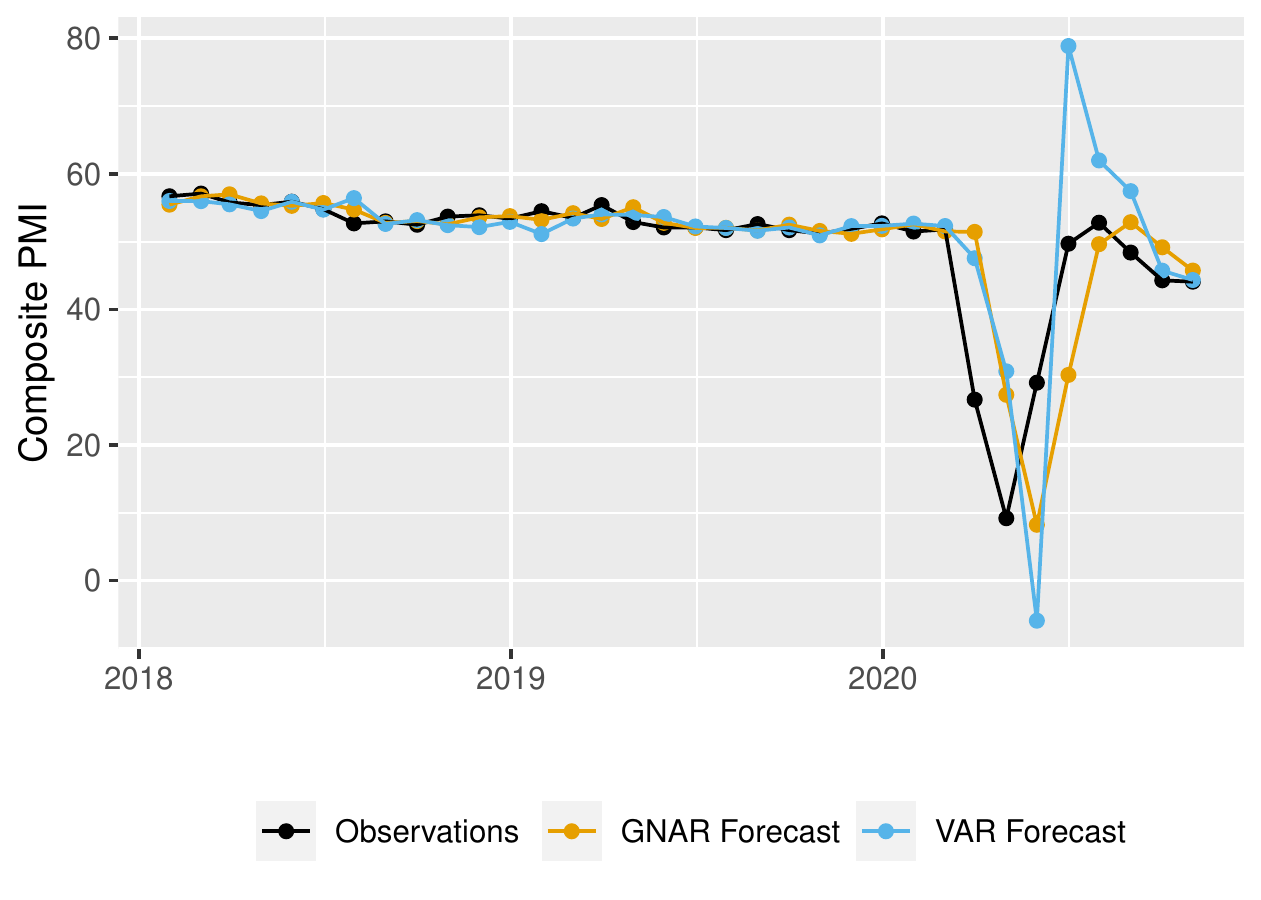}
    \caption{Spain: Out-of-sample one-step-ahead rolling composite PMI forecasts of the local-$\alpha$ GNAR($1, 1$) model and the VAR($2$) model.}
    \label{fig:Spain}
\end{figure}

\begin{figure}[p]
    \centering
    \includegraphics[width=0.5\textwidth]{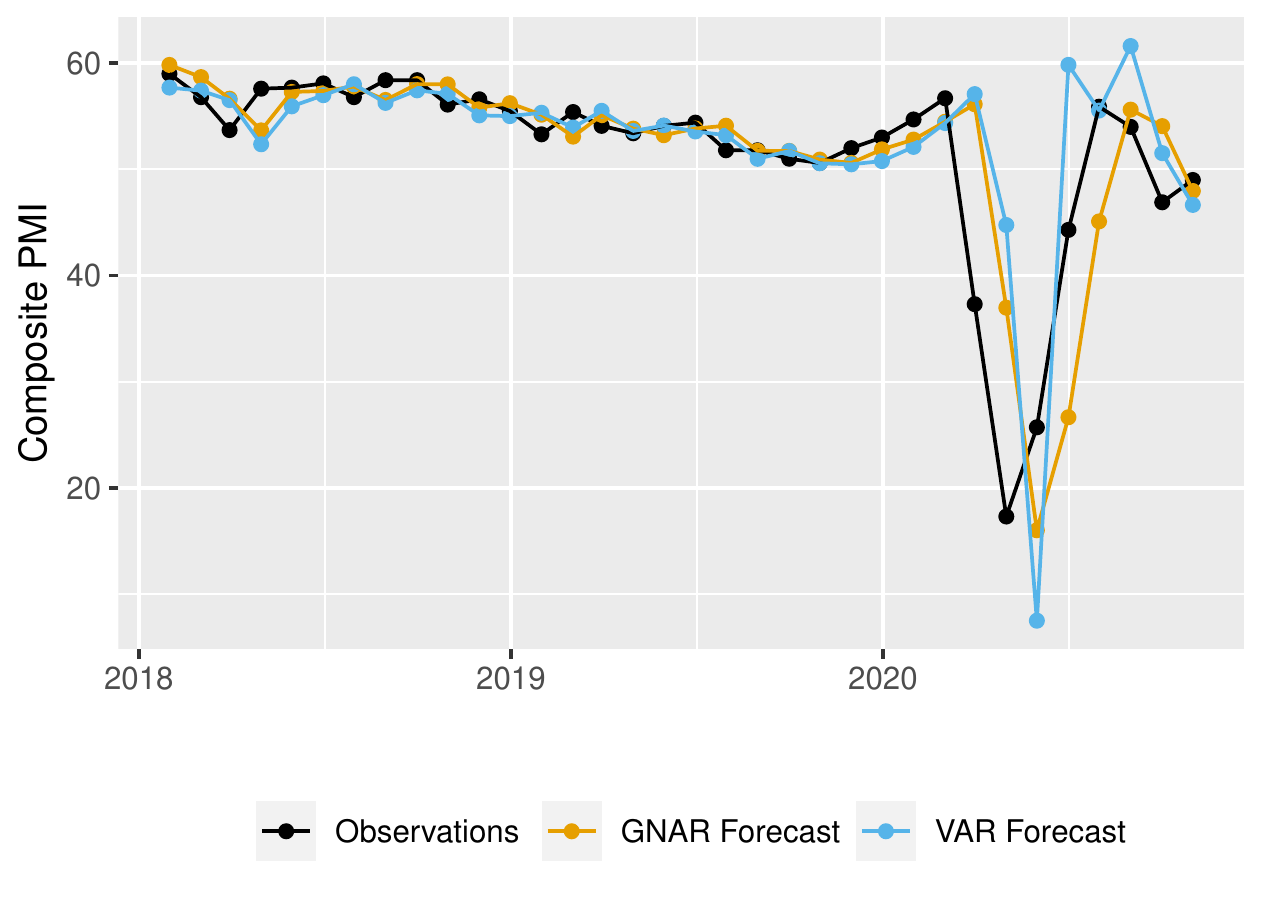}
    \caption{Ireland: Out-of-sample one-step-ahead rolling composite PMI forecasts of the local-$\alpha$ GNAR($1, 1$) model and the VAR($2$) model.}
    \label{fig:Ireland}
\end{figure}

Figures \ref{fig:USA}-\ref{fig:Ireland} plot the out-of-sample one-step-ahead rolling forecasts of the local-$\alpha$ GNAR($1, (1)$) model between Jan-18 and Oct-20 using the fully connected trade network (i.e. the GNAR specification with lowest mean-squared forecast error), together with the forecasts of the VAR($2$) model. The forecasts are arranged in order of 2019 country GDP.

\textcolor{black}{Table \ref{tab:GNARMSFESelection} shows the performance of the models selected by optimising for MSFE, rather than BIC. Specifically, models of all orders up to $p = 3$ were fit on the first 180 months of the in-sample period and their corresponding forecast errors were recorded for the remaining 60 months of the in-sample period. Model orders were selected to minimise the MSFE computed from these forecast errors. The resulting GNAR models using the fully-connected networks outperform both the VAR model and the univariate AR models in terms of MSFE.}

\begin{table}
\caption{\label{tab:GNARMSFESelection} Out-of-sample composite PMI forecast performance between Jan-18 and Oct-20 using GNAR, VAR and naive forecasting. MSFE=Out-of-sample mean-squared forecasting error; SE = standard error}
\centering
\framebox{%
\begin{tabular}{lrrr}
\hline
Model & Model order & Parameters & MSFE (SE) \\ 
\hline
Global-$\mathbf{\alpha}$ GNAR, fully-connected net & 1, (1) & 2 & 38.2 (6.8)\\
Global-$\mathbf{\alpha}$ GNAR, nearest-neighbour net & 1, (0) &  1 & 38.8 (7.0)\\
Local-$\mathbf{\alpha}$ GNAR, fully-connected net & 1, (1) & 13 & 38.5 (6.8)\\
Local-$\mathbf{\alpha}$ GNAR, nearest-neighbour net & 1, (0) & 12 & 39.1 (7.0)\\
\\
VAR & 1 & 144 & 40.7 (8.0)\\
AR & 1 & 12 & 39.1 (7.0)\\
Naive forecast & - & -  & 40.6 (7.2)\\
\hline
\end{tabular} }
\end{table}

%
%

\section{\textcolor{black}{Simulation results for GNARX model order selection using BIC optimisation}}\label{sec:modorder}

\textcolor{black}{We simulate network time series data of three different lengths $T$ from a local-$\alpha$ GNARX(1, (1), 1) model with the fiveNet network used in the simulation experiments of \cite{leeming:phd}. Parameters are set as $\alpha_{1,1} = 0.4, \alpha_{2,1} = \alpha_{4,1} = \alpha_{5,1} = 0.2, \alpha_{3,1} = 0.4, \beta_{1,1} = 0.5, \lambda_{1,1} = 0.4, \lambda_{1,2} = 0.2$. Both the GNARX innovations and the exogenous regressor are sampled i.i.d from a standard normal distribution. These settings guarantee stationarity of the resulting process.}

\textcolor{black}{Tables~\ref{tab:GNARXsim1} and~\ref{tab:GNARXsim2} show that, as the lengths of the time series increases, the correct model order is identified with increasing probability using both global BIC optimisation and our stagewise approach. The latter approach, which we adopt in our PMI forecasting experiments, involves first selecting autoregressive order $p$ while ignoring the network. Once $p$ has been selected, we optimise on the remaining components of the model order. To ensure computational feasibility, for global BIC optimisation we have set order limits of $p = p'= 3$. For stagewise optimisation, we are able to set a higher limit of $p = 12$.}

\begin{table}
\caption{\label{tab:GNARXsim1} Number of times each model order is selected by minimising the BIC out of 1000 simulations up to orders $p=p'=3$. The true model order has been bolded.}
\centering
\framebox{%
\begin{tabular}{lrrrrrrrrrr}
\hline
$T$ & \rotatebox[origin=c]{90}{(1, [0], 0)} & \rotatebox[origin=c]{90}{(1, [1], 0)} & \rotatebox[origin=c]{90}{\textbf{(1, [1], 1)}} & \rotatebox[origin=c]{90}{(1, [1], 2)} & \rotatebox[origin=c]{90}{(1, [1], 3)} & \rotatebox[origin=c]{90}{(1, [2], 0)} & \rotatebox[origin=c]{90}{(1, [2], 1)} & \rotatebox[origin=c]{90}{(1, [2], 2)} & \rotatebox[origin=c]{90}{(1, [2], 3)} & \rotatebox[origin=c]{90}{(1, [3], 1)} \\ 
\hline
32 & 1 & 413 & \textbf{517} & 35 & 6 & 7 & 17 & 1 & 1 & 2\\
64 & 0 &  165 & \textbf{789} & 17 & 2 & 2 & 21 & 1 & 0 & 3 \\
128 & 0 & 18 & \textbf{960} & 8 & 0 & 0 & 14 & 0 & 0 & 0 \\
\hline
\end{tabular} }
\end{table}

\begin{table}
\caption{\label{tab:GNARXsim2} Number of times each model order is selected by stagewise BIC optimisation out of 1000 simulations up to orders $p=12, p'=3$. The true model order has been bolded.}
\centering
\framebox{%
\begin{tabular}{lrrrrrrrrrrr}
\hline
$T$ & \rotatebox[origin=c]{90}{(1, [0], 0)} &
\rotatebox[origin=c]{90}{(1, [0], 1)} &
\rotatebox[origin=c]{90}{(1, [1], 0)} & \rotatebox[origin=c]{90}{\textbf{(1, [1], 1)}} & \rotatebox[origin=c]{90}{(1, [1], 2)} & \rotatebox[origin=c]{90}{(1, [1], 3)} & \rotatebox[origin=c]{90}{(1, [2], 0)} & \rotatebox[origin=c]{90}{(1, [2], 1)} & \rotatebox[origin=c]{90}{(1, [2], 2)} & \rotatebox[origin=c]{90}{(1, [2], 3)} & \rotatebox[origin=c]{90}{(1, [3], 1)} \\ 
\hline
32 & 3 & 1 & 467 & \textbf{486} & 13 & 3 & 7 & 14 & 0 & 1 & 5\\
64 & 0 & 0 &  170 & \textbf{796} & 12 & 3 & 2 & 15 & 0 & 0 & 2 \\
128 & 0 & 0 & 19 & \textbf{957} & 11 & 0 & 0 & 11 & 0 & 0 & 2 \\
\hline
\end{tabular} }
\end{table}

%
%
\section{GNARX Residual Analysis}
\begin{figure}[p]
    \centering
    \includegraphics[width=1\textwidth]{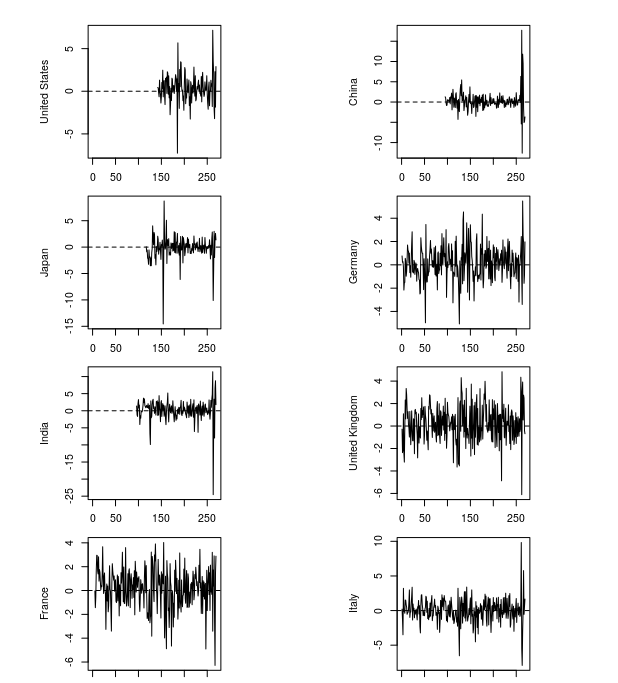}
    \caption{Time series plots of residuals from best GNARX model. (US, China, Japan, Germany, India, UK, France, Italy)}
    \label{fig:res1a}
\end{figure}

\begin{figure}[p]
    \centering
    \includegraphics[width=1\textwidth]{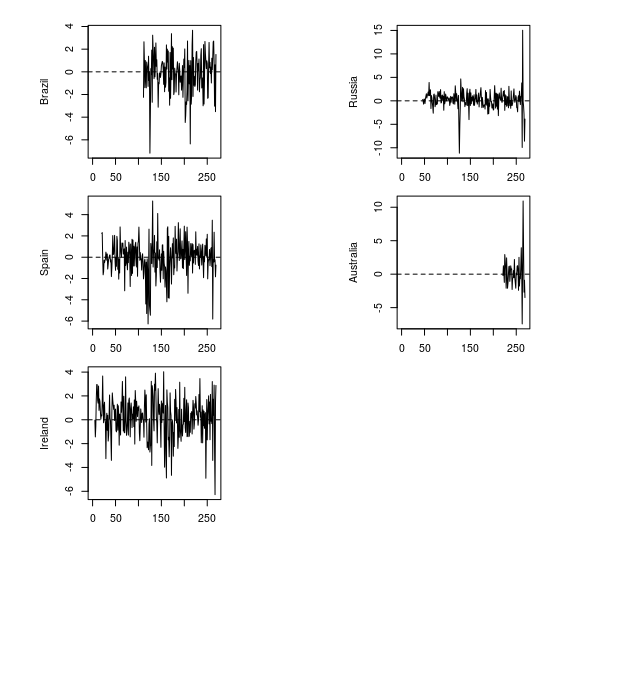}
    \caption{Time series plots of residuals from best GNARX model. (Brazil, Russia, Spain, Australia, Ireland)}
    \label{fig:res1b}
\end{figure}

\begin{figure}[p]
    \centering
    \includegraphics[width=1\textwidth]{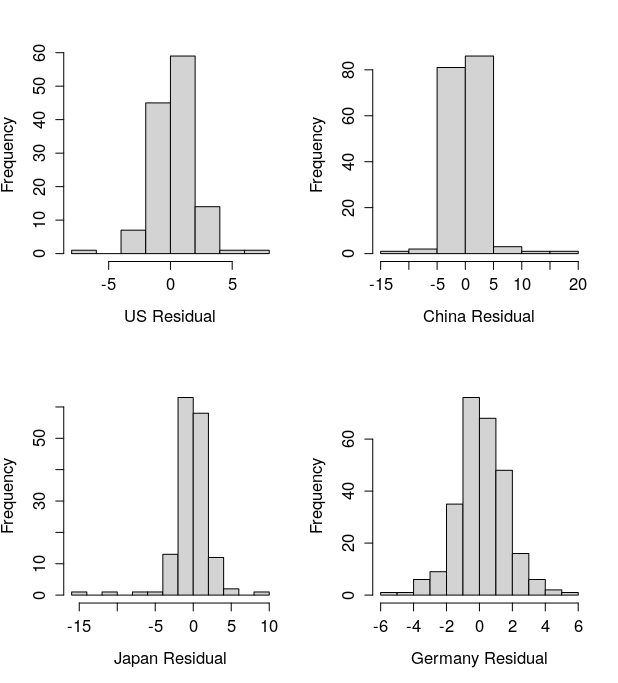}
    \caption{Histograms of  residuals from best GNARX model. (US, China, Japan, Germany)}
    \label{fig:res2a}
\end{figure}

\begin{figure}[p]
    \centering
    \includegraphics[width=1\textwidth]{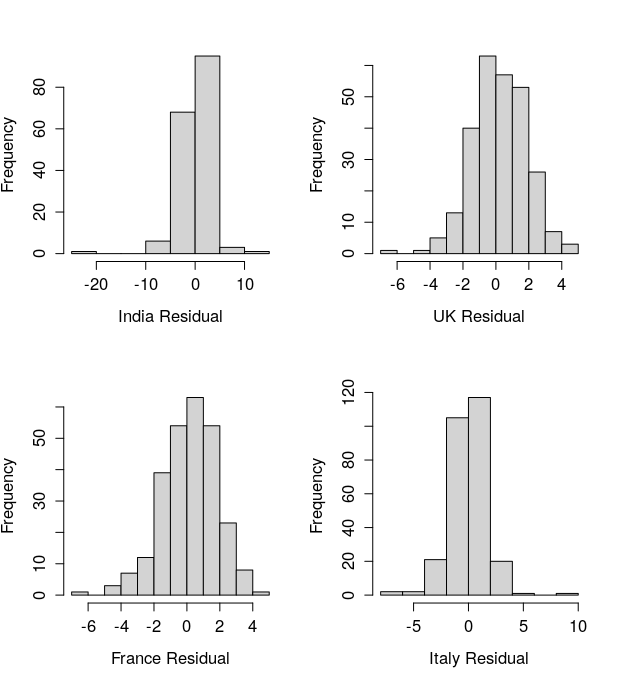}
    \caption{Histograms of  residuals from best GNARX model. (India, UK, France, Italy)}
    \label{fig:res2b}
\end{figure}

\begin{figure}[p]
    \centering
    \includegraphics[width=1\textwidth]{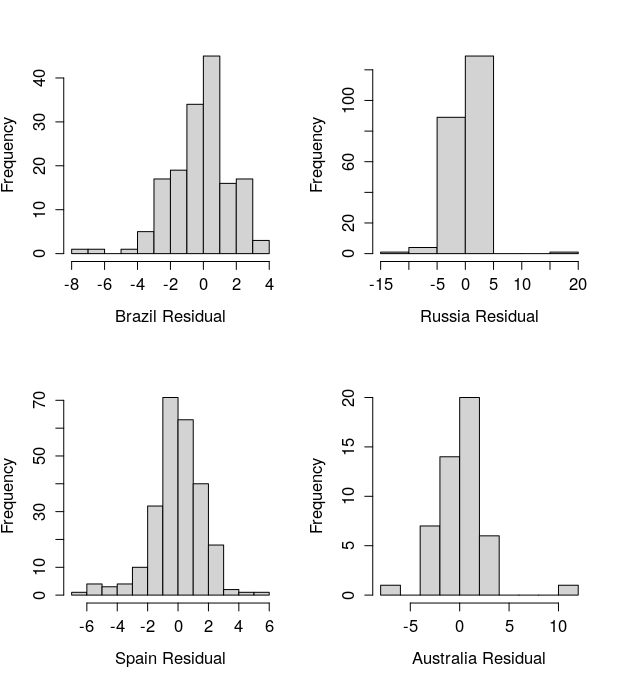}
    \caption{Histograms of  residuals from best GNARX model. (Brazil, Russia, Spain, Australia)}
    \label{fig:res2c}
\end{figure}

\begin{figure}[p]
    \centering
    \includegraphics[width=1\textwidth]{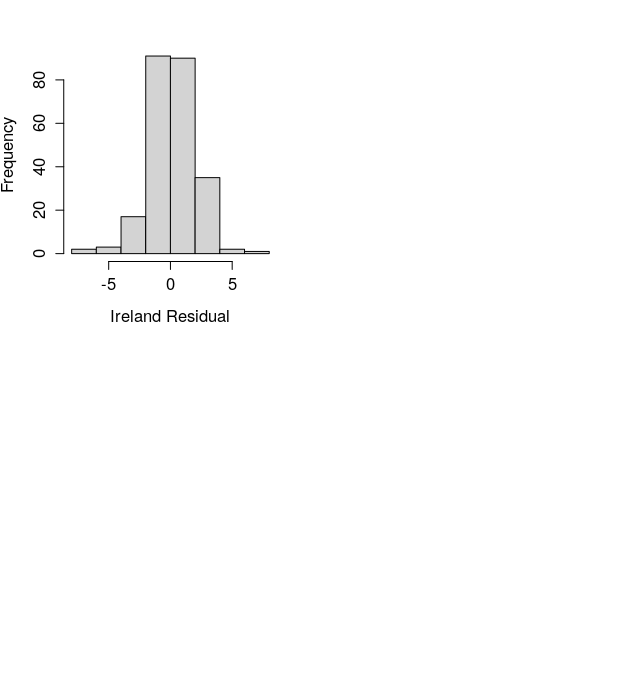}
    \caption{Histograms of  residuals from best GNARX model. (Ireland)}
    \label{fig:res2d}
\end{figure}

\begin{figure}[p]
    \centering
    \includegraphics[width=1\textwidth]{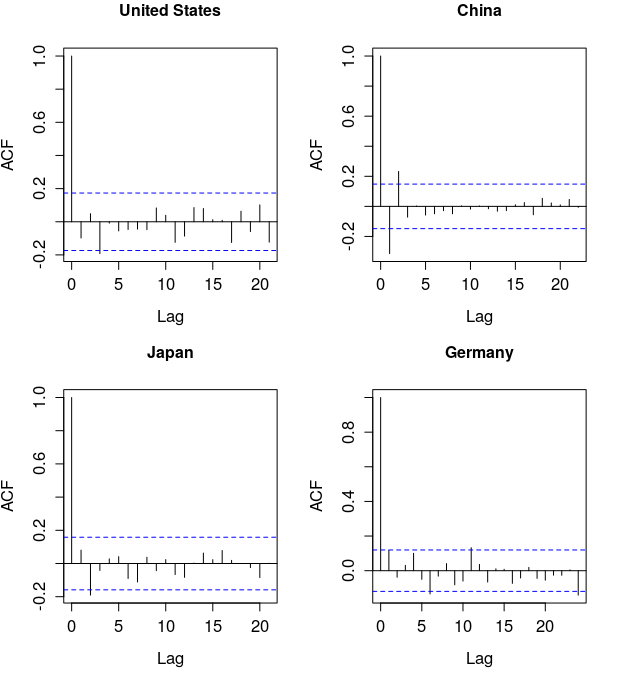}
    \caption{Autocorrelation plots of residuals from best GNARX model. (US, China, Japan, Germany)}
    \label{fig:res3a}
\end{figure}

\begin{figure}[p]
    \centering
    \includegraphics[width=1\textwidth]{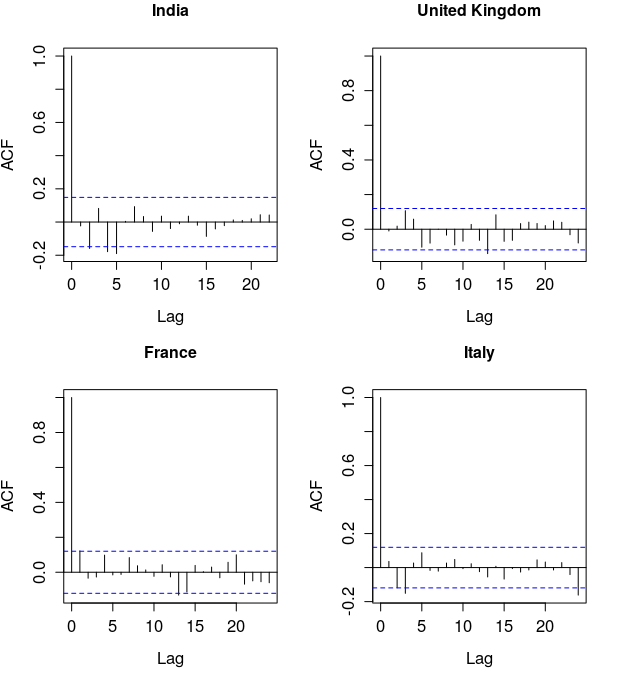}
    \caption{Autocorrelation plots of residuals from best GNARX model. (India, UK, France, Italy)}
    \label{fig:res3b}
\end{figure}

\begin{figure}[p]
    \centering
    \includegraphics[width=1\textwidth]{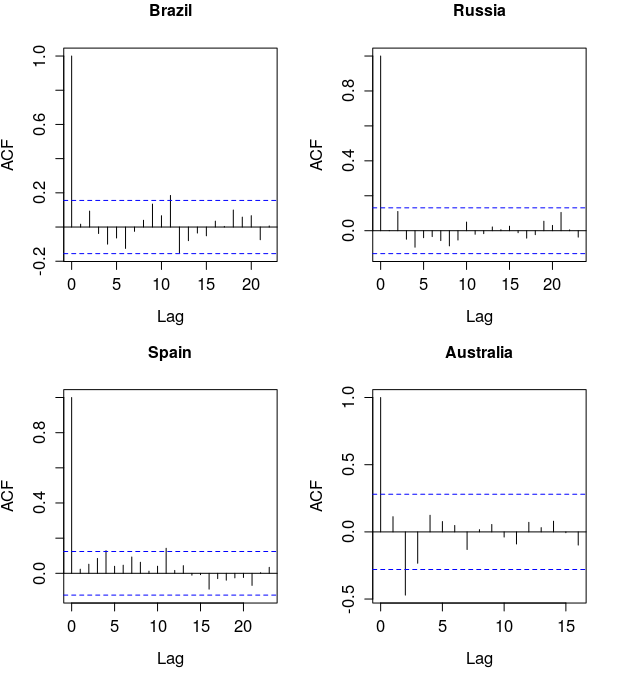}
    \caption{Autocorrelation plots of residuals from best GNARX model. (Brazil, Russia, Spain, Australia)}
    \label{fig:res3c}
\end{figure}

\begin{figure}[p]
    \centering
    \includegraphics[width=1\textwidth]{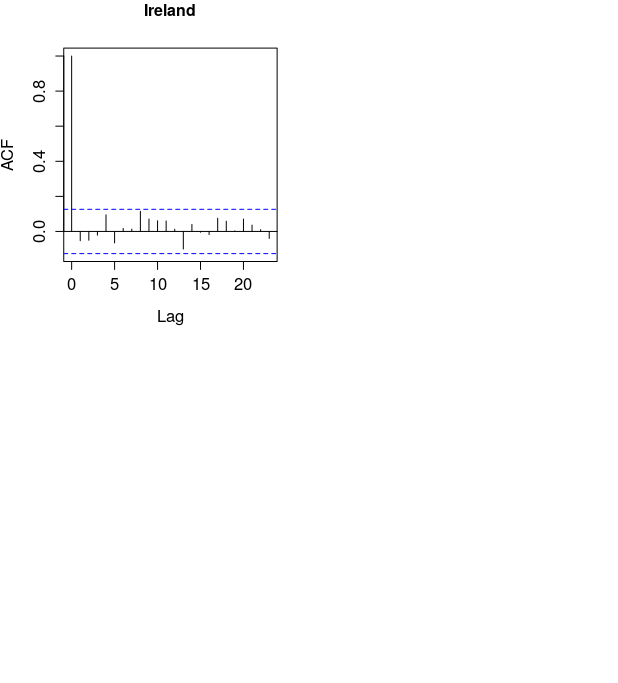}
    \caption{Autocorrelation plots of residuals from best GNARX model. (Ireland)}
    \label{fig:res3d}
\end{figure}

In Figures \ref{fig:res1a}-\ref{fig:res3d}, we show the \textcolor{black}{GNARX$(5, [1, 0, 1, 0, 1], 3, 2)$} residual analyses for all countries, arranged in order of 2019 country GDP.

%
%

\section{Additional GNARX experiment charts}

\begin{figure}[p]
    \centering
    \includegraphics[width=0.5\textwidth]{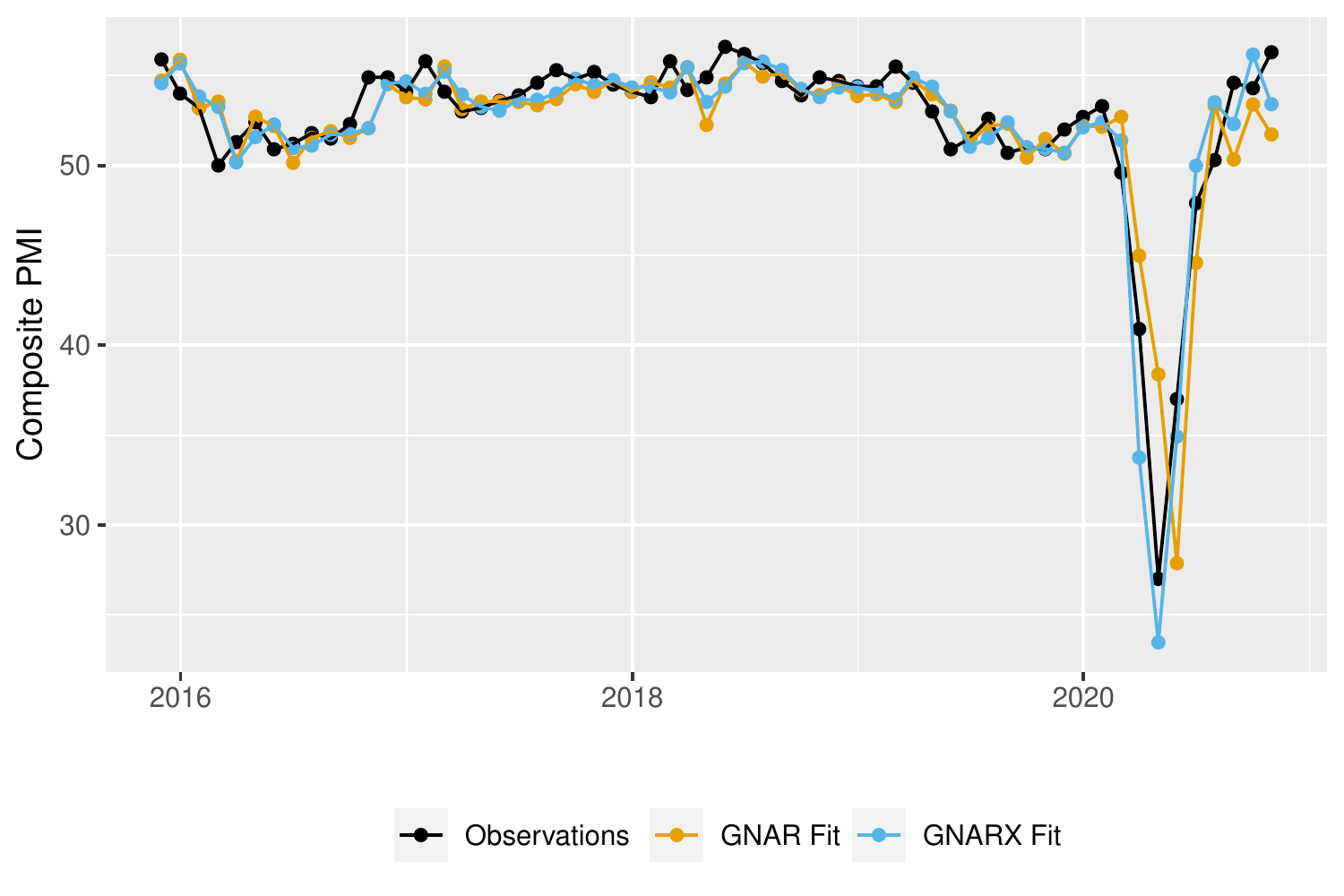}
    \caption{United States: Comparison of \textcolor{black}{GNARX$(5, [1, 0, 1, 0, 1])$} and \textcolor{black}{GNARX$(5, [1, 0, 1, 0, 1], 3, 2)$} fits. }
    \label{fig:USAFit}
\end{figure}

\begin{figure}[p]
    \centering
    \includegraphics[width=0.5\textwidth]{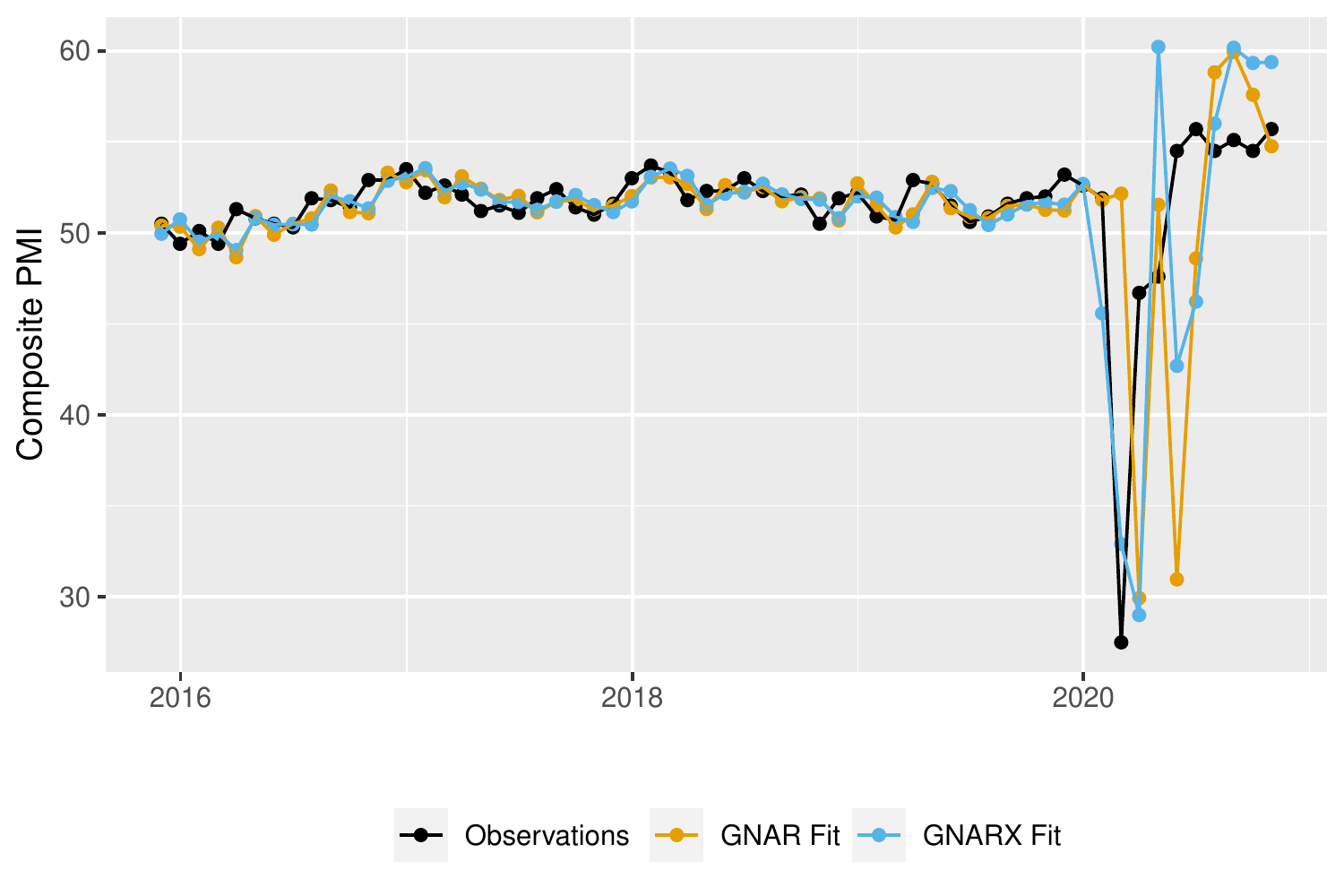}
    \caption{China: Comparison of \textcolor{black}{GNARX$(5, [1, 0, 1, 0, 1])$} and \textcolor{black}{GNARX$(5, [1, 0, 1, 0, 1], 3, 2)$} fits.  }
    \label{fig:ChinaFit}black\end{figure}

\begin{figure}[p]
    \centering
    \includegraphics[width=0.5\textwidth]{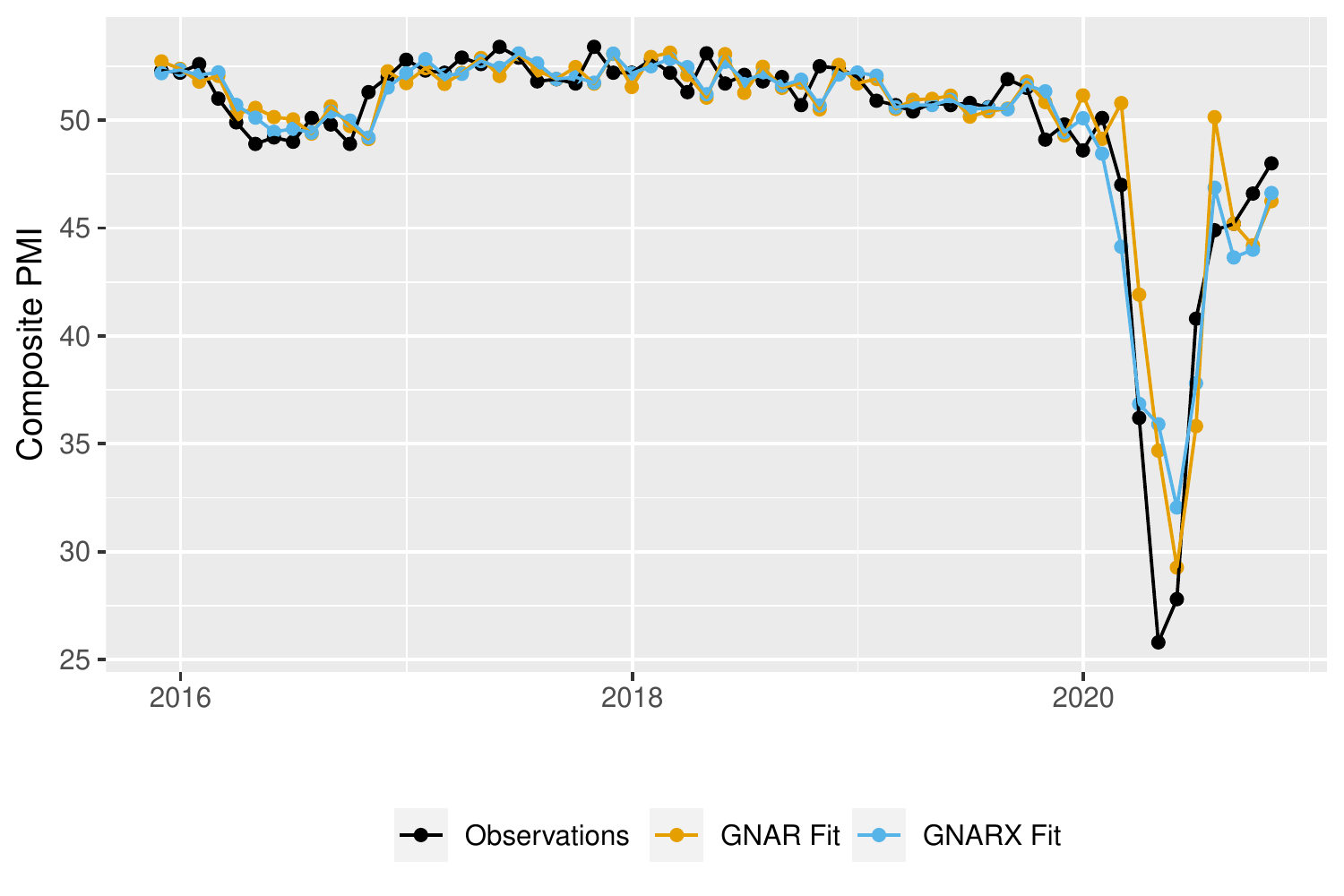}
    \caption{Japan: Comparison of \textcolor{black}{GNARX$(5, [1, 0, 1, 0, 1])$} and \textcolor{black}{GNARX$(5, [1, 0, 1, 0, 1], 3, 2)$} fits.  }
    \label{fig:JapanFit}
\end{figure}

\begin{figure}[p]
    \centering
    \includegraphics[width=0.5\textwidth]{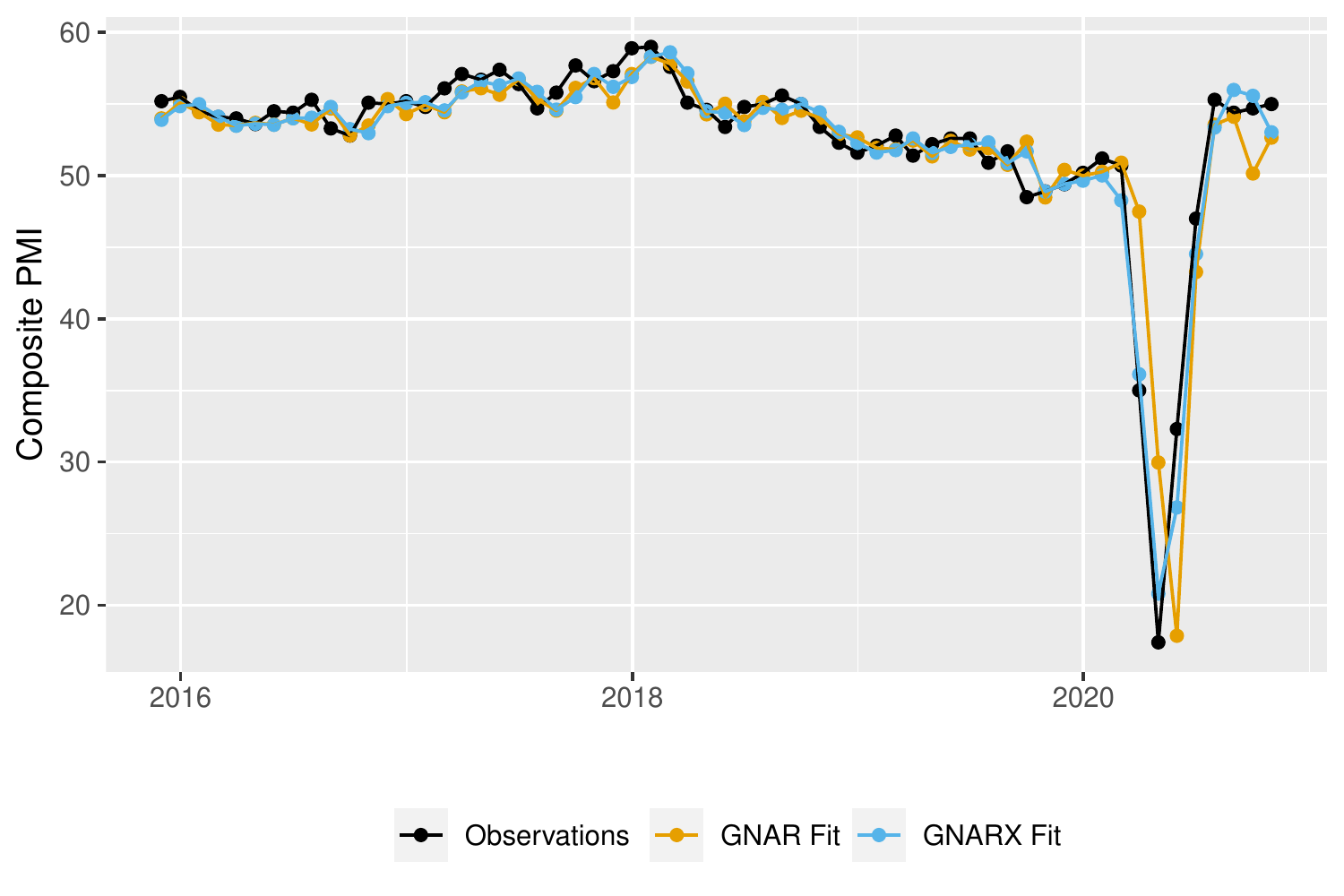}
    \caption{Germany: Comparison of \textcolor{black}{GNARX$(5, [1, 0, 1, 0, 1])$} and \textcolor{black}{GNARX$(5, [1, 0, 1, 0, 1], 3, 2)$} fits. }
    \label{fig:GermanyFit}
\end{figure}

\begin{figure}[p]
    \centering
    \includegraphics[width=0.5\textwidth]{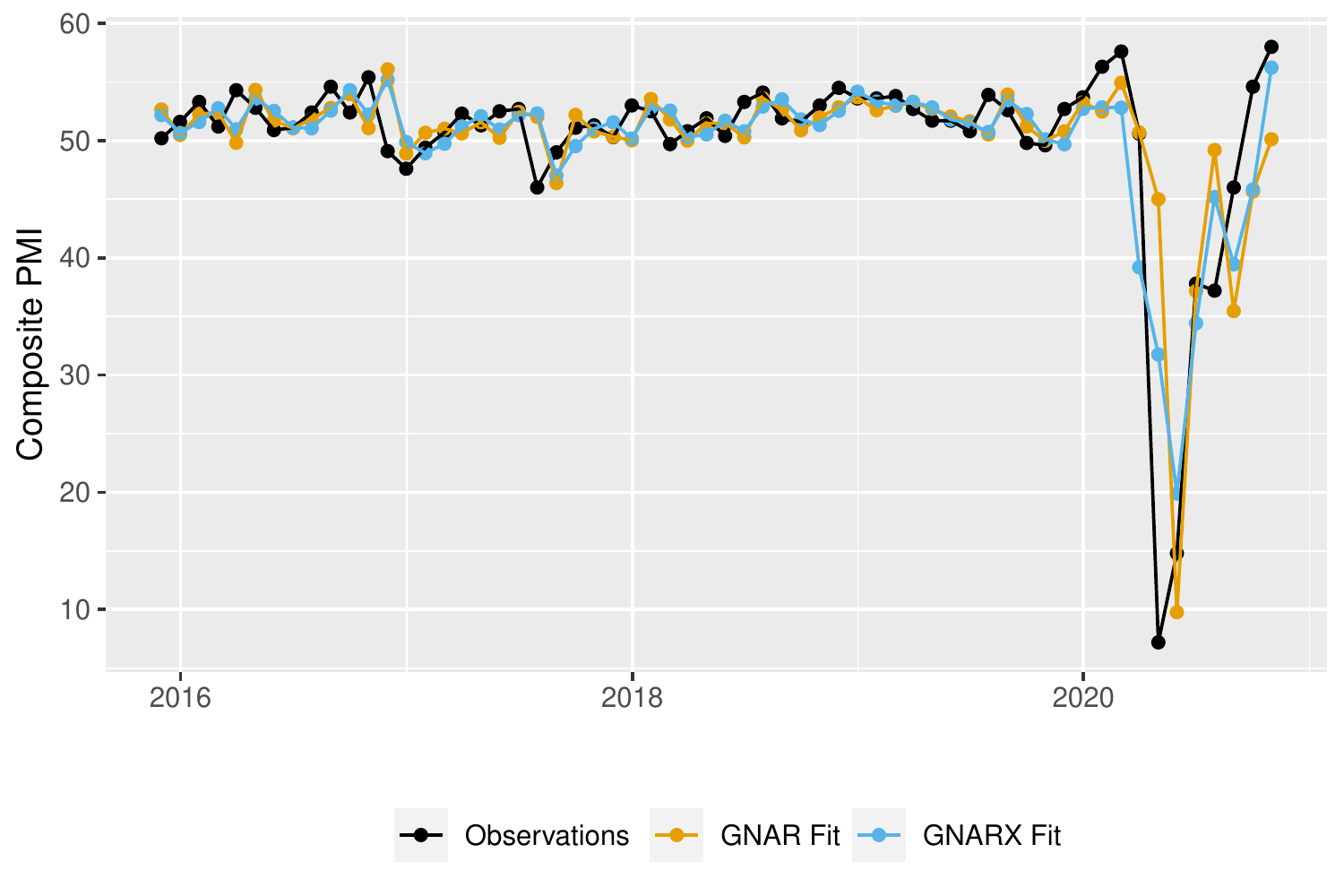}
    \caption{India: Comparison of \textcolor{black}{GNARX$(5, [1, 0, 1, 0, 1])$} and \textcolor{black}{GNARX$(5, [1, 0, 1, 0, 1], 3, 2)$} fits. }
    \label{fig:IndiaFit}
\end{figure}

\begin{figure}[p]
    \centering
    \includegraphics[width=0.5\textwidth]{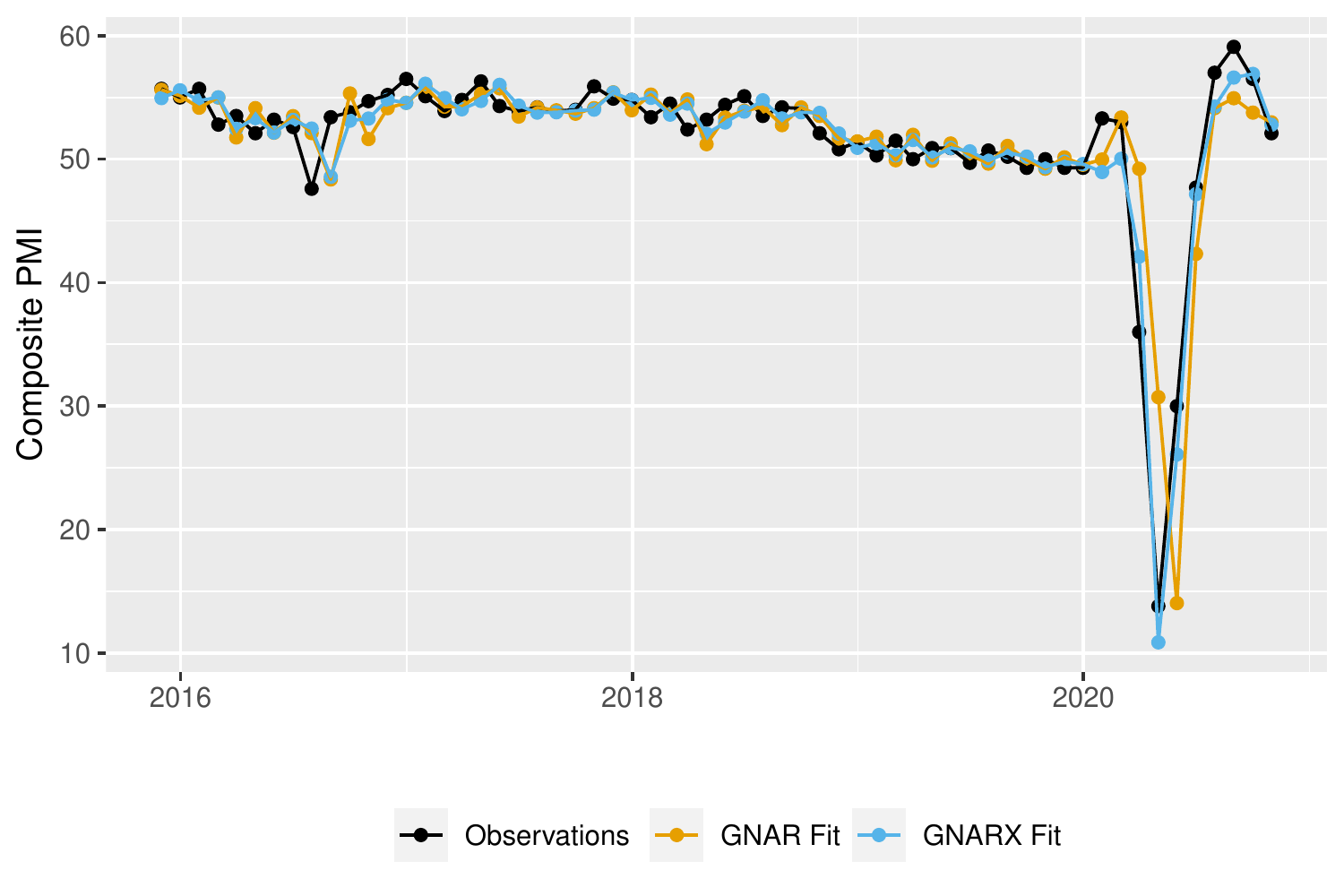}
    \caption{United Kingdom: Comparison of \textcolor{black}{GNARX$(5, [1, 0, 1, 0, 1])$} and \textcolor{black}{GNARX$(5, [1, 0, 1, 0, 1], 3, 2)$} fits. }
    \label{fig:UKFit}
\end{figure}

\begin{figure}[p]
    \centering
    \includegraphics[width=0.5\textwidth]{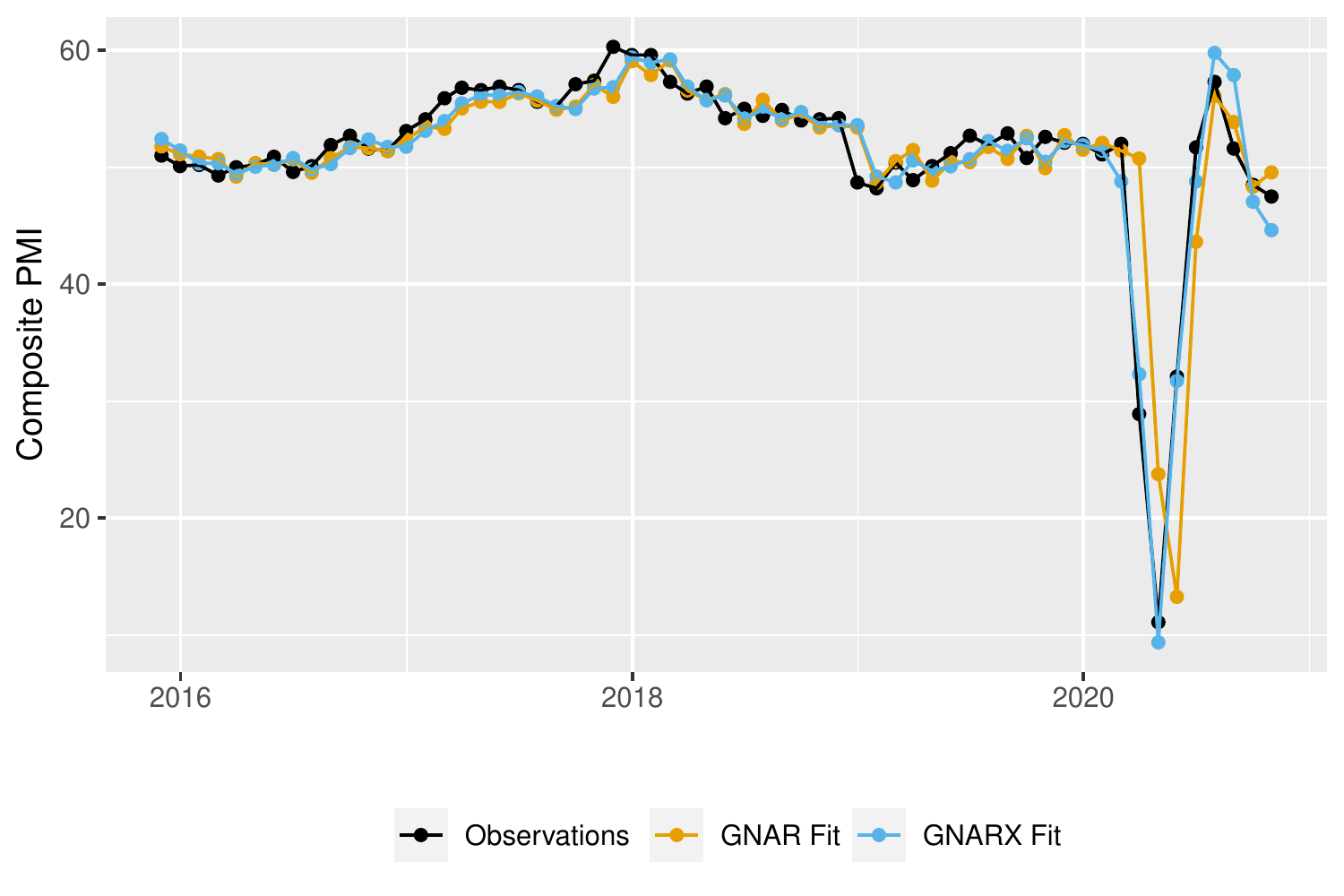}
    \caption{France: Comparison of \textcolor{black}{GNARX$(5, [1, 0, 1, 0, 1])$} and \textcolor{black}{GNARX$(5, [1, 0, 1, 0, 1], 3, 2)$} fits.}
    \label{fig:FranceFit}
\end{figure}

\begin{figure}[p]
    \centering
    \includegraphics[width=0.5\textwidth]{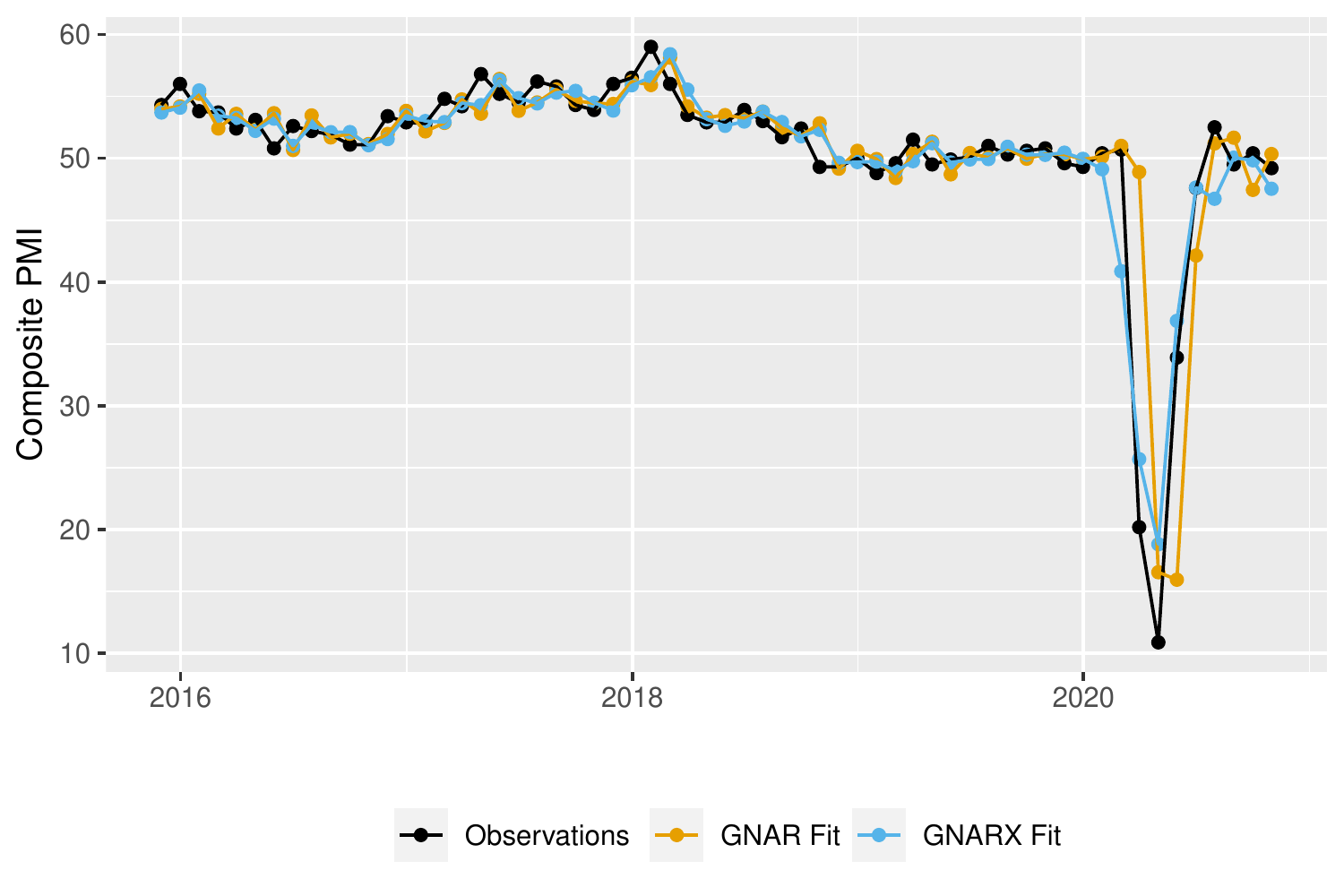}
    \caption{Italy: Comparison of \textcolor{black}{GNARX$(5, [1, 0, 1, 0, 1])$} and \textcolor{black}{GNARX$(5, [1, 0, 1, 0, 1], 3, 2)$} fits.  }
    \label{fig:ItalyFit}
\end{figure}

\begin{figure}[p]
    \centering
    \includegraphics[width=0.5\textwidth]{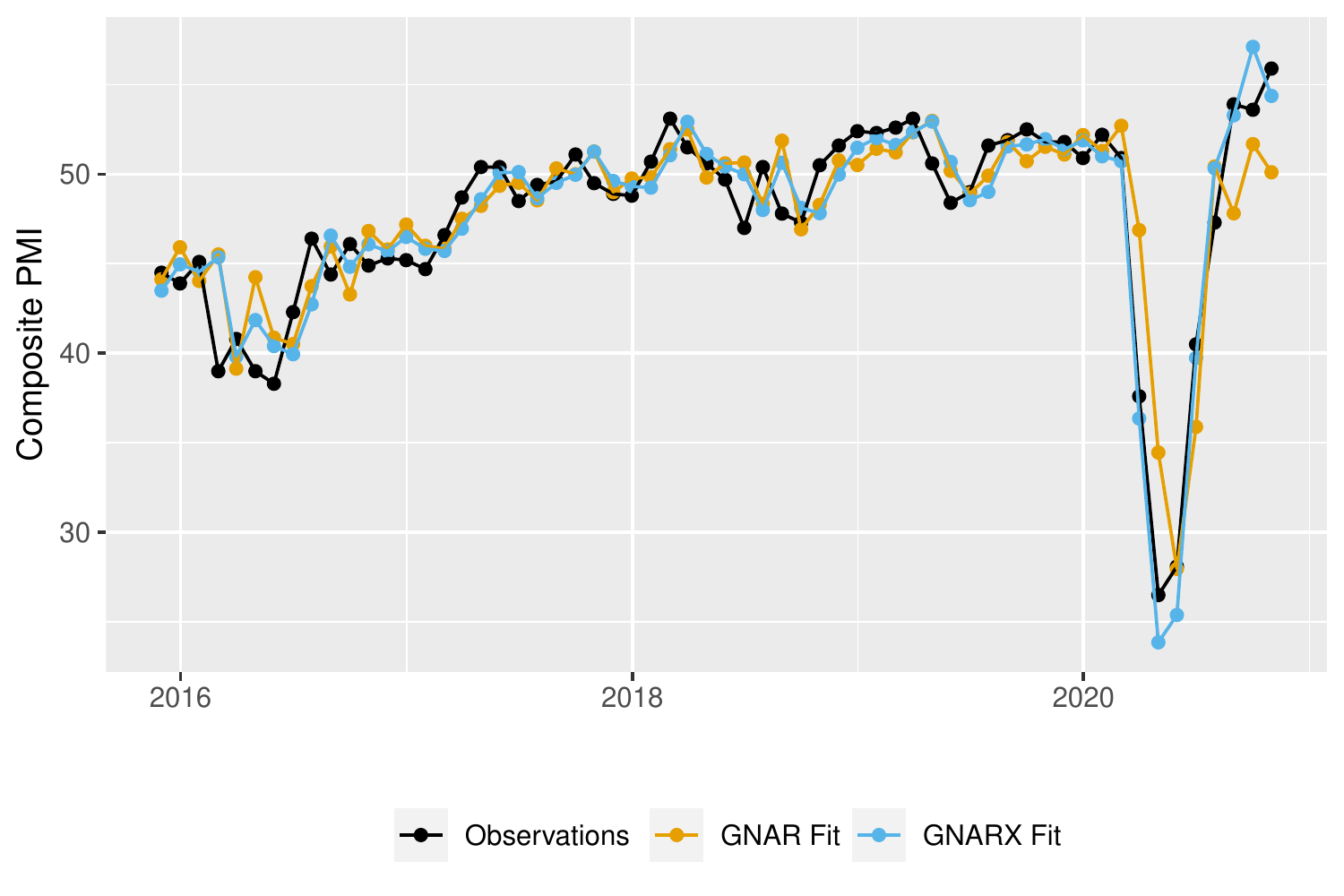}
    \caption{Brazil: Comparison of \textcolor{black}{GNARX$(5, [1, 0, 1, 0, 1])$} and \textcolor{black}{GNARX$(5, [1, 0, 1, 0, 1], 3, 2)$} fits.  }
    \label{fig:BrazilFit}
\end{figure}

\begin{figure}[p]
    \centering
    \includegraphics[width=0.5\textwidth]{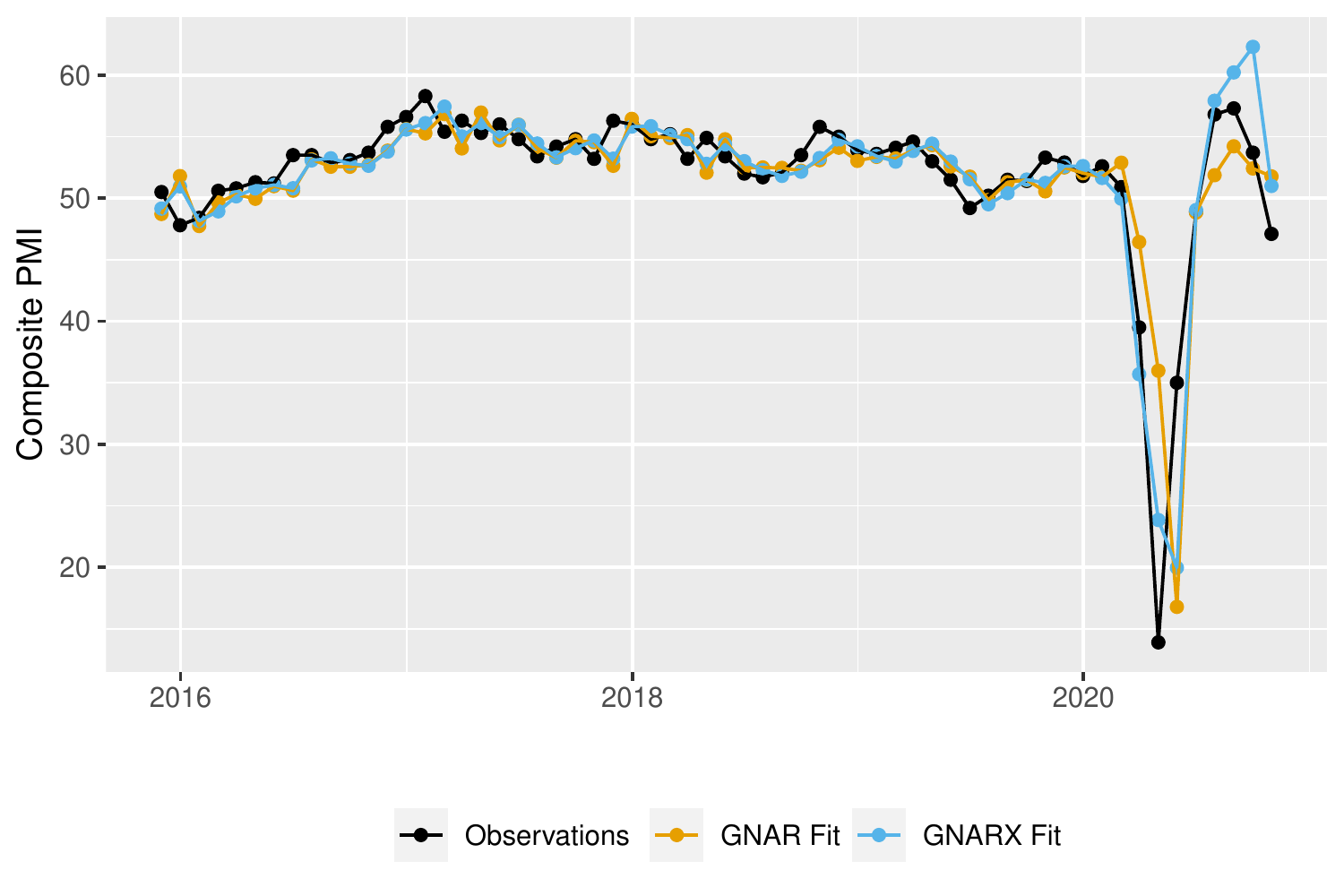}
    \caption{Russia: Comparison of \textcolor{black}{GNARX$(5, [1, 0, 1, 0, 1])$} and \textcolor{black}{GNARX$(5, [1, 0, 1, 0, 1], 3, 2)$} fits.}
    \label{fig:RussiaFit}
\end{figure}

\begin{figure}[p]
    \centering
    \includegraphics[width=0.5\textwidth]{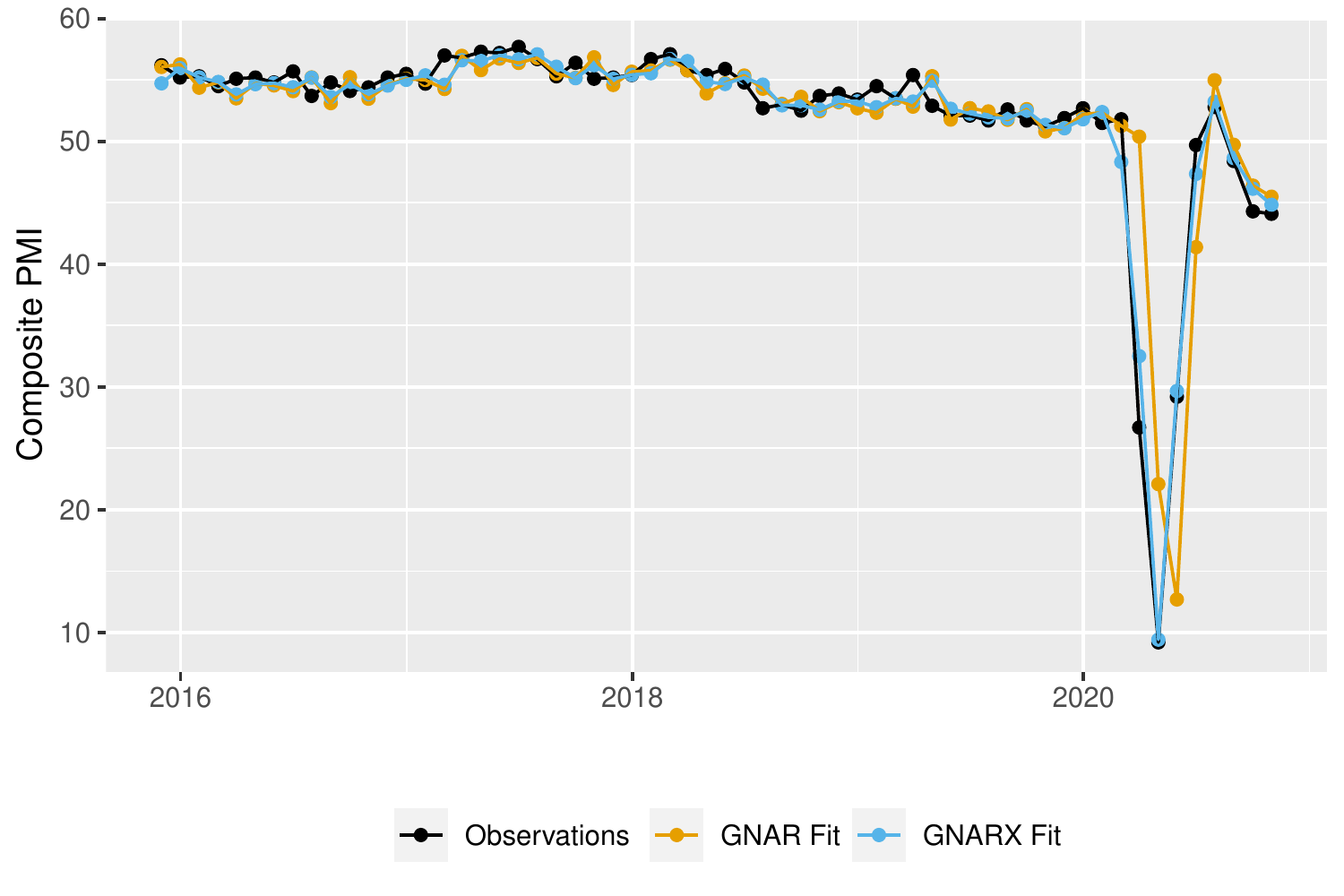}
    \caption{Spain: Comparison of \textcolor{black}{GNARX$(5, [1, 0, 1, 0, 1])$} and \textcolor{black}{GNARX$(5, [1, 0, 1, 0, 1], 3, 2)$} fits. }
    \label{fig:SpainFit}
\end{figure}

\begin{figure}[p]
    \centering
    \includegraphics[width=0.5\textwidth]{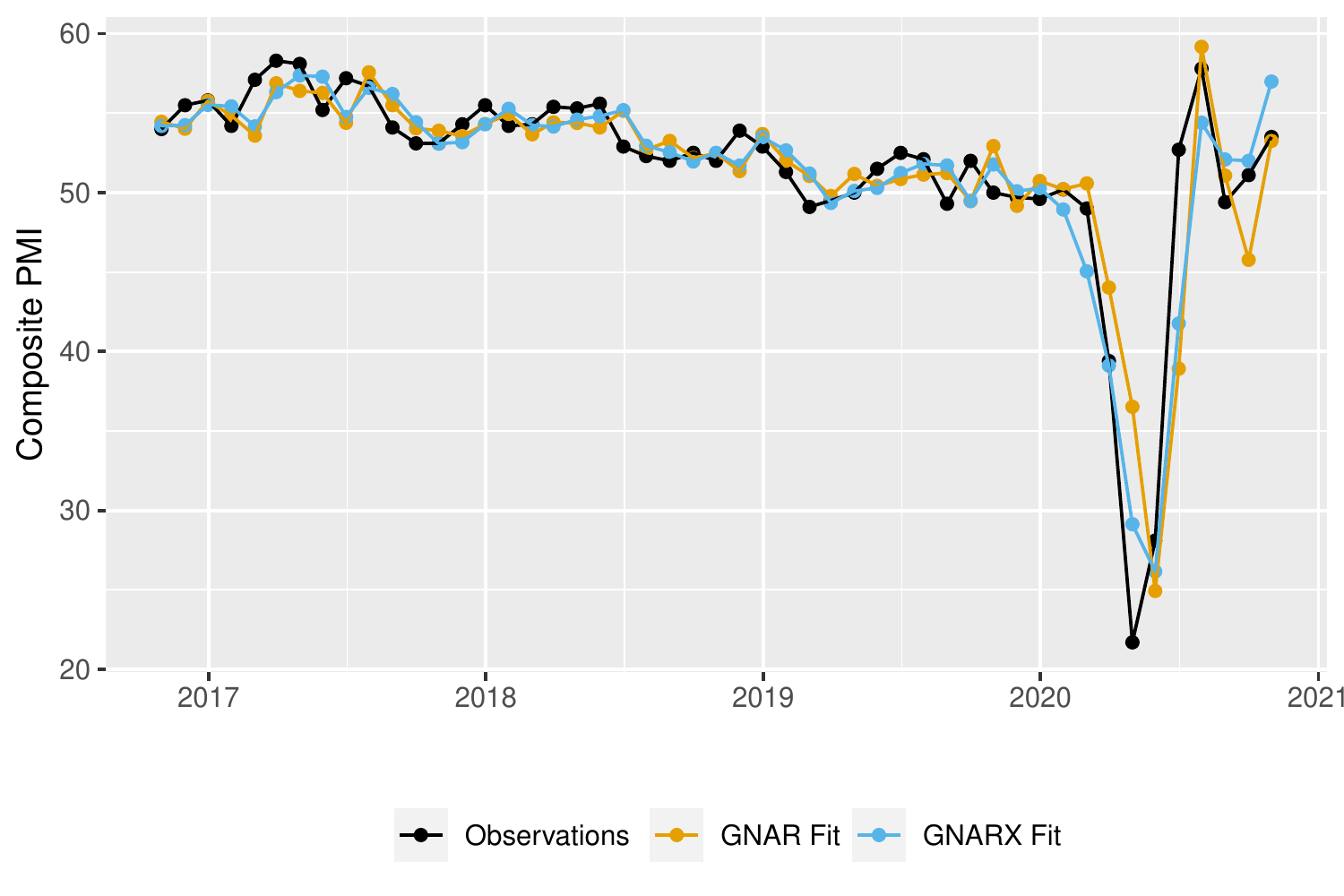}
    \caption{Australia: Comparison of \textcolor{black}{GNARX$(5, [1, 0, 1, 0, 1])$} and \textcolor{black}{GNARX$(5, [1, 0, 1, 0, 1], 3, 2)$} fits. }
    \label{fig:AustraliaFit}
\end{figure}

\begin{figure}[p]
    \centering
    \includegraphics[width=0.5\textwidth]{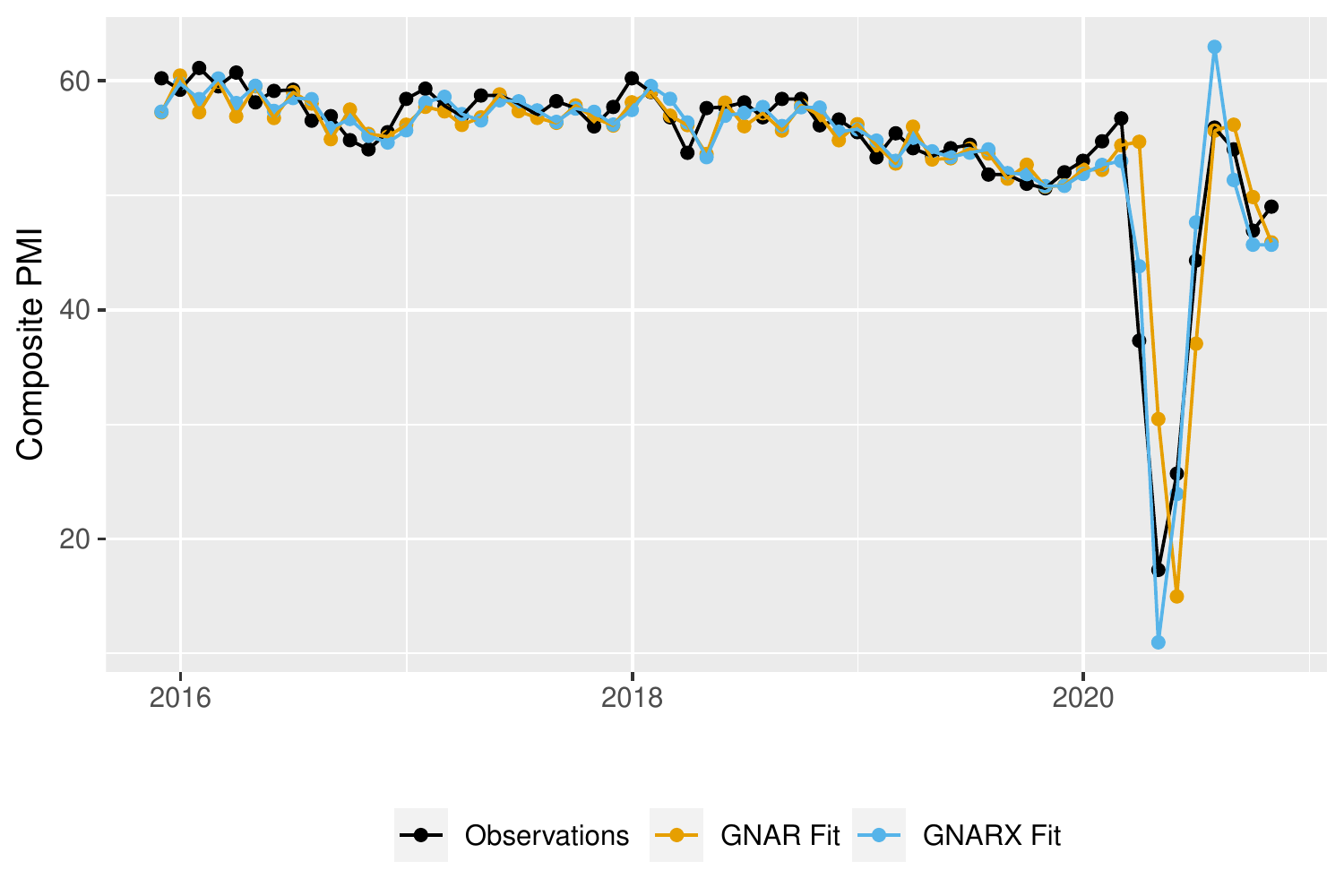}
    \caption{Ireland: Comparison of \textcolor{black}{GNARX$(5, [1, 0, 1, 0, 1])$} and \textcolor{black}{GNARX$(5, [1, 0, 1, 0, 1], 3, 2)$} fits.}
    \label{fig:IrelandFit}
\end{figure}

Figures \ref{fig:USAFit}-\ref{fig:IrelandFit} show the full set of fitted series for the global-$\alpha$ \textcolor{black}{GNARX$(5, [1, 0, 1, 0, 1], 3, 2)$} model using the fully-connected network and stringency indices only over the most recent five-year period, together with the fitted series of the GNAR model of corresponding model order.

%
%

\section{Full set of GNARX model forecasting results}
The complete set of GNARX model forecasting results is shown in Table~\ref{tab:experiment2}. \textcolor{black}{We have also included the performance of the univariate AR models with and without (current-period) external regressors, whose order was selected based on BIC optimisation without external regressors as mentioned in Section 3.2 of the main report. Again, each node may have different AR coefficients, comparable to the local-$\alpha$ GNARX models.}

As a side note, the chosen model orders suggest that the nearest-neighbour network is not an adequate representation of links between countries, losing valuable information by assuming a country's business confidence depends only on its closest trade partner. This finding holds regardless of which external regressors are included and whether or not the global-$\alpha$ assumption is invoked.

\begin{table}
\caption{\label{tab:experiment2}In-sample composite PMI forecast performance between Jan-98 and Oct-20 using GNARX models.
	MSFE=In-sample mean-squared forecasting error.}
\centering
\fbox{%
\begin{tabular}{lrrr}
\hline
Model & Model order & BIC & MSFE (SE) \\ 
\hline
\multicolumn{4}{|l|}{\textbf{Without external regressors}} \tabularnewline
\hline
Global-$\mathbf{\alpha}$ GNAR, fully-connected net & \textcolor{black}{4, (1, 1, 1, 0)} & \textcolor{black}{12653} & \textcolor{black}{7.1 (0.9)}\\
Global-$\mathbf{\alpha}$ GNAR, nearest-neighbour net & \textcolor{black}{4, (1, 1, 1, 0)} & \textcolor{black}{12895} & \textcolor{black}{7.8 (1.0)} \\
Local-$\mathbf{\alpha}$ GNAR, fully-connected net & 3, (1, 1, 1) & \textcolor{black}{12806} & \textcolor{black}{6.7 (0.9)}\\
Local-$\mathbf{\alpha}$ GNAR, nearest-neighbour net & \textcolor{black}{3, (1, 1, 1)} & \textcolor{black}{13036} & \textcolor{black}{7.4 (1.0)} \\
\textcolor{black}{AR} & \textcolor{black}{3} & \textcolor{black}{13049} & \textcolor{black}{7.5 (1.0)} \\
\hline
 & & & \\
\hline
\multicolumn{4}{|l|}{\textbf{Stringency indices only}} \tabularnewline
\hline
Global-$\mathbf{\alpha}$ GNARX, fully-connected net & \textcolor{black}{5, (1, 0, 1, 0, 1), 3} & \textcolor{black}{11323} & \textcolor{black}{4.2 (0.3)} \\
Global-$\mathbf{\alpha}$ GNARX, nearest-neighbour net & \textcolor{black}{2, (0, 0), 3} & \textcolor{black}{11409} & \textcolor{black}{4.4 (0.4)} \\
Local-$\mathbf{\alpha}$ GNARX, fully-connected net & \textcolor{black}{2, (1, 1), 3} & \textcolor{black}{11331} & \textcolor{black}{4.0 (0.3)} \\
Local-$\mathbf{\alpha}$ GNARX, nearest-neighbour net & \textcolor{black}{2, (0, 0), 3} & \textcolor{black}{11364} & \textcolor{black}{4.1 (0.3)} \\
\textcolor{black}{AR} & \textcolor{black}{2, 3} & \textcolor{black}{11364} & \textcolor{black}{4.1 (0.3)} \\
\hline
 & & & \\
\hline
\multicolumn{4}{|l|}{\textbf{COVID-19 death rates only}} \tabularnewline
\hline
Global-$\mathbf{\alpha}$ GNARX, fully-connected net & \textcolor{black}{4, (1, 1, 1, 0), 3} & \textcolor{black}{12020} & \textcolor{black}{5.5 (0.8)} \\
Global-$\mathbf{\alpha}$ GNARX, nearest-neighbour net & \textcolor{black}{4, (0, 0, 0, 0), 1} & \textcolor{black}{12324} & \textcolor{black}{6.3 (0.9)} \\
Local-$\mathbf{\alpha}$ GNARX, fully-connected net & \textcolor{black}{2, (1, 1), 3} & \textcolor{black}{12212} & \textcolor{black}{5.5 (0.8)} \\
Local-$\mathbf{\alpha}$ GNARX, nearest-neighbour net & \textcolor{black}{2, (0, 0), 3} & \textcolor{black}{12389} & \textcolor{black}{6.0 (0.8)} \\
\textcolor{black}{AR} & \textcolor{black}{2, 3} & \textcolor{black}{12389} & \textcolor{black}{6.0 (0.8)} \\
\hline
 & & & \\
\hline
\multicolumn{4}{|l|}{\textbf{With both external regressors}} \tabularnewline
\hline
Global-$\mathbf{\alpha}$ GNARX, fully-connected net & \textcolor{black}{5, (1, 0, 1, 0, 1), 3, 2} & \textcolor{black}{11175} & \textcolor{black}{4.0 (0.3)} \\
Global-$\mathbf{\alpha}$ GNARX, nearest-neighbour net & \textcolor{black}{2, (0, 0), 3, 0} & \textcolor{black}{11301} & \textcolor{black}{4.2 (0.4)} \\
Local-$\mathbf{\alpha}$ GNARX, fully-connected net & \textcolor{black}{2, (1, 1), 3, 1} & \textcolor{black}{11199} & \textcolor{black}{3.8 (0.3)} \\
Local-$\mathbf{\alpha}$ GNARX, nearest-neighbour net & \textcolor{black}{2, (0, 0), 3, 0} & \textcolor{black}{11264} & \textcolor{black}{3.9 (0.3)} \\
\textcolor{black}{AR} & \textcolor{black}{2, 3, 0} & \textcolor{black}{11264} & \textcolor{black}{3.9 (0.3)} \\
\hline
\end{tabular}}
\end{table}

%
%

\section{Generating bootstrap prediction intervals for the GNARX model}

In order to obtain asymptotically pertinent prediction intervals for GNARX model forecasts, we provide a forward bootstrap algorithm similar to the algorithm used for linear autoregressive models described in \citet{pan2016bootstrap}. These prediction intervals account for both innovations in the PMI series and estimation error in the GNARX parameters, which are assumed to be independent. The pseudo-code is provided in Algorithm \ref{alg:bootstrap}.

\begin{algorithm}
\caption{Generating $h$-step ahead bootstrap prediction intervals for the GNARX model}
\label{alg:bootstrap}
\SetKwProg{myproc}{procedure}{}{}
\myproc{}{
    \KwIn{GNARX model order ($p, \textbf{s}, \textbf{p}'$), observed data $\left\{(\textbf{X}_{1,t},...,\textbf{X}_{H,t}, \textbf{Y}_t), t=1,...,T\right\}$, forecast steps $h$, and scenario assumptions for exogenous regressors $\left\{(\textbf{X}_{1,t},...,\textbf{X}_{H,t}), t=T+1,...,T+h\right\}$, number of bootstrap samples $B$ and prediction interval coverage $(1-\alpha)100\%$.}
    \KwOut{$(1-\alpha)100\%$ prediction interval for $\textbf{Y}_{T+h}$.}
    
    $p^*\gets\max{(p,p'_1,..,p'_H)}$
    
    $\hat\gamma \gets$ GNARX parameter estimates using $\left\{(\textbf{X}_{1,t},...,\textbf{X}_{H,t}, \textbf{Y}_t), t=1,...,T\right\}$
    
    $\left\{\hat{\textbf{u}}_t, t= p^*+1,...,T\right\}\gets$ GNARX residuals using $\left\{(\textbf{X}_{1,t},...,\textbf{X}_{H,t}, \textbf{Y}_t), t=1,...,T\right\}$
    
    $b\gets0$
    
    \Repeat{$b=B$}{
        $\left\{\textbf{u}^*_t, t=p^*+1,...,T\right\}\gets$ samples with replacement from  $\left\{\hat{\textbf{u}}_t, t=p^*+1,...,T\right\}$
        
        $(\textbf{Y}_1^*,...,\textbf{Y}_p^*)\gets$ sample from $\left\{(\textbf{Y}_k,...,\textbf{Y}_{k+p^*-1}), k=1,...,T-p^*+1\right\}$
        
        $\left\{\textbf{Y}^*_t, t=p^*+1,...,T\right\}\gets$ simulations from the GNARX model with parameters $\hat\gamma$, starting values $(\textbf{Y}_1^*,...,\textbf{Y}_p^*)$, residuals $\left\{\textbf{u}^*_t, t=p^*+1,...,T\right\}$ and exogenous regressors $\left\{(\textbf{X}_{1,t},...,\textbf{X}_{H,t}), t=1,...,T\right\}$ (i.e. in the same order as the observed exogenous regressor time series)
        
        $\hat\gamma^* \gets$ GNARX parameter estimates using $\left\{(\textbf{X}_{1,t},...,\textbf{X}_{H,t}, \textbf{Y}_t^*), \forall t\right\}$
        
        $\left\{\hat{\textbf{Y}}^*_{T+1},...,\hat{\textbf{Y}}_{T+h}^*\right\}\gets$ predicted values from the GNARX model with parameters $\hat\gamma^*\text{, data }\left\{(\textbf{X}_{1,t},...,\textbf{X}_{H,t}, \textbf{Y}_t), t=1,...,T\right\}$ and scenario assumptions $\left\{(\textbf{X}_{1,t},...,\textbf{X}_{H,t}), t=T+1,...,T+h\right\}$ 
        
        $\left\{\textbf{Y}^*_{T+1},...,\textbf{Y}_{T+h}^*\right\}\gets$ simulated values from the GNARX model with parameters $\hat\gamma$, data $\left\{(\textbf{X}_{1,t},...,\textbf{X}_{H,t}, \textbf{Y}_t), t=1,...,T\right\}$,  scenario assumptions $\left\{(\textbf{X}_{1,t},...,\textbf{X}_{H,t}), t=T+1,...,T+h\right\}$ and residuals $\left\{\textbf{u}^*_t, t=T+1,...,T+h\right\}$ 
        
        $\textbf{e}_b\gets\textbf{Y}_{T+h}^*-\hat{\textbf{Y}}_{T+h}^*$
        
        $b\gets b+1$
    }
    
    $q(\alpha/2)\gets\alpha/2$-quantile of empirical distribution of $\left\{\textbf{e}_b, b=1,..,B\right\}$  (i.e. an $N$-length vector where the $i$-th entry is the quantile of the empirical distribution of the $i$-th entry of $\textbf{e}_b$ over all $b$)
    
    $q(1-\alpha/2)\gets(1-\alpha/2$)-quantile of empirical distribution of $\left\{\textbf{e}_b, b=1,..,B\right\}$
    
    $\left\{\hat{\textbf{Y}}_{T+1},...,\hat{\textbf{Y}}_{T+h}\right\}\gets$ predicted values from the GNARX model with parameters $\hat\gamma$, data $\left\{(\textbf{X}_{1,t},...,\textbf{X}_{H,t}, \textbf{Y}_t), t=1,...,T\right\}$ and scenario assumptions $\left\{(\textbf{X}_{1,t},...,\textbf{X}_{H,t}), t=T+1,...,T+h\right\}$ 
    
    \Return{$[\hat{\textbf{Y}}_{T+h}+q(\alpha/2),\hat{\textbf{Y}}_{T+h}+q(1-\alpha/2)]$}
}
\end{algorithm}

\newpage

\bibliographystyle{guy3}
\bibliography{bibliography}